\documentclass{ar-1col}
\usepackage[numbers]{natbib}
\usepackage{epsfig,rotate}
\usepackage{graphicx}
\usepackage{color}
\usepackage{ulem}
\usepackage{bm}
\usepackage[tbtags]{amsmath}
\usepackage{amssymb}
\usepackage{subfigure}
\usepackage[left=3cm,right=3cm,top=1.5cm, bottom=1.5cm]{geometry}

\jname{Xxxx. Xxx. Xxx. Xxx.}
\jvol{AA}
\jyear{YYYY}
\doi{10.1146/((please add article doi))}

\def\lsco{La$_{2-x}$Sr$_x$CuO$_4$}
\def\lbco{La$_{2-x}$Ba$_x$CuO$_4$}

\def\lnsco{La$_{1.6-x}$Nd$_{0.4}$Sr$_x$CuO$_4$}

\def\ybco{YBa$_2$Cu$_3$O$_{6+x}$}

\def\bscco{Bi$_2$Sr$_2$CaCu$_2$O$_{8+\delta}$}

\newcommand{\dq}[1][{}]{\ensuremath{\Delta_{\mbox{{\bm{Q}}}}}}

\newcommand{\rhoqs}{{\mbox{$\rho_{2{\bf P}}$}}}
\newcommand{\rhoq}{{\mbox{$\rho_{{\bf Q}}$}}}

\newcommand{\rhoqx}{{\mbox{$\rho_{2{\bf P}_x}$}}}
\newcommand{\rhoqy}{{\mbox{$\rho_{2{\bf P}_y}$}}}

\newcommand{\rhoqxyn}{{\mbox{$\rho_{{\bf P}_x-{\bf P}_y}$}}}

\newcommand{\rhoqxyp}{{\mbox{$\rho_{{\bf P}_x+{\bf P}_y}$}}}
\newcommand{\rhoqij}{{\mbox{$\rho_{{\bf P}_i-{\bf P}_j}$}}}

\newcommand{\sq}{{\mbox{$M^z_{\bf Q}$}}}
\newcommand{\sqxyn}{{\mbox{$M^z_{{\bf P}_x-{\bf P}_y}$}}}
\newcommand{\sqxyp}{{\mbox{$M^z_{{\bf P}_x+{\bf P}_y}$}}}
\newcommand{\sqij}{{\mbox{$M^z_{{\bf P}_i-{\bf P}_j}$}}}

\newcommand{\Dpx}{{\mbox{$\Delta_{{\bf P}_x}$}}}
\newcommand{\Dpy}{{\mbox{$\Delta_{{\bf P}_y}$}}}
\newcommand{\Dnx}{{\mbox{$\Delta_{-{\bf P}_x}$}}}
\newcommand{\Dny}{{\mbox{$\Delta_{-{\bf P}_y}$}}}
\newcommand{\Dpq}{{\mbox{$\Delta_{{\bf P}}$}}}
\newcommand{\Dnq}{{\mbox{$\Delta_{-{\bf P}}$}}}

\newcommand{\Dnqc}{{\mbox{$\Delta^*_{-{\bf P}}$}}}

\newcommand{\Dpqi}{{\mbox{$\Delta_{{\bf P}_i}$}}}
\newcommand{\Dnqi}{{\mbox{$\Delta_{-{\bf P}_i}$}}}
\newcommand{\Dpqj}{{\mbox{$\Delta_{{\bf P}_j}$}}}
\newcommand{\Dnqj}{{\mbox{$\Delta_{-{\bf P}_j}$}}}

\newcommand{\Dnqic}{{\mbox{$\Delta^*_{-{\bf P}_i}$}}}
\newcommand{\Dpqjc}{{\mbox{$\Delta^*_{{\bf P}_j}$}}}

\newcommand{\Dd}{{\mbox{$\Delta_d$}}}

\newcommand{\rv}{{\bf r}}

\newcommand{\eps}{{\varepsilon}}

\newcommand{\pv}{{\bf p}}

\newcommand{\jv}{{\bf j}}

\newcommand{\oh}{{\frac{1}{2}}}

\newcommand{\cH}{{\mathcal H}}

\def\rf#1{(\ref{#1})}

\newcommand{\as}{a_s}

\newcommand{\phdag}{{\phantom{\dagger}}}

\newcommand{\bk}{{\bf k}}
\newcommand{\bq}{{\bf q}}

\newcommand{\bp}{{\bf p}}

\newcommand{\grad}{{\bm{\nabla}}}

\newcommand{\ch}{\hat{c}}

\newcommand{\be}{\begin{equation}}
\newcommand{\ee}{\end{equation}}
\newcommand{\bea}{\begin{eqnarray}}
\newcommand{\eea}{\end{eqnarray}}
\newcommand{\bse}{\begin{subequations}}
\newcommand{\ese}{\end{subequations}}

\newcommand{\fermiint}{g}

\def\rf#1{(\ref{#1})}

\input{epsf}

\begin{document}

\markboth{Agterberg et al.}{The Physics of Pair Density Waves}

\title{The Physics of Pair Density Waves: Cuprate Superconductors and Beyond}

\author{Daniel F. Agterberg,$^1$ J.C. S\'eamus Davis$^{2,3}$, Stephen D. Edkins$^{4}$, Eduardo Fradkin,$^5$  Dale J. Van Harlingen$^6$, Steven A. Kivelson$^7$, Patrick A. Lee$^8$, Leo Radzihovsky$^9$, John M. Tranquada$^{10}$, and Yuxuan Wang$^{11}$
\affil{$^1$Department of Physics, University of Wisconsin-Milwaukee, Milwaukee, Wisconsin 53201, USA}
\affil{$^2$Clarendon Laboratory, University of Oxford, Oxford OX13PU, United Kingdom}
\affil{$^3$Department of Physics, University College Cork, Cork T12 K8AF, Ireland}
\affil{$^4$Department of Applied Physics, Stanford University, Stanford, California 94305,
USA}
\affil{$^5$Department of  Physics and Institute for Condensed Matter Theory, University of Illinois at Urbana-Champaign, Urbana, Illinois 61801, USA}
\affil{$^6$Department of  Physics, University of Illinois at Urbana-Champaign, Urbana, Illinois 61801, USA}
\affil{$^7$Department of  Physics, Stanford University, Stanford, California 94305,
USA}
\affil{$^8$Department of  Physics, Massachusetts Institute of Technology, Cambridge,
Massachusetts 02139, USA}
\affil{$^9$Department of  Physics and Center for Theory of Quantum Matter, University of Colorado, Boulder, Colorado 80309, USA}
\affil{$^{10}$Condensed Matter  Physics \& Materials Science Division, Brookhaven National Laboratory, Upton, New York 11973, USA}
\affil{$^{11}$Department of  Physics, University of Florida, Gainesville, Florida 32611, USA}
}

\begin{abstract}
We review the physics of pair density wave (PDW) superconductors. We begin with  a  macroscopic description that emphasizes order induced by PDW states, such as charge density wave, and discuss related vestigial states that emerge as a consequence of partial meting of the PDW order. We review and critically discuss the mounting experimental evidence for such PDW order in the cuprate superconductors, the status of the  theoretical microscopic description of such order, and  the current debate on whether the PDW is a ``mother order''  or another competing order in the cuprates. In addition, we give an overview of the weak coupling version of PDW order, Fulde-Ferrell-Larkin-Ovchinnikov states, in the context of cold atom systems, unconventional superconductors, and non-centrosymmetric and Weyl materials.  
\end{abstract}

\maketitle

\tableofcontents

\section{INTRODUCTION}
A pair density wave (PDW) is a superconducting state in which the order parameter varies periodically as a function of position in such a way that its spatial average vanishes.  It is a  phase of matter defined in terms of broken symmetries  \cite{Larkin-1964,Fulde-1964,Himeda-2002,Berg-2007,Agterberg-2008,Berg-2009,Lee-2014}.
In this review, we characterize the macroscopic properties of such a state precisely and discuss the status of the (incomplete) theoretical understanding of the sorts of lattice-scale interactions that give rise to it (the ``mechanism''), how it is distinct from other phases with which it shares certain features, and the way in which the  partial melting of the PDW can give rise to daughter phases with a variety of patterns of vestigial order.  In addition to its intrinsic interest, there is now   evidence 
that significant (although probably not long-range correlated) PDW order exists in at least some regions of the 
phase diagram of the cuprate high temperature superconductors. We critically review this evidence, and then speculate on the  broader significance of these sightings to the understanding of broader issues  in the physics of the cuprates.   We also discuss  other strongly correlated systems, including other unconventional superconductors, certain cold atom systems, and carefully engineered mesoscopic devices, in which PDW and PDW-related states play an important role.

In a time-reversal- and inversion-invariant  Fermi liquid (FL), the  superconducting (SC) susceptibilities $\chi_{sc}({\bm q}, T)$   at  ${\bm q} = {\bm 0}$ diverge as the temperature  lowers towards a critical value.   Under some circumstances, there can be a local maximum in $\chi_{sc}({\bm q})$ at non-zero ${\bm q}$, but it is always smaller than $\chi_{sc }({\bm 0})$. Weak breaking of time-reversal symmetry can change the situation.  For example, in the presence of a  Zeeman magnetic field, the $T\to 0$ divergence of $\chi_{sc}({\bm 0})$ is cut off with the result that the maximum of $\chi_{sc}$ can occur at a wavenumber with  $|{\bm q}| \sim \epsilon_Z/v_F$ where the Zeeman energy is $\epsilon_z=g\mu_B H$ and $v_F$ is the Fermi velocity. While there is no longer a strict instability to a superconducting state at arbitrary weak interactions, under some circumstances such a system can form a finite ${\bm q}$ superconducting phase at low $T$.  This is the long-sought  Fulde-Ferrell-Larkin-Ovchinnikov (FFLO) phase \cite{Larkin-1964,Fulde-1964}, which (as we  briefly review) has been plausibly shown to exist in certain  materials and cold-atom systems.  

However, it has been conjectured that, for systems in which  the interactions are not weak, a PDW could occur independent of any explicit time-reversal symmetry breaking.  In that it does not necessarily break time-reversal symmetry, the PDW is thermodynamically distinct from an FFLO state; that the spatial average of the superconducting order vanishes distinguishes the PDW from a less exotic phase with simply coexisting  charge density wave (CDW) and SC order.  However, while it is straightforward to examine the phenomenological consequences of the existence of a PDW in terms of the standard methods of effective field theory, developing a microscopic theory has proven challenging. It is not favored at weak coupling and a treatment of the strong coupling problem faces obvious challenges.  In this review we focus on the phenomenological aspects and comment briefly on the microscopic models in section 5.

Thus, this is  less a review of a well-settled subject than a progress report on an interesting, rapidly developing subject.  The experimental evidence -- especially in the cuprates -- is sufficiently dramatic that it cannot be ignored.  However, there is not yet the web of consistent evidence from a wide variety of experiments that one would ideally count on to establish the correctness of any particular perspective on the properties of such complex materials.  This, combined with the absence of a reliable microscopic theory means that a portion of the discussion is necessarily speculative.  I is part of what makes the subject so exciting.

\section{PHENOMENOLOGICAL THEORY}

\subsection{Phenomenological Theory: Ginzburg-Landau-Wilson formulation}
\label{GL}

The simplest cartoon of a PDW state is one in  which the gap function varies sinusoidally as $\Delta({\bf x}) \sim \cos({\bf P}\cdot {\bf x})$.  As reviewed in section \ref{sec:cuprates}, a state with this spatial dependence and a periodicity of a few lattice spacings was proposed to occur in the cuprate superconductors. This is  similar to the expected FFLO states when a Zeeman magnetic field is applied to a usual, weak-coupling, spin-singlet superconductor \cite{Fulde-1964,Larkin-1964}:  the  Larkin and Ovchinnikov (LO) state has a similar spatially modulated magnitude while  the  Fulde-Ferrell (FF) state has a constant magnitude but spatially varying phase, i.e. 
$\Delta({\bf x}) \sim  e^{i{\bf P}\cdot {\bf x}}$.\cite{Casalbuoni-2004,TormaReview2018}. As noted in the introduction, FFLO phases differ from other PDW states in that, reflecting their origin, they have finite magnetization and long, field dependent periodicities that are many times the superconducting coherence length.  However,  though it does not seem to have been emphasized in the previous literature, the same sort of induced subsidiary orders should be expected in an LO state as in a simple PDW, so we will treat the two sorts of states simultaneously here.

In general, a PDW state has secondary orders whose existence is dictated by symmetry. These induced orders have played an important role in understanding the relevance of PDW order in the cuprates \cite{Berg-2007,Agterberg-2008,Berg-2009,Berg-2009b,Lee-2014,Agterberg-2015,Wang-2018,Lee-2018,Edkins-2018}. Primary among these is charge density wave (CDW) order.
In addition to exhibiting an induced CDW order, this unidirectional PDW state also has nematic and spatially uniform charge $4e$ superconducting orders. These induced order parameters can develop long range order when the original PDW order does not; the induced order is then often called vestigial order \cite{Fradkin2015}. Observation and understanding of this induced order is central to understanding the underlying PDW state. In this section, using a  Ginzburg-Landau-Wilson (GLW) approach, we illuminate different types of PDW ground states, the accompanying induced order, and the topological excitations of these states. Because the PDW order breaks translation symmetry in addition to the usual particle number conservation, the topological excitation spectrum is richer than in usual superconductors.  Key to this section is that we assume the existence of an underlying lattice that breaks rotational symmetry. This condition is not true in the context of cold atoms and this gives rise to different physical properties which are discussed in section 5. 

\begin{figure}[hbt]
\subfigure[]{\includegraphics[width=0.45\textwidth]{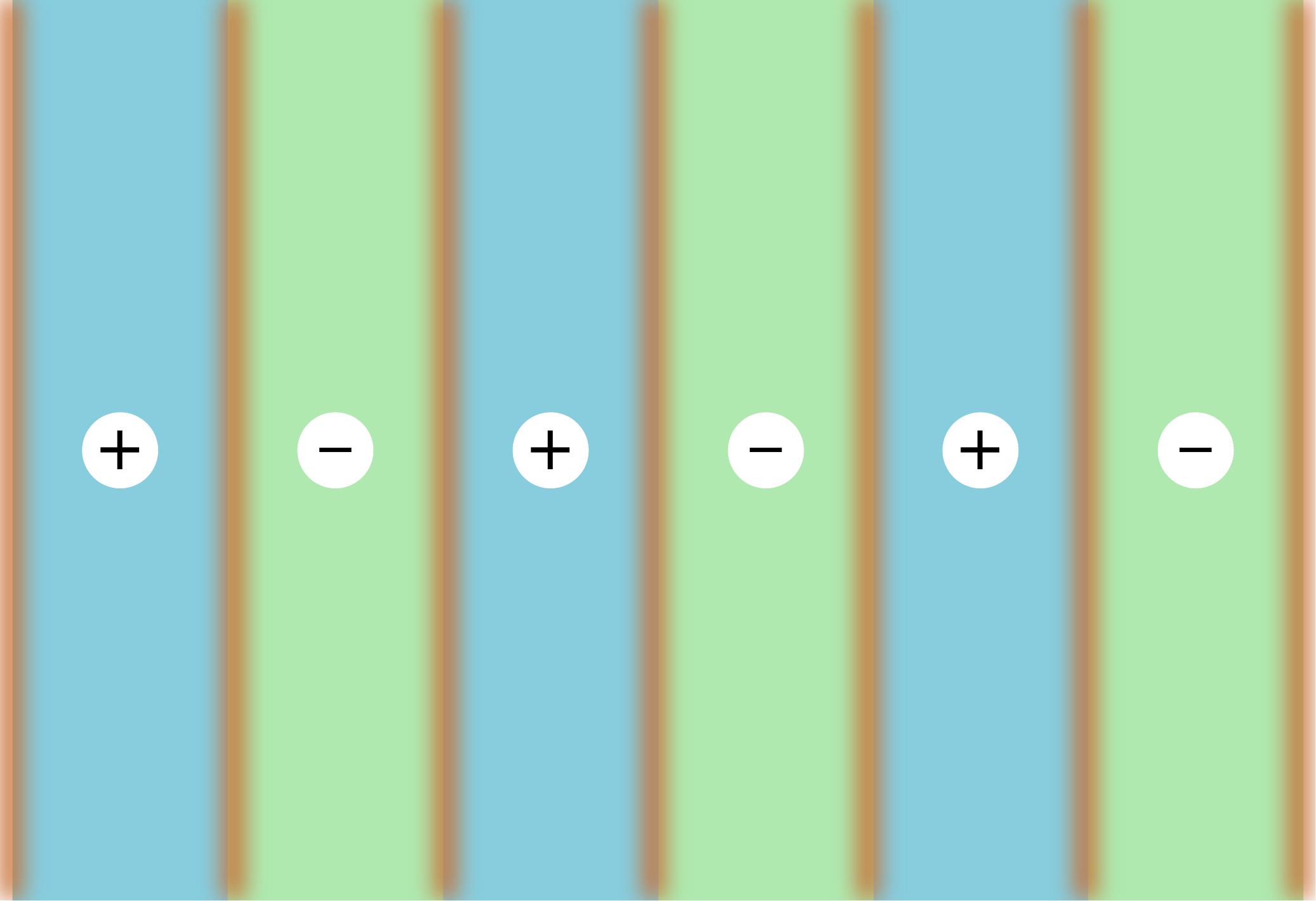}}
\subfigure[]{\includegraphics[width =0.45\textwidth]{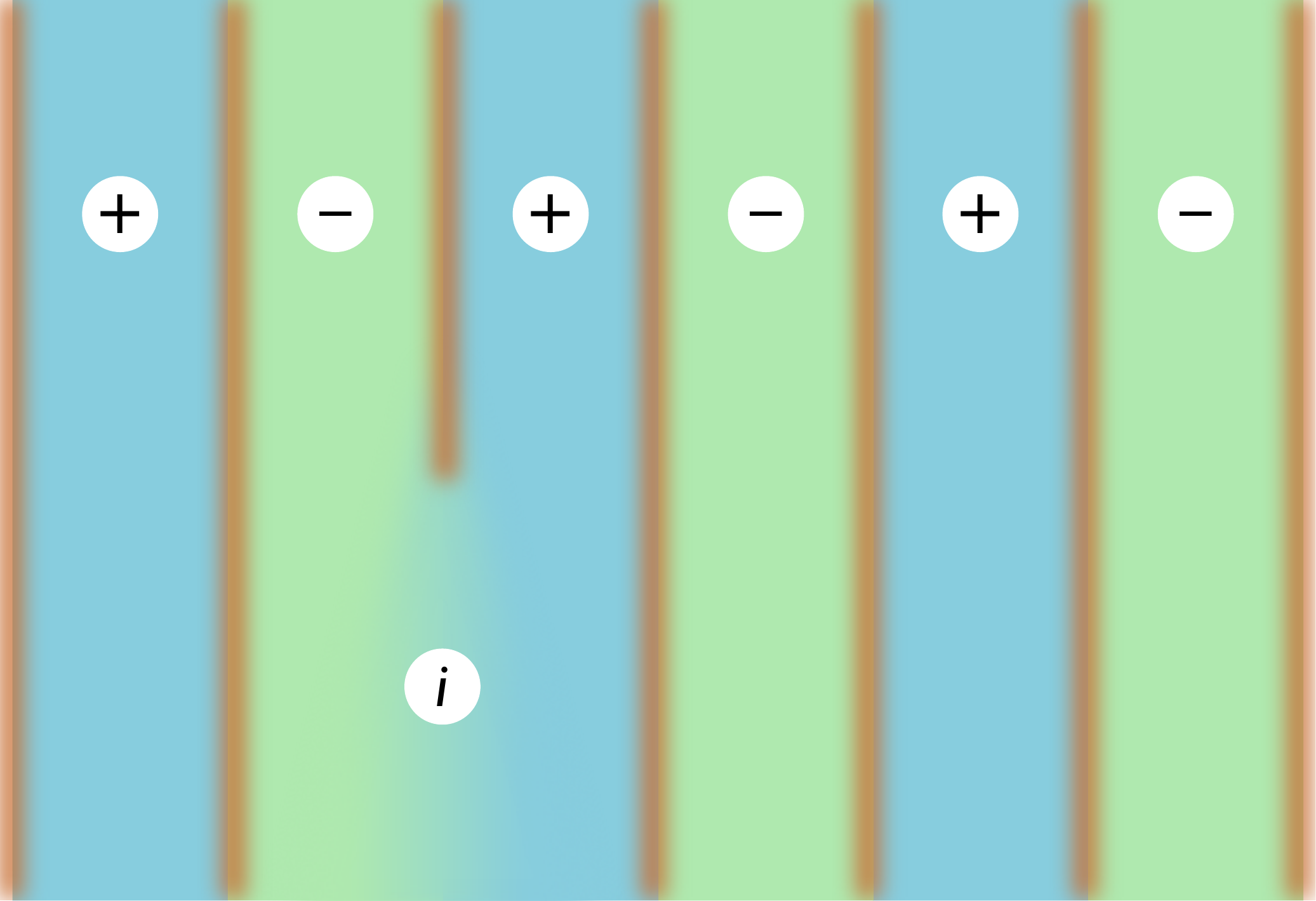}}
\caption{{Panel (a): Schematic real-space illustration of a unidirectional PDW state (blue and green stripes) with $\Delta_{\bf \pm P}$ its induced CDW state (orange stripes) $\rho_{\bf Q=2P}$. The PDW order parameter change sign in alternate domains and vanishes at the domain walls, while the local density of states is enhanced at the domain walls.  The local density of states has a CDW pattern with half the wavelength of the PDW. Panel (b): The PDW half vortex, around which the local SC phase  winds by $\pi$, bound with a CDW dislocation. Abbreviations: CDW, charge-density wave; PDW: pair-density wave; SC: superconducting.}}
\label{fig: pdwcdw}
\end{figure}

To be concrete, motivated by the cuprates, we consider a tetragonal system with square lattice sheets stacked along the $z$-axis. We will consider PDW order that exhibits spatial modulations along the in-plane $\hat{x}$ direction.  The modulations will be assumed to be incommensurate with the lattice. Tetragonal symmetry dictates that the PDW order parameter has four complex degrees of freedom with momenta $\pm {\bf P}_x$ and $\pm {\bf P}_y$, the corresponding order parameter is written as $\Delta_i=\{\Dpx,\Dpy,\Dnx,\Dny\}$. 
These order parameters are coupled to fermions via 
\begin{equation}
H_{\bm P}=\Delta_{\bm P}\int d{\bm k} F(\bm{k})c^\dagger_{{\bm k}+{\bm P}/2}c^\dagger_{-{\bm k}+{\bm P}/2}
\end{equation}
where $F(\bm{k})$ is an internal form factor. Note that the form factor is not a representation of a symmetry group and different form factors, e.g. $s$-wave, $d$-wave, can mix~\cite{Wang-2018}. Here, for concreteness, we take a $s$-wave form factor, so that $\Delta_{\bf P} = \sum_{{\bf k}} g\langle c_{-{\bf k}\downarrow}c_{{\bf k} + {\bf P}\uparrow}\rangle$ (where $g$ is the interaction constant) since different form factors do not alter the essential results discussed in this section, as discussed in Ref. \cite{Agterberg-2009}. The form of the GLW theory then follows from how these order parameters transform under various internal and spatial symmetries.
Here we do not write all such symmetry operations and refer to Ref. \cite{Agterberg-2009} for a more detailed discussion of these. However, it is worthwhile highlighting some key symmetries that help to understand the additional orders that are induced by the PDW order. In particular, under a lattice translation
${\bm T}$, the PDW order transforms as  $\Dpq\rightarrow e^{i{\bm T}\cdot {\bm P}}\Dpq$. 
Under time-reversal ($\mathcal{T}$) and parity ($\mathcal{P}$) symmetries  we have:
\begin{equation}
\Dpq\xrightarrow{\mathcal{T}} (\Dnq)^* \qquad \qquad \Dpq\xrightarrow{\mathcal{P}} \Dnq.
\end{equation} 

The GLW energy density consistent with time-reversal, parity, space group, and gauge symmetries is \cite{Agterberg-2008,Fradkin2015}
\begin{eqnarray}
{\mathcal H}=&\alpha\sum_{i}|\Dpqi|^2+\beta_1(\sum_{i}|\Dpqi|^2)^2 +
\beta_2\sum_{i<j}|\Dpqi|^2|\Dpqj|^2+\beta_3(|\Dpx|^2|\Dnx|^2 \label{free}\\&+|\Dpy|^2|\Dny|^2) 
+\beta_4[\Dpx\Dnx(\Dpy\Dny)^*+(\Dpx\Dnx)^*\Dpy\Dny].
\nonumber
\end{eqnarray}
  The parameters $\beta_i$ depend upon the specific microscopic model. Depending on which values are found for these, one of five possible ground states can be realized. These five phases include the following: the FF-type phase with only one momentum component, the FF$^*$  phase which is a bidirectional version of the FF-type phase, the LO type which include pairing with opposite momentum components: these include the unidirectional  phase, and  the bidirectional-I (II) phases which have a phase factor of $0$ $(\pi/2)$ between the two unidirectional components. These five states give rise to different patterns of induced order, providing a means to distinguish them. We now turn to these induced orders. 
\begin{table}
\begin{tabular}{|c|c|c|}
  \hline
  Phase & $(\Dpx,\Dpy,\Dnx,\Dny)$ & Induced Orders\\
  \hline
  FF-type& $(e^{i\phi_1},0,0,0)$ &  $l_x$, $\epsilon_{x^2-y^2}$ \\
  FF*-type & $(e^{i\phi_1},e^{i\phi_2},0,0)$&$l_x=l_y$\\&&  $\rhoqxyn$, $\sqxyn$  \\
 Unidirectional & $(e^{i\phi_1},0,e^{i\phi_2},0)$& $\epsilon_{x^2-y^2}$, $\Delta_{4e}$\\ &&$\rhoqx$\\ 
  Bidirectional-I & $(e^{i\phi_1},e^{i\phi_2},e^{i\phi_3},e^{i[\phi_1+\phi_3-\phi_2]})$&$\Delta_{4e}$ \\&& $\rhoqx,\rhoqy$,$\rhoqxyn,\rhoqxyp$\\
 Bidirectional-II & $(e^{i\phi_1},ie^{i\phi_2},e^{i\phi_3},ie^{i[\phi_1+\phi_3-\phi_2]})$& $\Delta_{4e}$\\ &&$\rhoqx,\rhoqy$ $\sqxyn,\sqxyp$\\
  \hline
\end{tabular}
\caption{{\bf PDW Ground States.} Distinct PDW
ground states and accompanying induced orders. In
the third column, other modes can be found by using the
relationships $\rho_{\bf Q}=(\rho_{-{\bf Q}})^*$ and $M^z_{\bf
Q}=(M^z_{-{\bf Q}})^*$.} 
\label{tab1}
\end{table}
\vglue 0.2 cm 

\noindent {\it Induced Order Parameters}

In the context of the cuprates,  Ising nematic and CDW order have been the most important of the induced orders. Since the induced CDW order appears often in this review,  we discuss it first. CDW order ($\rhoq$) preserves time-reversal and breaks translation symmetry and, given two superconducting order parameters with different pair density momenta, the CDW order is generally induced as   \cite{Agterberg-2008,Berg-2009,Lee-2014,Radzihovsky-2009}
\begin{equation}
\rhoqij \propto (\Dpqi \Dpqjc + \Dnqj \Dnqic)
\end{equation}
where the second term follows from time-reversal invariance.  
Two important examples of this are i) in the unidirectional PDW state $\rhoqs \propto \Dpq\Dnqc$ ({see Fig.~\ref{fig: pdwcdw}a for a real space illustration; }note time-reversal symmetry transforms this term into itself) and ii) when one of the superconducting order parameters is translation invariant, then $\rho_{{\bf P}}\propto (\Delta_0 \Dnqc + \Dpq \Delta_0^*)$ where $\Delta_0$ is a usual translation invariant superconducting order. The PDW ground states that are most often discussed with respect to the cuprates are the unidirectional  and the bidirectional-II PDW states. This is because these states exhibit CDW order at momentum 
$2{\bf P}_x$ and none at momentum ${\bf P}_x+{\bf P}_y$, which is consistent with experiment. 

Now we turn to the other induced orders. Ising nematic order ($\epsilon_{x^2-y^2}$) that breaks rotational 4-fold symmetry is given by 
$\epsilon_{x^2-y^2} \propto (|\Dpx|^2 + |\Dnx|^2 - |\Dpy|^2 -|\Dny|^2)$ \cite{Berg-2009,Radzihovsky-2009}. Magnetization-density wave (MDW) order ($\sq$),  which arises from spatially modulated orbital loops currents 
is given by $\sqij \propto i(\Dpqi \Dpqjc - \Dnqj\Dnqic)$ \cite{Agterberg-2008,Lee-2014}.  Translation invariant charge-4e superconducting order ($\Delta_{4e}$) is given by $\Delta_{4e}\propto \Dpq \Dnq$ \cite{Radzihovsky-2009, Berg-2009b}. Uniform charge-4e superconducting order, a state with $hc/4e$ flux quantization, can occur if the CDW stiffness of the PDW state is sufficiently weak \cite{Berg-2009b}, or in continuum systems (where the stripe order melts at any temperature) \cite{Radzihovsky_2011,Barci-2011}. It may also arise in PDW states with quenched disorder \cite{Senthil-2015}.
Loop current order ($l_i$), which is odd under both time-reversal and parity symmetries, but preserves  translation symmetry is given  by $l_i \propto (|\Dpqi|^2-|\Dnqi|^2)$ \cite{Agterberg-2015}. The induced orders appearing in the different PDW ground states are shown in Table~\ref{tab1}.

\vglue 0.2 cm 

\noindent {\it Topological excitations}

Just as gauge invariance of the action  implies the existence of vortices in usual superconductors,  gauge and  translation invariance of the Hamiltonian in Eq.~\eqref{free} imply topological excitations in PDW superconductors. Due to breaking of translation symmetry, these PDW topological excitations can exhibit properties that are quite different from usual superconducting vortices. This is reflected in the appearance of multiple $U(1)$ symmetries that appear in most PDW ground states. Gauge and translation symmetries allow PDW ground states to have up to three $U(1)$ symmetries (one from usual gauge invariance and one each from translation invariance along the $\hat{x}$ and $\hat{y}$ directions). In Table~\ref{tab1}, these $U(1)$ symmetries are given by the phase factors $\phi_i$, which are are not determined by minimizing the action.  These undetermined phases lead to Goldstone modes, in particular new modes associated with translational symmetry breaking. These modes are discussed in more detail in section \ref{sec:fflo}.  A winding in one or some of these $\phi_i$ gives rise to the topological excitations.

Physics associated with topological excitations is discussed in detail in Refs. \cite{Agterberg-2008,Lee-2014,Berg-2009b,Radzihovsky-2009}. Here we focus on the unidirectional (LO) state, which is sufficient to capture the key new physical properties associated with these vortices. Writing 
$(\Dpx,\Dpy,\Dnx,\Dny)=\Delta(e^{i\phi_1},0,e^{i\phi_2},0)/\sqrt{2}$, we allow
$(\phi_1,\phi_2)$ to have a phase winding of $(n,m)$ times $2\pi$
respectively. These we  call these $(n,m)$ vortices.
These vortices are most simply described by the following London theory
\begin{equation}
\mathcal{H}_L=\frac{1}{2} \sum_{i=x,y,z}\Big\{\rho_{s,i}[(\nabla_i
\phi_1-2eA_i)^2+(\nabla_i \phi_2-2eA_i)^2]+B_i^2\Big\}
\label{free2}
\end{equation}
where $\rho_{s,i}$ give the superfluid stiffness along these three directions (it is generally anisotropic) and the magnetic field $B=\nabla \times A$. Eq.~\eqref{free2} gives rise to a supercurrent with components {$J_i\propto\rho_{s,i}[\nabla_i
(\phi_1+\phi_2)/2-2eA_i]$}.  Far from the core of the vortex, the
minimum energy configuration has non-zero supercurrent, and a contour
integration of this supercurrent then implies that the flux enclosed in  a $(n,m)$
vortex is $(n+m)\Phi_0/2$, {$n$ and $m$ being the winding number in $\phi_1$ and $\phi_2$}. Consequently, $(1,0)$ vortices enclose half the usual flux quantum.~\cite{Agterberg-2008} In this description, $(1,1)$ vortices are the usual single flux quantum Abrikosov vortices. Eq.~\eqref{free2} shows
that these usual superconducting vortices typically have the lowest energy because
the phase winding can be completely screened by the vector
potential, implying that they have a finite energy per unit length. Fractional or zero-flux vortices have an energy per unit length that diverges as the
logarithm of the cross sectional system size.

An examination of the induced CDW order near a $\Phi_0/2$ $(1,0)$ vortex sheds insight into their physical origin. In particular, the relationship $\rhoqs \propto \Dpq\Dnqc$  reveals that a dislocation appears in the CDW order due to  the phase winding in $\Dpx$. Since this CDW order has half the periodicity of the
PDW order, a dislocation in the CDW order corresponds to half a dislocation in the PDW order. Consequently,  the half-flux vortex can be seen as a half-dislocation combined with a $\pi$ phase winding in the PDW order {[see Fig.~\ref{fig: pdwcdw}(b)]}. The
$\Phi_0/2$ vortices  of the other PDW  phases have a similar origin.
A generic prediction is that a PDW superconducting vortex
containing a half-flux quantum will be pinned to a
dislocation in the induced CDW order.

These $(n,m)$ vortices can have some notable physical consequences \cite{Agterberg-2008,Lee-2014,Berg-2009b,Radzihovsky-2009}. One example is in fluctuation driven vortex physics in two dimensions \cite{BKT-1,BKT-2,Jose-1977}. In particular, it possible that the lowest energy vortex is not a $\Phi_0/2$ $(1,0)$ vortex but either a $(1,1)$ Abrikosov vortex or a $(1,-1)$ PDW dislocation.  At a vortex unbinding transition, the lowest energy vortices will proliferate, leading to a phase that no longer has PDW order, but has vestigial order in one of the induced order parameters. In the case that the $(1,1)$ Abrikosov vortices proliferate (these can be the lowest energy vortices due the presence of the vector potential), the resultant phase will have only CDW order with wavevector twice the PDW wavevector \cite{Chen-2004-2,Agterberg-2008}. In the case that $(1,-1)$ PDW dislocations proliferate, the resultant phase will be a charge $4e$ superconductor \cite{Berg-2009b,Radzihovsky-2009}. As discussed in section \ref{sec:fflo}, this case is particularly relevant in the context of cold atoms because of rotational symmetry.
\vglue 0.2 cm 

\subsection {Coupling of PDW to uniform superconducting order}

The interpretation of a key cuprate experimental result \cite{Edkins-2018} in the context of PDW order requires an understanding of the coexistence of PDW order with usual superconducting d-wave order ($\Dd$). Here we present the simplest coexistence term that allows this to be addressed in zero magnetic field (an extension to finite field will be discussed later).  In zero field, 
the lowest order coupling term is
given by 
\begin{align}
{\mathcal H}_c=&\beta_{c1} \sum_i |\Dd|^2|\Dpqi|^2 +
\beta_{c2}[\Dd^2(\Dpx\Dnx+\Dpy\Dny)^*\nonumber\\
&+(\Dd^2)^*(\Dpx\Dnx+\Dpy\Dny)].
\label{PDWdc}
\end{align}
A key feature of this coupling is that
the $\beta_{c2}$ term can always be made negative by the correct choice of the relative phases between the PDW and d-wave orders. This has two consequences.  The first is that this coupling prefers the unidirectional (LO) or bidirectional-I PDW states. The second relates  to  the observation that coexisting PDW and uniform $d$-wave orders imply either the appearance of CDW $\rho_{\bf P}$ or MDW $M^z_{\bf P}$ order at the {\it same} momentum  as the PDW order. Whether it is CDW or MDW order depends upon the sign of the coefficient $\beta_{c2}$; if this is positive, then MDW will appear, if it is negative then CDW order will appear. Note that the coupling term $\beta_{c2}$ locks the phase of the uniform $\Delta_d$ order to the unidirectional (LO) and bidirectional PDW phases and consequently, half-flux quantum vortices will not longer exist.  Even if in the ground state $\Delta_d$ and $\Delta_{\bf P}$ do not coexist due to their competition, it is still possible that they coexist near vortices of either order, where the ground state order parameter is locally suppressed. We refer the reader to Sec. 3.2 for a review of recent experimental evidence of PDW near SC vortex halos.

There are two further topics that are not discussed here but deserve some attention. The first is the phenomenological response of the PDW order to static impurities and the second is the consequence of commensurate PDW order.  Non-magnetic impurities do not couple directly to the PDW order but couple indirectly to the PDW through the induced CDW order \cite{Berg-2009,Senthil-2015}. The primary consequence of this is that for weak disorder, one expects a destruction of the PDW order   since the induced CDW order will be disordered on the Imry-Ma length scale (thereby removing the PDW periodicity on long scales), resulting in vestigial nematic order \cite{Nie-2013} and  charge $4e$ superconducting order $\Delta_{4e}$ \cite{Senthil-2015}. In addition, disorder can locally nucleate induced orders that do not generically appear in the PDW ground state of interest \cite{Chan-2016}.  In contrast to the above considerations for incommensurate PDW order, in a PDW commensurate with the lattice potential, the latter gaps out the PDW phonons (the $\phi_1-\phi_2$ mode), thereby precluding half-vortex defects \cite{Agterberg-2008}.

\vglue 0.2 cm 

\subsection {Bogoliubov spectrum of a pair density wave}

Next we review some properties of the momentum-space Bogoliubov spectrum associated with a PDW and point out a number of features which makes it distinct from uniform pairing. As an illustration we initially discuss what happens to the cuprate band structure if we impose a {unidirectional} PDW order with ${\Dpx,\Dnx}$ on it. 

\begin{figure}
\begin{center}
\includegraphics[width=2in]{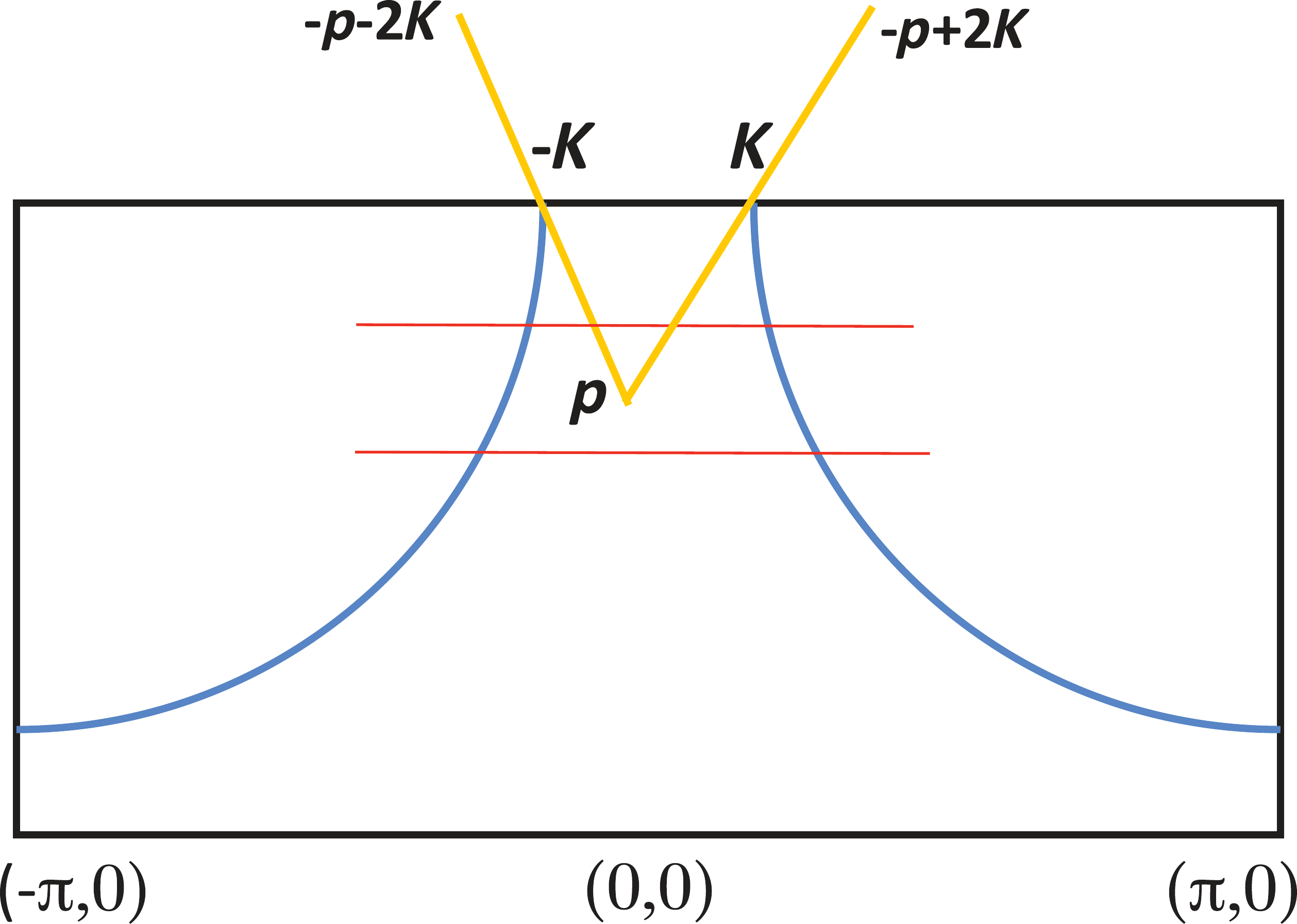}
\caption{A sketch of the top half of the Brillouin zone for the Cuprates with the Fermi surface shown in blue. Pairing of the electrons as indicated form PDW's with momenta $\pm {2\bf K}$ after Umklapp. Figure adapted from Reference \cite{{Lee-2014}}.}
\label{Fig: BZ_top}
\end{center}
\end{figure}

\begin{figure}
\begin{center}
\includegraphics[width=2in]{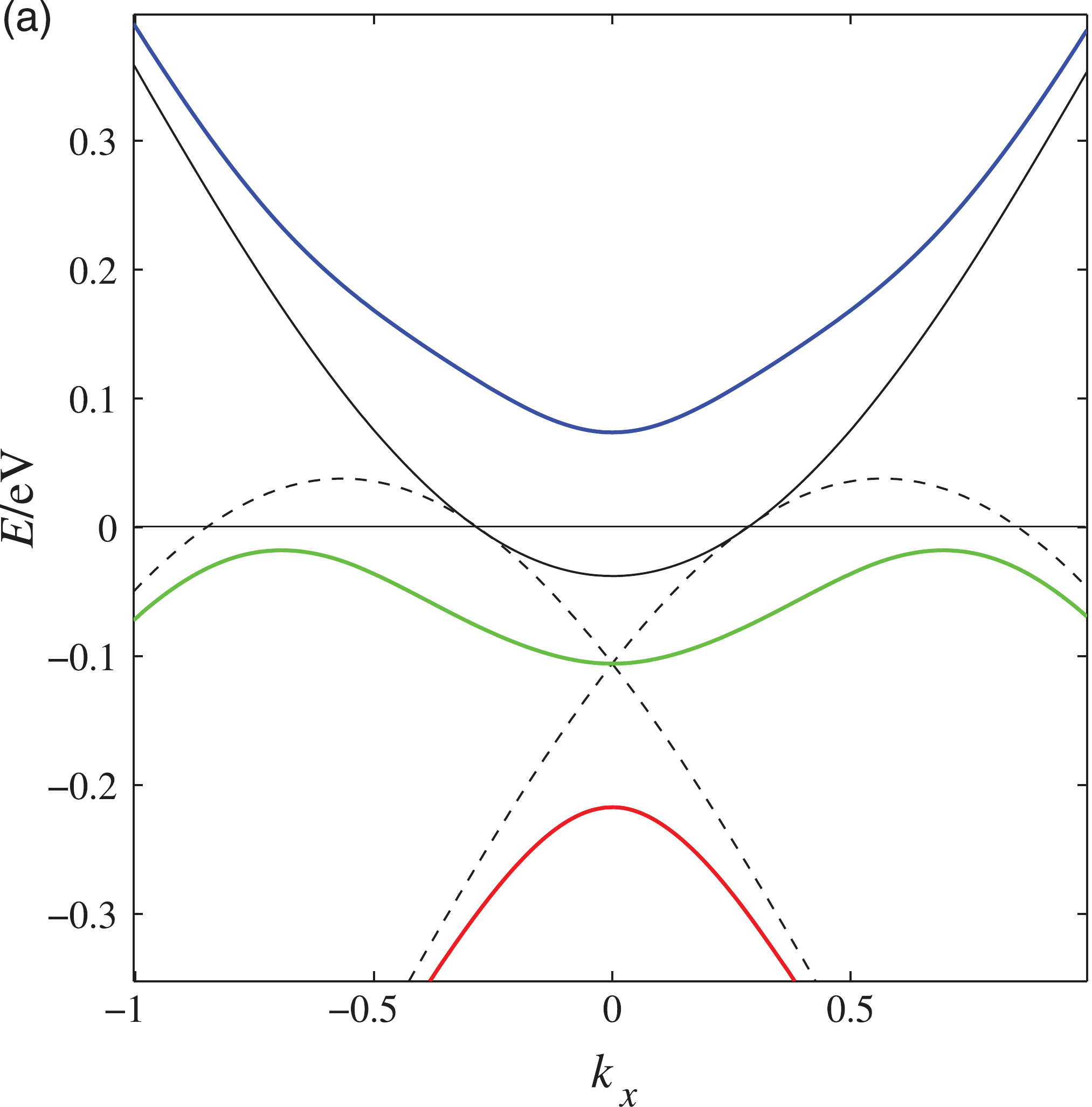}
\caption{A ``semiconductor'' picture of the formation of the Bogoliubov band as a function of $k_x$ for $k_y=\pi$. Solid black line is the original electron dispersion. Dashed lines are the hole bands. Note the shift by the PDW momenta by $\pm {\bf P}_x$.  The blue, green and red lines are the resulting hybridized Bogoliubov bands. Figure adapted from Reference \cite{{Lee-2014}}.}
\label{Fig: dispersion}
\end{center}
\end{figure}

\begin{figure}
\begin{center}
\includegraphics[width=3in]{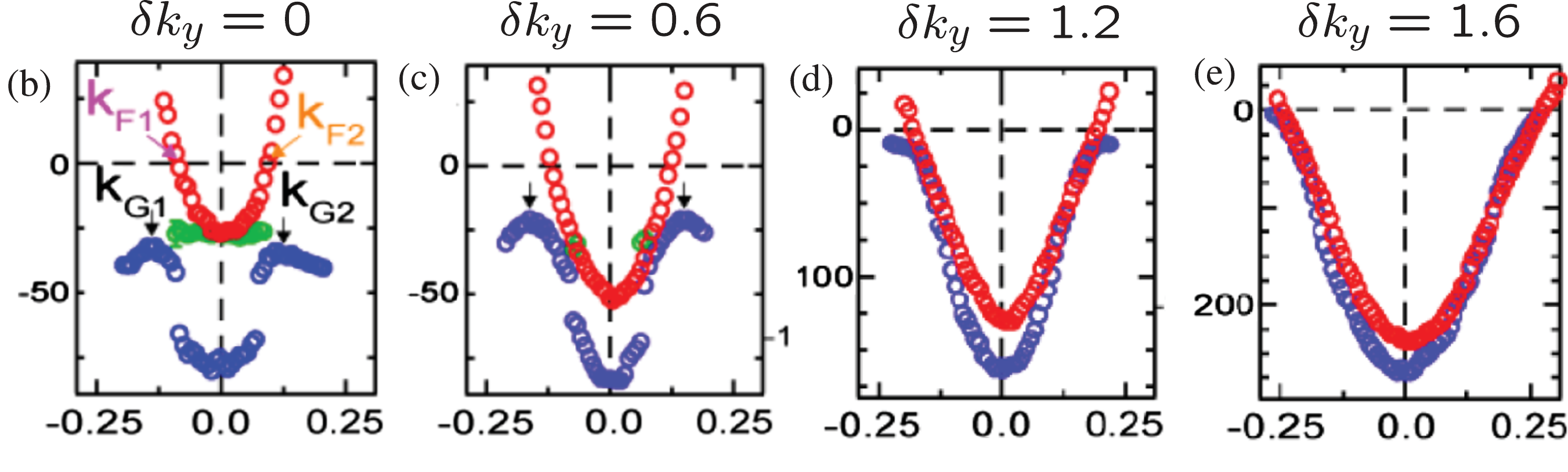}
\caption{Plots of the ARPES spectra as a function of $k_x$ for several $k_y$ measured from $\pi$. The red data points are taken in the high temperature metallic state. The blue data are in the pseudogap phase and the green dots represent additional features that appeared below the superconducting $T_c$. Figure adapted from Reference \cite{{Lee-2014}}.}
\label{Fig: He_spectrum}
\end{center}
\end{figure}

In Fig~ \ref{Fig: BZ_top}, we pair electrons with momenta $\bf p$ and $2\bf K-\bf p$ to form a PDW with momentum $\bf {P}_x=2\bf K$ and similarly for $-\bf{P}_x$. In this figure $\bf K$ was chosen to be at the Fermi surface, but the features we discuss will be similar if $\bf K$ moves away from the Fermi surface.  Fig~ \ref{Fig: dispersion}  shows a cut of the spectrum at $k_y=\pi$ and the solid black line shows the electron band $\epsilon ({\bf k}_x)$. We  illustrate the formation of the Bogoliubov spectrum with the  usual ``semiconductor'' representation where the dashed line represent the hole spectrum $-\epsilon (-{\bf k}_x + {\bf P}_x)$. The hybridization of the two bands form the blue,  green and red bands. We can estimate the weight of these bands in a photo-emission experiment by tracking how much of the original solid black band is admixed. For example, the red band is made up mainly of hole bands and will be almost invisible in angle-resolved photoemission (ARPES). The first thing to note is that unlike the uniform superconductors, the spectrum is not particle hole symmetric. This is due to the shift of the hole band by $\bf{P}_x$. An immediate consequence of this is that the top of the green band does not line up with the Fermi momentum $k_F$. Fig~ \ref{Fig: He_spectrum} shows several of scans for different $k_y$ in an experiment performed in Bi-2201.~\cite{He-2011} This material is unique in that there is a quite clear onset of the pseudogap at about 140K while the superconducting $T_c$ is rather low, so that the spectrum can be mapped out over a wide temperature range covering the high temperature metallic phase, the pseudogap phase and the SC phase.  An important point noticed by the experimentalists is that in the scan for $k_y=\pi$ , the top of the low temperature band marked by $K_G$ does not line up with the Fermi momentum $K_F$ observed a higher temperature. This was used as an argument against a fluctuating pairing phase as the origin of the pseudogap, but now we see that this objection does not apply to the PDW. Nevertheless, it is worth noting that upon averaging over k space to compute the spectrum which is observed by STS experiments,  an approximate particle-hole symmetry can be restored.

\begin{figure}[htb]
\begin{center}
\includegraphics[width=3in]{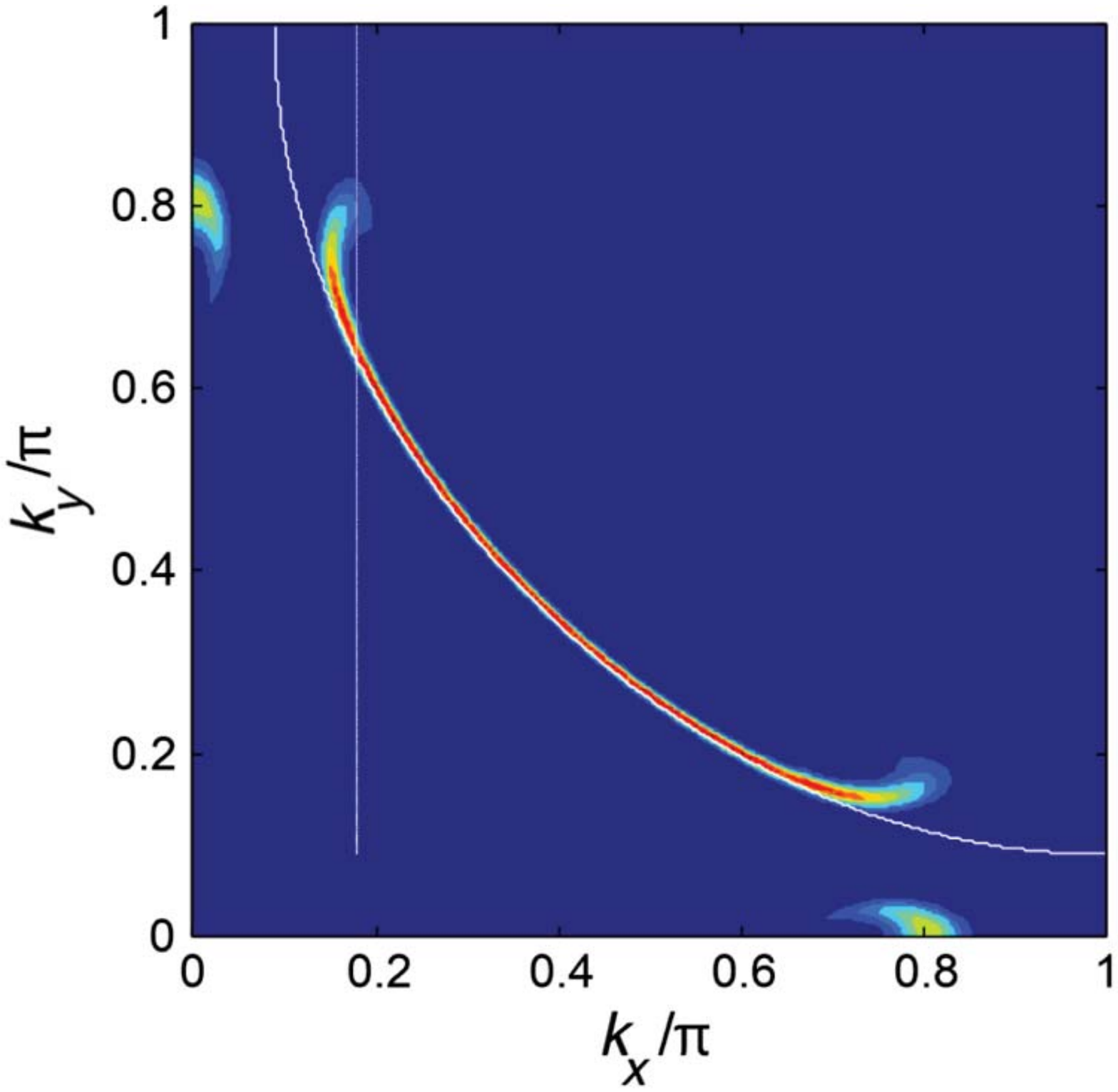}
\caption{A picture of the Fermi arc of gapless excitations in the PDW state. Figure adapted from Reference \cite{{Lee-2014}}.}
\label{Fig: fermi_arc}
\end{center}
\end{figure}

A second important observation is that the PDW spectrum naturally has lines of gapless excitations in 2D. This is in contrast with the uniform SC that can only have nodal points in 2D. While these lines of zero's form closed contours, ARPES is dominated by the electron like segments which resemble  lines of zero crossing.  This is commonly referred to as ``Fermi arcs" and was first discovered by ARPES experiments in the pseudogap regime. 
The existence of these arc-like features in the PDW spectrum was pointed out by Baruch and Orgad  \cite{Baruch-2008} and by Berg et al  \cite{Berg-2009}. Here we explain  how it comes about. Returning to Fig.~ \ref{Fig: dispersion}, imagine gradually moving away from $k_y=\pi$, following the scans indicated by the horizontal lines in Fig. \ref{Fig: BZ_top}. 
The black line in Fig. \ref{Fig: dispersion} will move down in energy, but the dashed line will move up. Upon hybridization, the top of the green line moves up in energy and eventually cross zero, resulting in a gapless excitation. With further decrease of $k_y$ from $\pi$ these crossings continue and form a closed contour of zero crossings, Most of the electron spectral weight lies on the crossing for $k_x$ closer to the origin. This results in the Fermi arc shown in Fig. \ref{Fig: fermi_arc}, the back side of the closed loop having almost no weight and being invisible invisible. Here the arc has been symmetrized assuming the co-existence of PDW along both the ${\bm P}_x$ and ${\bm P}_y$ directions. 

Finally, let us look at how the zero crossings appear at the ends of the Fermi arc. From our discussion above it is clear that in the PDW spectrum the zero crossing is formed by an occupied  band moving up in energy to meet the Fermi level. This is also what is seen experimentally in Fig~ \ref{Fig: He_spectrum}. Another candidate for the pseudogap is a CDW at momentum $\bf Q$. While this can certainly open a gap at the anti-node near $(0, \pi)$, the gap sits at a fixed momentum $k_x$ and as $k_y$ moves away from $\pi$, the occupied states only move to lower energy and remain occupied. Therefore, the only way a Fermi arc of gapless excitations can form is for an unoccupied state to move down in energy. This contrasting behavior between a PDW and CDW driven anti-nodal gap and Fermi arc was emphasized in Refs. \cite{Berg-2009,Lee-2014}. We also point out that if the PDW is bi-directional and the effect of  a composite bi-directional CDW of the type discussed in the last section is added, the Fermi arcs can be connected by the CDW's momenta to form electron-like closed Fermi pockets ~\cite{Lee-2014,Lee-2018}. This mechanism to produce an electron pocket that can explain what is seen in quantum oscillations was first proposed by Harrison and Sebastian.~\cite{Harrison-2011} The advantage of the PDW picture is that the hole pockets that remain near the anti-nodes in their picture are automatically gapped-out. Similar features in the Bogoliubov spectrum were obtained with a different version of the PDW state \cite{Wang-2015}, indicating the robustness of the PDW interpretation of the ARPES spectra. A recent paper studied the spectrum of the PDW in a $t-t'-J$ model using the self-consistent Gutzwiller approximation, and reached similar conclusions \cite{Tu-2019}.

In summary, the Bogoliubov spectrum associated with the PDW has several features which stand in strong contrast to our intuition based on that of the uniform superconductor. These include the lack of particle-hole symmetry in the spectrum and nodal lines (surfaces) in two (three) dimensions. In the context of Cuprate superconductors, many of these features are consistent with ARPES results on Bi-2201 which are very difficult to explain based on CDW, fluctuating d-wave SC, or other conventional pictures \cite{He-2011}.


\section{CUPRATES}
\label{sec:cuprates}

\subsection{\lbco\ and related cuprates}
\begin{figure}[t]
\begin{minipage}{.5\linewidth}
\centering
\includegraphics[width=2.45in]{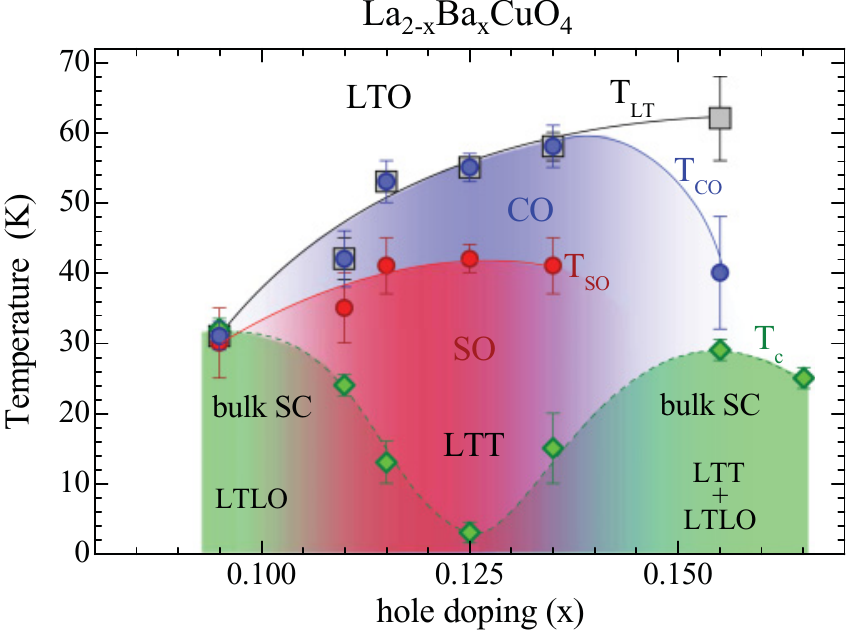}
\end{minipage}%
\begin{minipage}{.5\linewidth}
\centering
\includegraphics[width=2.45in]{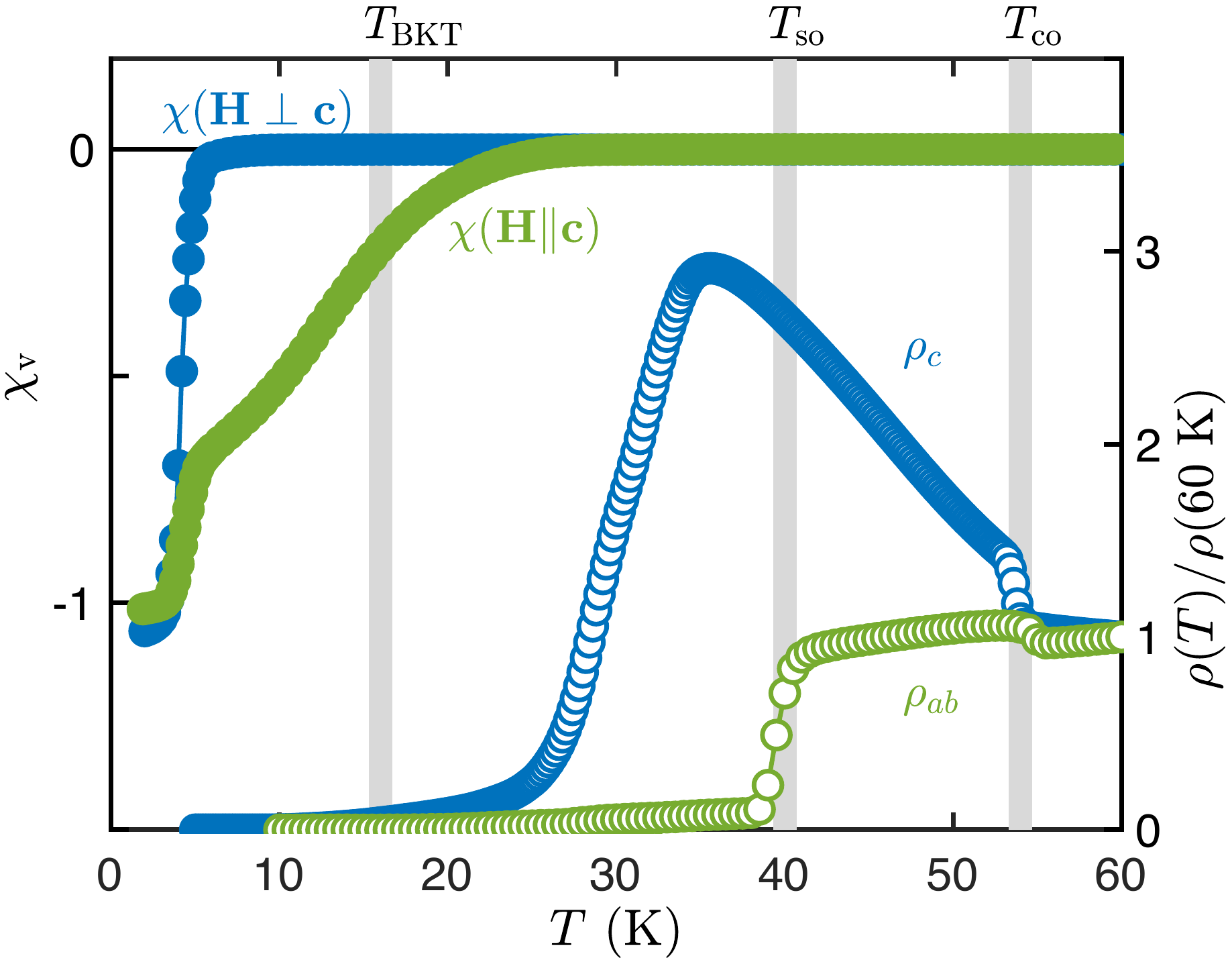}
\end{minipage}
\caption{(Left) Phase diagram for \lbco, indicating charge order (CO) below $T_{\rm co}$ (limited by the structural transition at $T_{\rm LT}$), spin order (SO) below $T_{\rm so}$, and bulk superconductivity \cite{huck11}.  (Right) Zero-field-cooled volume susceptibility (left axis, filled symbols) measured with a field of 2 G oriented parallel to the $c$-axis (probing in-plane screening) and perpendicular to $c$ (probing interlayer screening). In-plane resistivity, $\rho_{ab}$, and $c$-axis resistivity, $\rho_c$, (open symbols, right axis), in \lbco\ with $x=1/8$. Abbreviations: CO, charge order; LTLO, low-temperature less orthorhombic; LTO, low-temperature orthorhombic; LTT, low-temperature tetragonal; SC, superconducting; SO, spin order. Replotted from References \cite{Li-2007} and \cite{tran08}.}
\label{fg:lbco}
\end{figure}

The phase diagram of \lbco, shown in Fig. \ref{fg:lbco} (LBCO), has an anomalous dip in the bulk superconducting $T_c$ at $x=1/8$ \cite{mood88}, which is correlated with the appearance of charge and spin stripe orders, as detected by neutron and x-ray diffraction on single crystals \cite{fuji04,huck11}.  Pinning of the stripes to the lattice is enabled by structural  distortion within the CuO$_2$ layers, 
such that there is a preferred axis along one of the Cu-O directions which rotates by 90$^\circ$ from one layer to the next \cite{axe89,axe94}.  The relevant  distortion appears below a structural phase transition labelled $T_{\rm LT}$ in the figure.

For LBCO with $x=1/8$, careful measurements of the resistivity within the planes, $\rho_{ab}$, and perpendicular to the planes, $\rho_c$, revealed a surprising anisotropy \cite{Li-2007,tran08}.  As shown in Fig. \ref{fg:lbco}, cooling below the spin ordering temperature, $T_{\rm so}$, leads to a drop in $\rho_{ab}$ by an order of magnitude.  In contrast, $\rho_c$ continues to rise through $T_{\rm so}$, eventually turning down at a lower temperature.  The drop in $\rho_{ab}$ at $\sim40$~K appears to correspond to the onset of 2D superconducting correlations within the CuO$_2$ layers, as confirmed by measurements of anisotropic diamagnetism \cite{tran08}.  Below the transition, $\rho_{ab}$ continues to decrease, extrapolating to zero at $T_{\rm BKT}=16$~K, where nonlinear transport is observed \cite{Li-2007}, consistent with a Berezinskii-Kosterlitz-Thouless transition \cite{BKT-1,BKT-2}; $\rho_c$ remains finite down to $\sim10$~K, while bulk susceptibility indicates a bulk superconducting transition at $\sim5$~K.

Related behavior is observed in rare-earth-doped \lsco, such as \lnsco\ \cite{tran95a}, associated with the same low-temperature structural phase as in LBCO.  In particular, a study of the superconducting state via the $c$-axis optical reflectivity, measured as a function of Nd concentration (which controls the structural phase and the degree of stripe order), demonstrated a sharp decrease in interlayer Josephson coupling corresponding to the rise of stripe order \cite{taji01}.  (At low enough frequency, the layers behave as a coherent superconductor, resulting in a reflectivity of unity.  At the frequency of the Josephson plasma resonance (JPR), the interlayer coherence breaks down, and charge oscillates between the layers; the reflectivity drops below the normal-state response before recovering at higher frequency \cite{baso05}.)  The JPR results on LNSCO, and the studies of LBCO, motivated proposals that PDW order associated with stripes would cause a cancellation of the interlayer Josephson coupling, consistent with the observation of 2D superconductivity \cite{Himeda-2002,Berg-2007}.  

Somewhat weaker charge stripe \cite{crof14,tham14} and spin stripe \cite{suzu98,kimu00} order  is found in \lsco\ with $x\sim0.12$ in zero field.   For $x=0.10$, it has been observed that applying a $c$-axis magnetic field induces spin-stripe order \cite{lake02}.  Measurements of $c$-axis reflectivity on a similar sample indicate that the field causes a rapid decrease in the coherent interlayer coupling \cite{Schafgans-2010}.  Similar behavior is found in LBCO $x=0.095$ \cite{wen12}, where superconductivity in the decoupled planes survives to at least 35~T \cite{steg13}.  Evidence for bilayer decoupling has recently been reported for La$_{2-x}$Ca$_{1+x}$Cu$_2$O$_6$ \cite{zhon18}.  The field-induced decoupling is similar to the behavior found in LBCO $x=0.125$, suggesting that PDW order is present in these sample and that it is less sensitive to magnetic field than is the uniform $d$-wave order.

The loss of coherent coupling between superconducting layers can be explained by the presence of PDW order, but what do other experiments tell us?  The superconducting gap for PDW order is predicted, in weak-coupling analysis \cite{Baruch-2008}, to be large in the antinodal regions of reciprocal space, but zero along finite arcs centered on the $d$-wave nodal points.  The gap structure can be tested by ARPES.  Such measurements on LBCO $x=1/8$ do show the antinodal gap, but they also suggest a $d$-wave-like dispersion near the nodal region \cite{he09,vall06}.  At the node, however, the spectral function is broad in energy and shifted to $\sim20$~meV below the chemical potential, similar to a recent observation of a nodal gap in \lsco\ with $x=0.08$ \cite{razz13}.  Mean-field calculations of the PDW state that take account of the spin-stripe order predict such a gap \cite{Loder-2011}.   While there remains a large degree of uncertainly concerning its interpretation, the ARPES data should not be ignored and will receive further discussion in section 6.

One picture of the PDW state involves superconducting stripes that are phase locked by Josephson coupling.  That suggests that optical reflectivity measurements with the polarization in-plane but perpendicular to the stripes might yield a response similar to the JPR behavior found in $c$-axis reflectivity.  Indeed, such behavior has been observed for LBCO $x=0.125$, with reflectivity approaching one at low frequency and crossing below the normal  state response above 20~meV, for temperatures of 40 K and below \cite{home12}.  Another test is to apply a $c$-axis magnetic field strong enough to destroy the interstripe Josephson coupling.  Such an experiment on LBCO $x=0.125$ finds that, for $T<1$~K, fields above 30~T are required to eliminate all 2D coherence, resulting in a highly-resistive metallic state with approximate particle-hole symmetry, consistent with pair correlations surviving in decoupled charge stripes \cite{li19}.

The cancellation of the Josephson coupling for PDW orders that are orthogonal in adjacent layers applies for linear interactions.  If one can drive large ac currents, then it may be possible to induce a nonlinear coupling.  Cavalleri's group has demonstrated this on LBCO with $x$ slightly away from 1/8.  For $x=0.115$, the JPR is at a frequency too low  to detect; nevertheless, use of a high-intensity terahertz beam generates a JPR response at the third harmonic, even at temperatures far above the bulk $T_c$ (but below the charge-ordering temperature) \cite{raja18}.  This experiment provides intriguing evidence for the PDW state.

One of the most distinctive features of the PDW state is the spatial variation of  the order parameter between stripes in the same and adjacent planes. 
The resulting phase structure in turn suggests several approaches for exploring this exotic phase based on probing the nature of quasiparticle and Josephson tunneling into a PDW material.  Several such measurements have been carried out that exhibit evidence for the predicted PDW phase. 

Scanning tunneling microscopy (STM) on \lsco\ $x=0.12$ with the tunneling current along the $c$ axis revealed an unexpected zero-bias anomaly \cite{yuli_scanning_2007}.  Subsequent calculations in a $t-J$ model presented an explanation for this observation from a PDW state exhibiting antiphase domains \cite{Yang-2009}. Additional experiments reveled an anisotropic spatial modulation of the zero-bias peak consistent with this model \cite{yuli_spatial_2010}.

\begin{figure}[hbt]
\begin{minipage}{1\linewidth}
\centering
\includegraphics[width=3.4in]{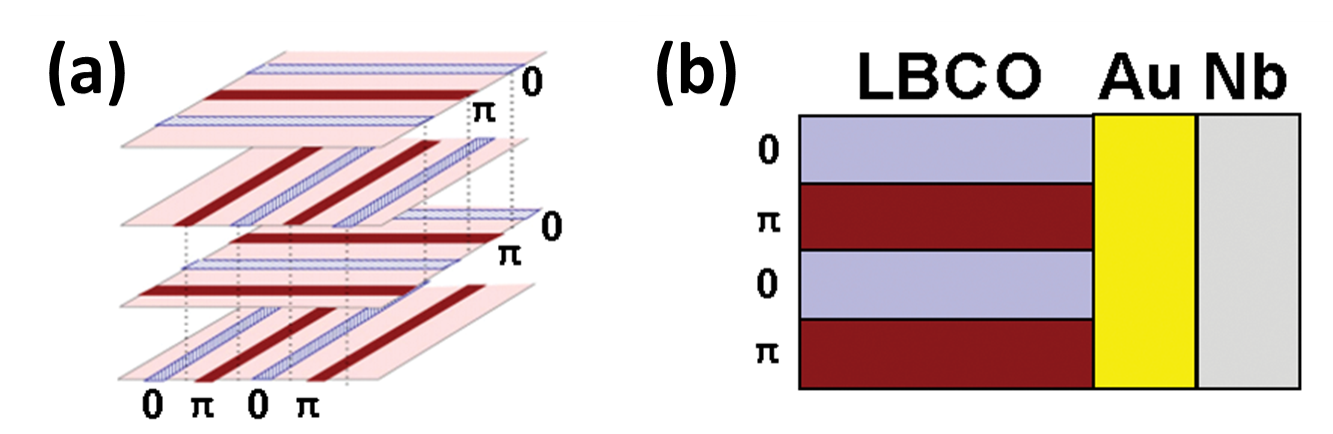}
\end{minipage}%
\caption{(a) The predicted modulation of the phase of the charge stripes in the PDW state.  This can be probed  by measuring the current-phase relation of a Josephson junction fabricated between the crystal (LBCO) and a conventional superconductor (Nb), using a Au tunneling barrier.}
\label{fg:cpr sample}
\end{figure}

For a stripe-ordered cuprate such as LBCO, where the interlayer Josephson coupling is frustrated by the orthogonal stripe orientation in neighboring layers [Fig. \ref{fg:cpr sample}(a)],  a proposed test is to reduce the cancellation by applying a magnetic field parallel to the planes; when oriented at 45$^\circ$ to the stripes, the field partially compensates for the momentum mismatch between the layers, causing an enhancement of the Josephson tunneling \cite{yang13}.  A test of this type has been done through transport measurements at low temperature and high magnetic field on single crystals of Eu- and Nd-doped \lsco\ \cite{shi19}.  In a highly dissipative regime with evidence for in-plane superconducting correlations, the ratio of the $c$-axis to the in-plane resistivity decreased as the in-plane field was increased, consistent with the predicted scenario.

An even more direct test of the PDW state is to probe the Josephson current-phase relation (CPR) of junctions between the candidate crystal and a conventional superconductor.  In the presence of a PDW state, the rapid spatially-modulated sign changes in the Josephson coupling will suppress the first-order Josephson coupling and manifest itself as a significant sin(2$\phi$) harmonic in the CPR of a junction containing LBCO (see Fig. \ref{fg:cpr sample}b).  This phenomenon has been predicted and observed in other junctions with spatially alternating critical current density  \cite{buzdin_periodic_2003,moshe_shapiro_2007,stoutimore_second-harmonic_2018,schneider_half-h/2e_2004}. Additionally, we expect the fraction of Josephson current exhibiting a sin(2$\phi$) CPR to increase with temperature as the interlayer Josephson coupling and conventional 3D superconductivity are suppressed within LBCO, giving way to an increasing proportion of spatially varying 2D superconductivity within the CuO$_2$ planes \cite{Berg-2007}.

\begin{figure}[hbt]
\centering
\includegraphics[width=0.8 \textwidth]{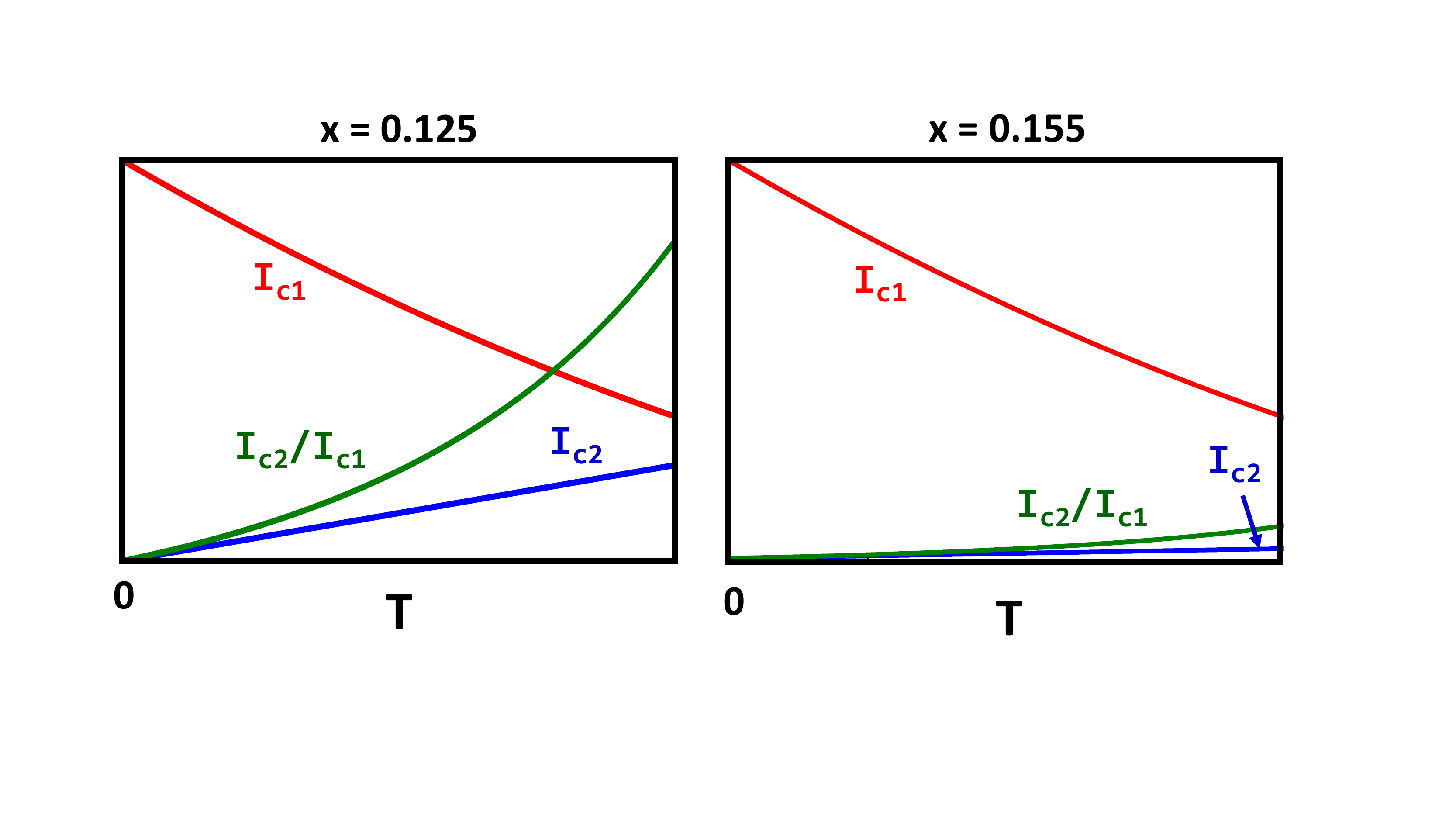}
\caption{Expected experimental signatures of the PDW state in the current-phase relation of Josephson junctions between La$_{2-x}$Ba$_x$CO$_4$ and a conventional superconductor. At x=0.125 in the PDW regime, with increasing temperature as the first-order Josephson effect decreases, we expect a growing $\sin(2\phi)$ component of the CPR arising from the phase modulation and a corresponding increase in the ratio of the $\sin(2\phi)$ to $\sin \phi$ component. In contrast, at x=0.155 near the maximum $T_c$, the $\sin (2\phi)$ and its ratio to the $\sin \phi$ component should be very small. Abbreviation: CPR, current-phase relation; FFT, fast Fourier transform.}
\label{fg:cpr}
\end{figure}

This experiment has recently been carried out by the Van Harlingen group in Urbana on crystals from Genda Gu at Brookhaven \cite{hamilton_signatures_2018}. Using both dc Josephson interferometry and anisotropic SQUID measurement techniques, they compared the CPR of \lbco-Au-Nb Josephson junctions for $x=0.155$ where the T$_c$ is maximum and at $x=0.125$. As shown in  Fig. \ref{fg:cpr}, at $x=0.155$, the CPR is nearly sinusoidal, with a nearly negligible sin(2$\phi$) component. However, at $x=0.125$, the non-sinusoidal shape of the CPR arising from the onset of a sin(2$\phi$) component is apparent. The expected temperature dependence shown in Figure 8 was observed. By using junctions fabricated straddling the corner of the crystal, it was also demonstrated that the order parameter symmetry of the crystal remains $d_{x^2-y^2}$ in the region where the T$_c$ is suppressed and the PDW state is present. 


\subsection{\bscco: Vortex Halos and Josephson Microscopy}

In this subsection we describe spectroscopic imaging scanning tunneling microscopy (SISTM) and scanned Josephson  tunneling microscopy (SJTM) studied of PDW states in the high-$T_c$ superconductor  Bi$_2$Sr$_2$CaCu$_2$O$_{8+\delta}$.

\subsubsection{Spectroscopic Imaging and Josephson STM}

SISTM \cite{Fujita2015} has become a key technique for determining electronic structure of quantum materials. At the surface of each sample, the tip-sample differential conductance for single-electron tunnelling, $\left. \frac{dI}{dV} \right|_{{\bm r},V} \equiv g({\bm r},E)$, is measured versus voltage $V=\frac{E}{e}$ and location ${\bm r} = (x,y)$. The resultant array of $g({\bm r},E)$ images is related to the density of electronic states $N({\bm r},E)$ as  
\begin{equation}
    N({\bm r},E) \propto \frac{ g({\bm r},E)}{\int^{e V_S} g({\bm r}) dE}
    \end{equation}
    where $V_{S}$ is a fixed but arbitrary voltage used to establish each tunnel junction. SJTM is a quite different technique in which 
    the magnitude of the maximum Cooper-pair tunnelling current from a superconducting tip, $|I_c ({\bm r})|$, is measured versus location ${\bm r}$, yielding an image of the density of electron pairs in the superconducting condensate.

\subsubsection{Evidence for a PDW in {\bscco} in a Magnetic Field}
In the of PDW studies, focus has recently turned to modulations of the density of single-electron states $N({\bm r},E)$ within the vortex halos \cite{Hoffman-2002,Matsuba2007,Yoshizawa2013,Machida2016} - regions of suppressed but non-zero superconductivity that surround vortex cores. Whether these modulations stem from a field-induced PDW may be studied using Ginzburg-Landau (GL) analysis. Consider a homogeneous \textit{d}-wave superconductor $\Delta_{SC}({\bm r}) = F_{d}\; \Delta_{SC}$ (where $F_d$ is  
a $d$-wave 
 a form factor) coexisting with a
 uniform PDW $\Delta_{PDW}^{{\bm P}}({\bm r}) = F_{p}\Delta_{\bm P}
 \left[ \exp(i{\bm P} \cdot {\bm r}) +  \exp(-i{\bm P} \cdot {\bm r}) \right]$ 
 with form factor 
 $F_{\rm PDW}$. The symmetry allowed $N({\bm r})$ modulations generated by interactions occur as products of these two order parameters that transform as density-like quantities. Specifically, the product $\Delta_{{\bm P}} \Delta_{SC}^* \Rightarrow N({\bm r})\propto \cos({\bm P}\cdot {\bm r})$ results in $N({\bm r})$ modulations at the PDW wavevector ${\bm P}$, while  $\Delta_{{\bm P}} \Delta_{-{\bm P}}^* \Rightarrow N({\bm r}) \propto \cos(2{\bm P}\cdot {\bm r})$ produces N$({\bm r})$ modulations occurring at ${\bm P}$. 
 Consequently, in the case in which PDW order arises 
 in a halo surrounding a vortex in a \textit{d}-wave superconductor,   there should be two sets of $N({\bm r})$ modulations at ${\bm P}$ and at $2 {\bm P}$ within each halo, with those at $2{\bm P}$ decaying with distance from the core at twice the rate as those at ${\bm P}$ (if $\Delta_{PDW}^{{\bm P}} = \Delta_{PDW}^{{\bm P}}(|{\bm r}| = 0) \; \exp(-|{\bm r}|/\epsilon)$)\cite{Agterberg-2015,Wang-2018,Lee-2018,Edkins-2018}. 

To explore these predictions, single-electron tunneling conductance $g({\bm r},E)$ was measured by Edkins \textit{et al.} \cite{Edkins-2018} for {\bscco} samples ($T_{c} \sim 88$K; $p \sim$ 17\%) at $T = 2K$. The $g({\bm r},E)$ is first measured at zero field and then at magnetic field $B = 8.25$T, in the identical field of view (FOV) using an identical STM tip. The $g({\bm r},E,B)$ and $g({\bm r},E,0)$ are registered to each other with picometer precision, and then subtracted to yield $\delta g({\bm r},E,B) = g({\bm r},E,B) - g({\bm r},E,0)$. This result is the field-induced perturbation to the density of states $\delta N({\bm r},E,B) \propto \delta g({\bm r},E,B)$. 

\begin{figure}
    \centering
    \includegraphics[width=\columnwidth]{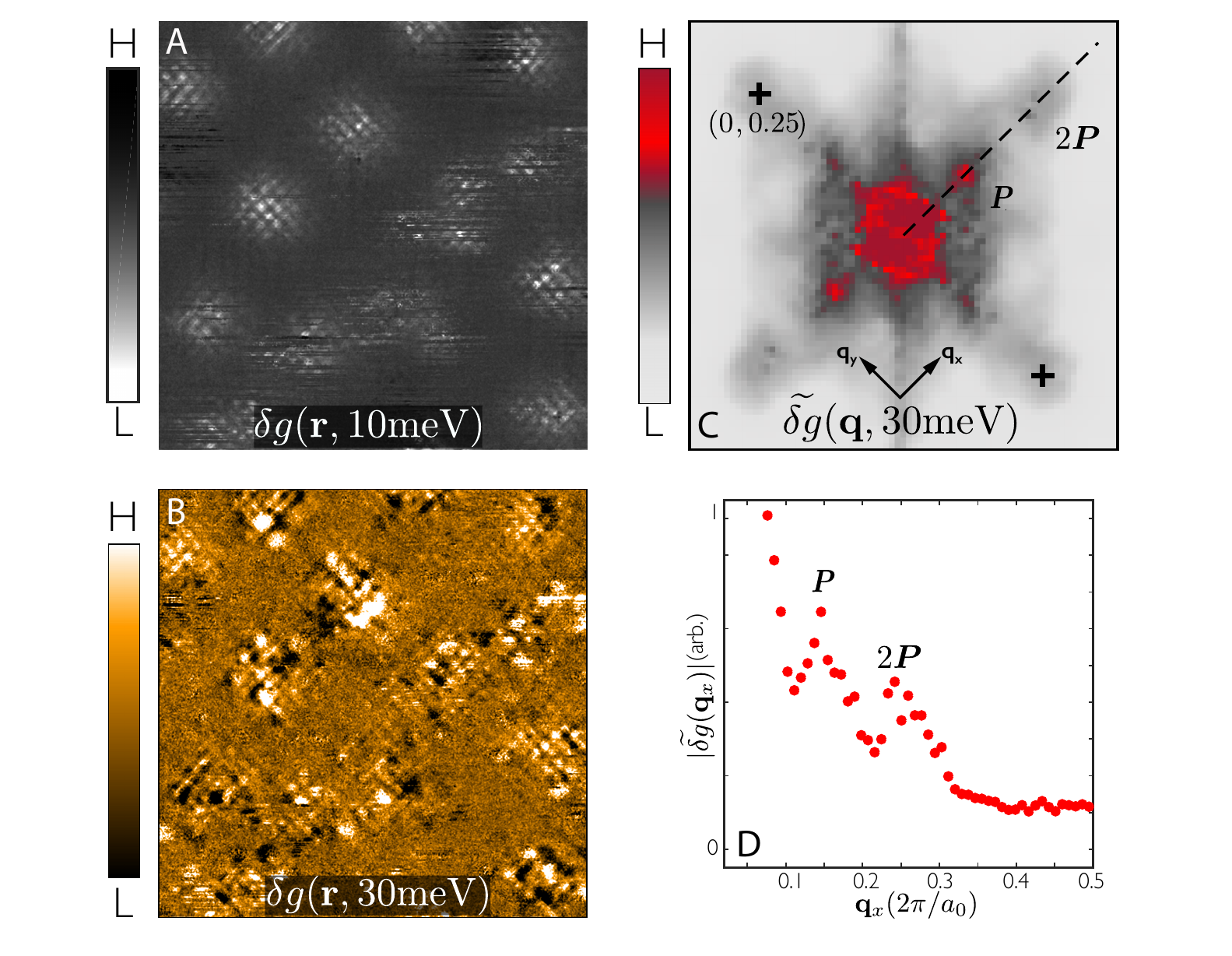}
    \caption{{\bf A} Measured $\delta g({\bm r},E,B) = g({\bm r},E = 10 \textrm{meV},B=8.25\textrm{T}) - g({\bm r},E = 10 \textrm{meV},B = 0\textrm{T})$ in a 58nm $\times$ 58nm FOV, showing typical examples of the low-energy Bogoliubov quasiparticle modulations within halo regions surrounding four vortex cores in \bscco. {\bf B} Measured field-induced modulations $\delta g({\bm r},E = 30 \textrm{meV},B) = g({\bm r},E= 30 \textrm{meV},B=8.25\textrm{T}) - g({\bm r},E = 30 \textrm{meV},B = 0\textrm{T})$ in the same 58nm $\times$ 58nm FOV. Although the vortex halos are clearly seen to occur at exactly same locations as in A, the modulations therein are radically different. {\bf C} Amplitude Fourier transform $\left |  \widetilde{\delta g}({\bm q},30 \textrm{meV}) \right| $  (square root of power spectral density) of $\delta g({\bm r},E = 30 \textrm{meV},B)$ data in B. The $\vec{q}\approx[(\pm \frac{1}{4},0);(0,\pm \frac{1}{4})]2\pi/a_0$ points are indicated by black crosses. Four sharp maxima, indicated by ${\bm P}$, occur at ${\bm q}=[{\bm P}_{x};{\bm P}_y]\approx[(\pm \frac{1}{8},0);(0,\pm \frac{1}{8})]2\pi/a_0$ whereas four broader maxima, indicated by ${2\bm P}$ occur at ${\bm q}\approx[(\pm \frac{1}{4},0);(0,\pm \frac{1}{4})]2\pi/a_0$. {\bf D} Measured $\left |  \widetilde{\delta g}({\bm q},30 \textrm{meV}) \right| $ along (0,0)-(1/2,0) showing the two distinct maxima in the field induced $N({\bm r})$ modulations, occurring at by ${\bm P} = 0.117 \pm 0.01$ and $2{\bm P} = 0.231 \pm 0.01$. Figure from Reference \cite{Edkins-2018} with permission.}
    \label{fig:bscco_fig_1}
\end{figure}

Figure \ref{fig:bscco_fig_1}A shows measured $\delta g({\bm r},E=10\textrm{meV},B)$ exhibiting the classic 'halo' of modulations in the density of Bogoliubov quasiparticles at ${\bm q} \approx [(\pm 1/4,0);(0,\pm1/4)\frac{2\pi}{a_0}]$. However, for the energy range $25 < |E| < 50$meV which is $|E| \approx \Delta_{SC}$, the measured $\delta g({\bm r},E=30\textrm{meV},B)$ shown in Fig. \ref{fig:bscco_fig_1}B contains distinct modulations within each halo. Fourier analysis yields $\left |\widetilde{\delta g}({\bm q},30\textrm{meV})\right |$ as shown in Fig. \ref{fig:bscco_fig_1}C, reveals a set of eight maxima at ${\bm q}=[{\bm P}_{x};{\bm P}_y]\approx[(\pm \frac{1}{8},0);(0,\pm \frac{1}{8})]2\pi/a_0$ which we label ${\bm P}$, and at ${\bm q}\approx[(\pm \frac{1}{4},0);(0,\pm \frac{1}{4})]2\pi/a_0$ which we label $2{\bm P}$. The inset to Fig. \ref{fig:bscco_fig_1}C shows the measured amplitude $\left |\widetilde{\delta g}({\bm q},30\textrm{meV})\right |$ along (1,0) indicating that the field-induced $N({\bm r},E)$ modulations occur, with both $\lambda \approx 8a_0$ and $\lambda \approx 4a_0$, along both the (1,0);(0,1) directions within every vortex halo. The fitted widths $\delta({\bm P})$ of all $|{\bm P}| \approx (1/8)2\pi / a_0$ peaks are close to half that of the $|2{\bm P}| \approx 1/4(2\pi / a_0)$ peaks: $\delta(2{\bm P}) = 1.8 \pm 0.2) \delta({\bm P})$. These phenomena occur in a particle-hole symmetric manner for $ 25  < |E| < 45$ meV and exhibit predominantly \textit{s}-symmetry form factor modulations at ${\bm P}$ and $2{\bm P}$. Finally, measured field-induced energy gap modulations $\delta\Delta({\bm r}) = \Delta({\bm r},B)-\Delta({\bm r},0)$ yield a Fourier transform $\widetilde{\delta\Delta}({\bm q})$ that exhibits energy-gap modulation at ${\bm P}$ but not at $2{\bm P}$.  

In the context of Ginzburg-Landau theory \cite{Agterberg-2015,Wang-2018,Lee-2018,Edkins-2018},these data indicate that, in \bscco, a field-induced pair density wave state emerges within the halo region surrounding each quantized vortex core. 

\subsubsection{Evidence for a PDW in {\bscco} at Zero Magnetic Field}
Atomic-resolution superconducting STM tips \cite{Pan1998} have also been applied for the study of the PDW state in underdoped cuprates, but at B=0. Ideally, if both the tip and sample are superconducting, with identical superconducting energy gaps $\Delta({\bm r})$ and quantum phase difference $\phi$, a Josephson current $I(\phi) = I_{J}\sin(\phi)$ of Cooper pairs can ensue. However, for nanometer scale junctions with normal-state resistance in the gigaohm range, thermal fluctuations will overwhelm stable-phase Josephson tunneling until the sub-millikelvin temperature range. Instead, phase-diffusion dominated Josephson tunneling is usually achieved, in which the measured $I(V)$ exhibits a maximum current $I_{c} \propto I_{J}^{2}$ \cite{Naaman2001}. Measuring $|I_{c}({\bm r})|$ has therefore become the established approach for visualizing the variation of Josephson tunnelling \cite{Naaman2001,Rodrigo2004,Proslier2006,Randeria2016} and thus of superfluid  (electron-pair) density $\rho_{s}({\bm r})$.

\begin{figure}
    \centering
    \includegraphics[width=\columnwidth]{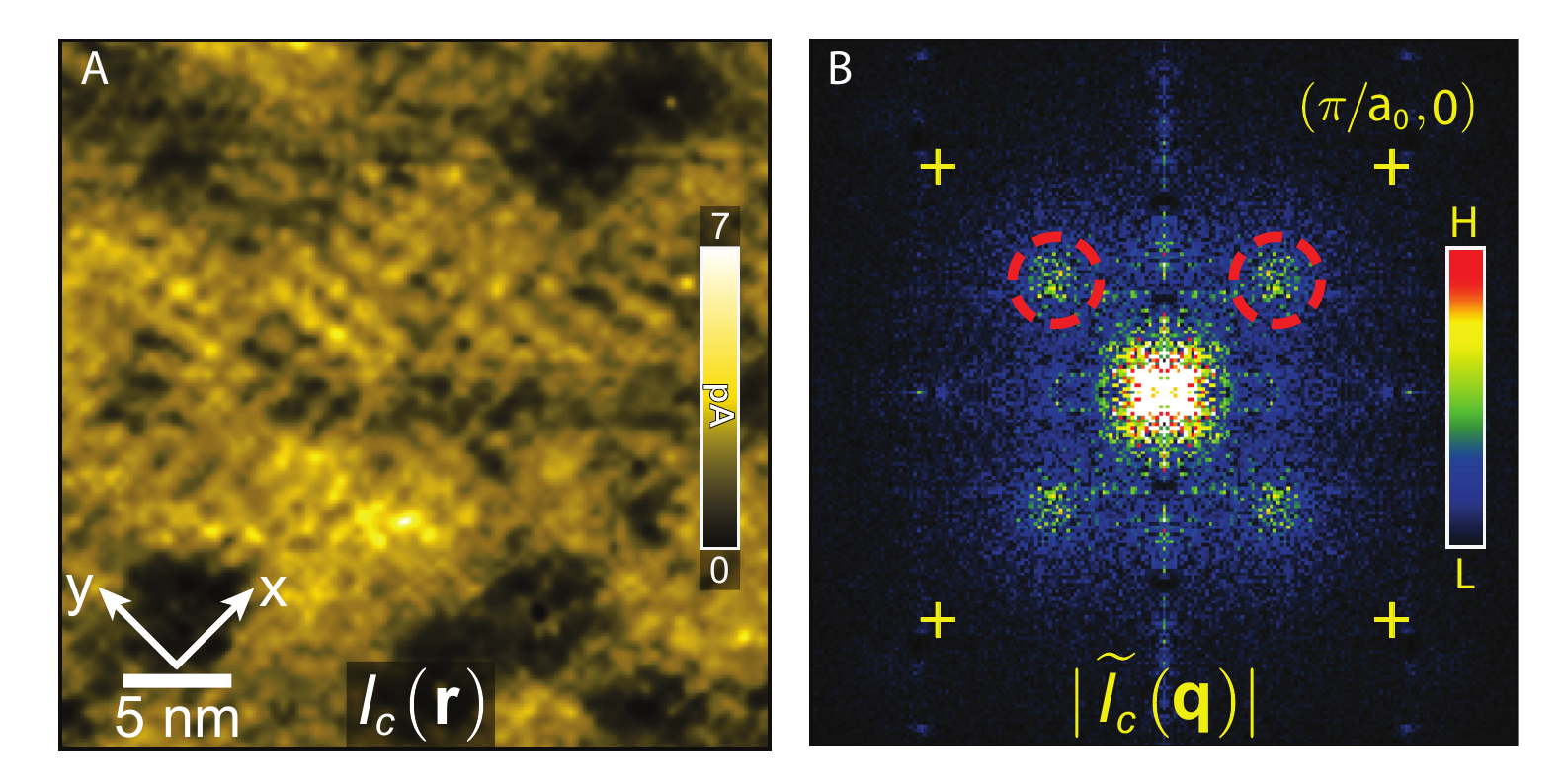}
    \caption{{\bf A} Typical $I_{c}({\bm r})$ image from \bscco at $p=0.17$\%. The $I_{c}({\bm r})$ modulations are parallel to the CuO$_2$ directions (structural supermodulation-induced $I_{c}({\bm r})$ modulations along the (1,1) directions are removed). {\bf B} $|(I_{c}({\bm q})|$, the Fourier transform of $I_{c}({\bm r})$ (crosses at ${\bm q}=(\pi/a_{0},0);(0,\pi/a_{0}))$. Maxima from $I_{c}({\bm r})$ modulations (dashed red circles) occur at ${\bm q}=(0.25,0)2\pi/a_0;(0,0.25)2\pi/a_0$. Figure adapted  from Reference \cite{Hamidian-2016}).  }
    \label{fig:bscco_fig_2}
\end{figure}

In a PDW of the type $\Delta_{PDW}^{{\bm P}}({\bm r}) = F_{pdw} \Delta_{{\bm P}} ({\bm r}) \left[ \exp(i{\bm P} \cdot {\bm r}) +  \exp(-i{\bm P} \cdot {\bm r}) \right]$, the superfluid density $\rho_s({\bm r})$ modulates spatially. Searches for such phenomena in cuprates required visualizing $|I_{c}({\bm r})|$ with nanometer resolution, high $I_{J}$, low $R_{N}$ and low operating temperatures. For the \bscco  samples ($T_{c} = 88K, p=0.17)$ studied by Hamidian \textit{et al.} \cite{Hamidian-2016}, the STM operates below 50mK, and high $I_{J}$ achieved using an exfoliated nanometer-sized flake of \bscco with spatial resolution $\sim 1$nm adhering to the end of each tungsten STM tip. Figure \ref{fig:bscco_fig_2}A shows a typical $|I_{c}({\bm r})|$ image measured under those conditions, clearly exhibiting periodic modulations in $\rho_s({\bm r})$ along the CuO$_{2}$ axes (1,0);(0,1). Figure \ref{fig:bscco_fig_2}B shows $|\tilde{I}_{c}({\bm q})|$, the magnitude of the Fourier transform of $|I_{c}({\bm r})|$, indicating that the wavevectors of $\rho_s({\bm r})$ modulations in \bscco are at ${\bm q} = (0.25 \pm 0.02,0)2\pi/a_0$; $ (0,0.25 \pm 0.02)2\pi/a_0$. 

Single-electron tunneling SISTM studies on equivalent \bscco crystals reveal intense electronic structure modulations \cite{Hanaguri2004a,McElroy2005,Kohsaka2007} that are locally commensurate \cite{Mesaros2016,Zhang2018} and unidirectional \cite{Kohsaka2007,Hamidian2015}, exhibit 4a$_0$ periodicity \cite{Mesaros2016,Zhang2018} with a \textit{d}-symmetry form factor \cite{Hamidian2015,Fujita2014}, and are concentrated at particle-hole symmetric energies $|E| \approx \Delta_{1}$ where $\Delta_{1}$ is the pseudogap energy scale. \cite{Fujita2015,McElroy2005,Kohsaka2007,Hamidian2015}.

At values of $p$ where simultaneous data exist, the wavelengths of these modulations are indistinguishable from those in $|I_{c}(\bm r)|$ within joint uncertainty. While this is consistent with a composite PDW order formed out of a CDW with the same wave-vector and the uniform \textit{d}-wave superconductivity, whether the order parameter of the fundamental state underpinning these single-electron signatures is a CDW or a PDW remains to be determined. The relationship of the $|I_{c}(\bm r)|$ modulations observed at zero magnetic field to the PDW in vortex halos also remains unclear, as it is yet to be established whether the $\lambda \approx 4a_0$ $|I_{c}(\bm r)|$ modulations could be consistent with an underlying $\lambda \approx 8a_0$ PDW, as expected. 
Nevertheless, the $|I_{c}({\bm r})|$ imaging data (e.g Fig. \ref{fig:bscco_fig_2}) provide strong direct evidence for the existence of an PDW coexisting with a robust homogeneous Cooper-pair condensate in underdoped \bscco. 


\section{OTHER SYSTEMS}
\label{sec:other}
\subsection{Organics, Fe-based superconductors, and heavy fermion materials}


A variety of materials  have been argued to be candidates for the originally proposed FFLO state, in which a Zeeman field shifts the energy of the up-spin and down-spin partners of the Cooper pairs in opposite directions, such that it becomes energetically beneficial to create Cooper pairs with finite momentum.  The materials discussed in this section have been reviewed previously \cite{Agosta-2018,Matsuda-2007,Kenzelmann-2017}, so here we highlight the main results and refer to these reviews for more detail.

The most compelling case for realizing the original FFLO-like PDW superconductor \cite{Larkin-1964,Fulde-1964} is in organic materials. A detailed overview of the experimental evidence for this in quasi-2D organic materials with in-plane magnetic fields is given in Ref.~\cite{Agosta-2018}. Organic materials have proven to be ideal for realizing the FFLO-like PDW phase for three reasons: they are quasi-2D, suppressing the creation of vortices and allowing the high fields needed to create FFLO-like PDW  states to be reached; they are clean with mean free paths typically a factor of 10-100 greater than the superconducting coherence length; and they have weak spin-orbit coupling.  The primary evidence for the existence of the FFLO state is largely of three types: the observation of the characteristic upturn of the upper critical field at low temperatures; the observation of two high-field phase transitions at low temperatures; and the observation of an inhomogeneous magnetic field distribution consistent with that expected for a FFLO-like phase. $\kappa$-(BEDT-TTF)$_2$Cu(NCS)$_2$ presents the strongest case in which: magnetic torque measurements suggest a first order transition line at high fields inside the SC state \cite{Bergk-2011,Tsuchiya-2015}, specific heat measurements observing the same first order transition \cite{Lortz-2007,Agosta-2017}; NMR measurements consistent with the observation of spin-polarized quasiparticles localized near the spatial nodes of the FFLO order parameters \cite{Wright-2011,Mayaffre-2014}, and evidence of multiple phase transitions in rf-penetration depth measurements \cite{Agosta-2012}. Among the quasi-2D organics there is also evidence for a FFLO-like PDW state in $\lambda$-(BETS)$_2$GaCl$_4$ \cite{Tanatar-2002,Coniglio-2011} and $\beta''$-(ET)$_2$SF$_5$CH$_2$CF$_2$SO$_3$ \cite{Koutroulaki-2016,Cho-2009}. In addition, resistivity measurements observe an upper critical field and field anisotropy behavior consistent with a FFLO-like PDW state in the quasi-1D organic (TMTSF)$_2$ClO$_4$ \cite{Yonezawa-2008}. 

Fe-based superconductors represent a likely class of materials in which to realize a FFLO-state \cite{Gurevich-2010} because of their high upper critical fields. To date, experimental evidence has been found for a FFLO-like PDW state in KFe$_2$As$_2$ \cite{Cho-2017} where magnetic torque and specific heat measurements observe two superconducting transitions at high fields and observe a characteristic upturn of the upper critical-field at low-temperatures.

Finally, there were  reports that FFLO phases also appear in the heavy fermions superconductors UPd$_2$Al$_3$ \cite{Gloos-1993,Modler-1996}, CeRu$_2$ \cite{Modler-1996,Yamashita-1997}, and CeCoIn$_5$  \cite{Radovan-2003,Bianchi-2003}. However,  the phase transition attributed to the FFLO-like PDW phase in both UPd$_2$Al$_3$ and CeRu$_2$  has been argued to a consequence of a vortex related transition \cite{Matsuda-2007}. The case for CeCoIn$_5$ is much more interesting. Subsequent to the original discovery of a new low-temperature, high-field superconducting phase  that was argued to be a FFLO-like PDW phase\cite{Radovan-2003,Bianchi-2003}, this phase was found to have  spin-density wave (SDW) order \cite{Kenzelmann-2008}. This SDW order exists only within the superconducting state. Due to the coexistence of usual superconductivity and SDW order, PDW order will also generically exist \cite{Agterberg-2009}, making it difficult to identify a primary order parameter. This has led to many proposals that are still being experimentally untangled \cite{Kenzelmann-2017}.


\subsection{Pair Density Wave in degenerate atomic gases: FFLO}

Experimental progress in trapping, cooling and coherently manipulating
Feshbach-resonant atomic gases opened unprecedented opportunities to
study degenerate strongly interacting quantum many-body systems in a
broad range of previously unexplored regimes
\cite{BlochReview,KetterleZwierleinReview,GRaop,GiorginiRMP,RSreview}.
This has led to a realization of paired fermionic superfluids
\cite{Regal2004prl,Zwierlein2004prl,Kinast2004prl}
and the associated Bardeen-Cooper-Schrieffer (BCS) to Bose-Einstein condensation (BEC)
crossover \cite{Eagles,Leggett,NSR}.

These neutral atomic systems are particularly well-suited to imposing (pseudo-) magnetization, corresponding to the number imbalance
$P\equiv (N_\uparrow-N_\downarrow)/N$ in the pairing hyperfine $\uparrow-\downarrow$
species, circumventing challenges of charged electronic superconductors realized in solid state, as discussed in much of this review 
\cite{Zwierlein06Science,Partridge06Science,Shin2006prl,Navon2009prl}.
The imbalance and the associated Fermi surface mismatch frustrate the
singlet paired state
\cite{Combescot01,Liu03,Bedaque03,Caldas04} 
driving quantum phase transitions out of the gapped BCS superfluid to a variety of putative ground states and thermodynamic phases 
\cite{Castorina05,Sedrakian05nematic,SRprl,Pao06,Son06,Bulgac06pwavePRL}. One of the most interesting is the FFLO finite-momentum paired
state \cite{Larkin-1964,Fulde-1964}.

\begin{figure}
\begin{center}
\includegraphics[width = 3.5in]{{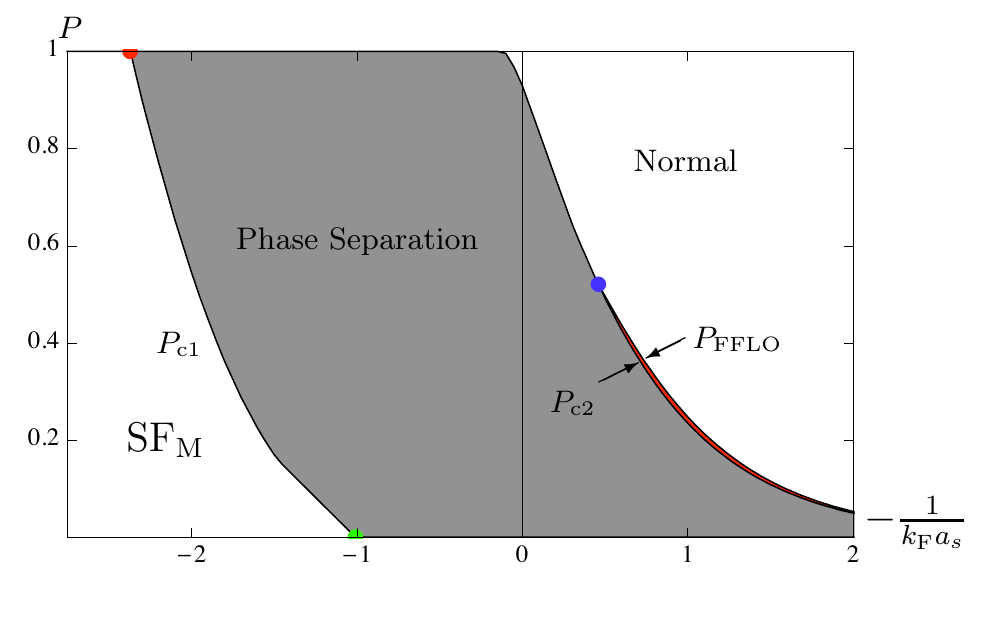}}
\end{center}
\caption{A mean-field zero-temperature 
phase diagram of an imbalanced Fermi gas, as a function of the inverse scattering length and normalized species imbalance (dimensionless magnetization) $P=(N_\uparrow-N_\downarrow)/N\equiv\Delta N/N$, showing the magnetized (imbalanced) superfluid ($SF_{\rm M}$), the FFLO state (approximated as the simplest FF state, confined to a narrow red sliver bounded by $P_{\rm FFLO}$ and $P_{{\rm c}2}$) and the
imbalanced normal Fermi liquid. Data from References \cite{SRprl} and \cite{SRaop}.}
\label{SRphasediagram}
\end{figure}

Experiments \cite{Zwierlein06Science,Partridge06Science,Shin2006prl,Navon2009prl} on trapped atomic systems have extensively explored and established the predicted interaction-imbalance phase
diagram, illustrated in Fig.\ref{SRphasediagram}\;\cite{SRprl,SRaop,Parish07nature},
dominated by the superfluid to (polarized) Fermi-liquid first-order
phase transition, that manifests in phase
separation \cite{Bedaque03,SRaop,Pao06}. However,
outside of one dimension, so far, no signatures of the enigmatic FFLO Pair Density Wave state \cite{Larkin-1964,Fulde-1964} have been seen. This is consistent with the narrowness of the FFLO sliver in the predicted phase diagram
\cite{SRprl,SRaop}, much remains to be understood about FFLO's stability, beyond mean-field analyses of simplest FFLO states \cite{RVprl,Rpra}.

In contrast, in one dimension (where it is robust and generic at any nonzero imbalance) the FFLO state has been experimentally realized in
a two-dimensional array of decoupled one-dimensional traps, generated via a two-dimensional optical periodic potential\cite{Hulet1dLO}. Although spin-resolved density profiles in
these experiments shows consistency with the FFLO interpretation, they
still lack the ``smoking gun'' observation of e.g., a finite momentum
condensate peak in the momentum distribution function, precluded in thermodynamic limit in one dimension by strong quantum and thermal fluctuations and by the inhomogeneous atomic density special to trapped gases. 

Experimental efforts are under way to move toward the
quasi-one-dimensional limit of coupled tubes, and in fact the 1d-3d
crossover signatures have been experimentally
demonstrated \cite{1d3dCrossoverPRL2016,ParishQuasi1dPRL2007}. This is
done by reducing the strength of the periodic optical potential,
thereby allowing the 1d PDWs of neighboring tubes to lock through
inter-tube coupling. Cooling and equilibration, particularly for pseudo-spin remains a challenging experimental problem.

\section{MECHANISM}

\label{sec:strongly-correlated}

\subsection{Evidence of PDW in Models of Strongly Correlated Systems}
\label{sec:evidence}

Condensates with finite momentum are problematic in conventional BCS theory. 
In the first place, so long as either time-reversal or inversion symmetry is preserved, the Fermi surface is always perfectly nested for some form of ${\bm P}={\bm 0}$ pairing, {\it i.e.} the superconducting susceptibility is peaked and logarithmically divergent as $T\to 0$, while it remains finite at all non-vanishing   ${\bm P}$.  
If the  ${\bm P}={\bm 0}$ divergence is quenched, as it is in any singlet channel by a finite Zeeman field, as recognized Fulde and Ferrell \cite{Fulde-1964} and 
by Larkin and Ovchinnikov \cite{Larkin-1964} 
the pair susceptibility can be peaked at a non-zero  ${\bm P}$, but in that case it remains  finite even as $T\to 0$.
In the Hartree-Fock approximation, used in BCS theory, 
lack of nesting leads to a finite critical coupling for the condensate to occur even at zero temperature. In the case of FFLO states, 
the Zeeman coupling to the external magnetic field acts as a small tuning parameter,  {\it i.e.} the critical coupling can still be parametrically small.

In contrast, for the  putative PDW states of high $T_c$ superconductors (HTSC) which occur in the absence of an external Zeeman coupling, 
no such small tuning parameter exists.  
Even the naive application of BCS theory to PDW states  typically  requires a critical coupling  
of strength comparable to the band-width.
More importantly,  
the superconducting states in these materials, 
 uniform or not, arise in strongly correlated systems whose
 normal state  is a strange metal, a metallic state without well defined fermionic quasiparticles.

\begin{figure}[hbt]
\begin{center}
\includegraphics[width=0.7\textwidth]{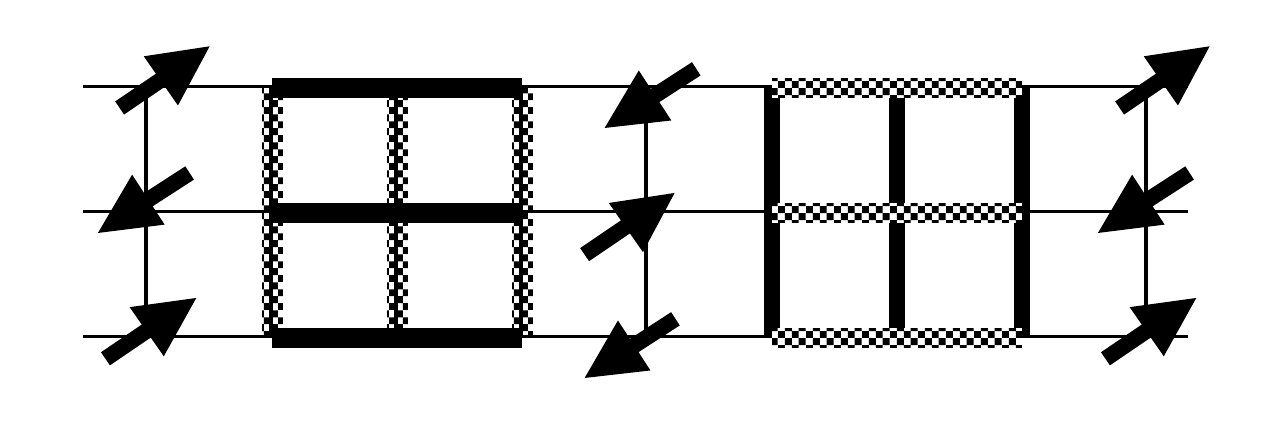}
\caption{Qualitative picture of the PDW. The local superconducting order parameter is $d$-wave: positive (negative) in the bold (shaded) links, and axes rotate after a half-period. In this example, presumably applicable to LBCO, the PDW  
 is intertwined with spin stripe order.  Notice that the charge density has half the period of the PDW order. Figure adapted from Reference \cite{Berg-2007}). }
\label{Fig:magnetic-stripe}
\end{center}
\end{figure}

The PDW  
looks locally like a $d$-wave superconductor. It breaks translation symmetry  
in such a way that the order parameter changes sign upon translation by half a period
(see Fig.\ref{Fig:magnetic-stripe}). Therefore, 
it 
is reasonable to suppose that it is  a close competitor of the uniform $d$-wave superconductor  under strong-coupling circumstances in which correlation lengths are short and the important physics is correspondingly local.  Indeed, in the context of a proposed SO(5) theory of intertwined antiferromagnetic and d-wave superconducting order, it was proposed in 1998 by Zhang \cite{Zhang-1998} some time ago that a ``SO(5) spiral'' state consisting of alternating stripes of N\'eel order and d-wave superconducting order -- precisely the sort of intertwined PDW and spin stripe shown in  Fig.\ref{Fig:magnetic-stripe} -- might arise in some circumstances.  
A variety of subsequent studies employing variational  wave functions 
have  also found such states, starting with the 2002 study of  
Himeda, Kato and Ogata \cite{Himeda-2002} who used variational Monte Carlo calculations of Gutzwiller-projected wave functions  
for the ground state of the  $t-t'-J$ model. 
These  states are inhomogeneous versions of the doped resonating valence bond (RVB) wave functions widely used as candidate ground states for the $t-J$ model.
These early results, as well as later variational and renormalized mean field studies \cite{Raczkowski-2007,Capello-2008,Yang-2009}, 
found three states whose variational ground state energies are very close: 
the uniform $d$-wave superconductor, the PDW state, and a striped $d$-wave superconductor. 
For doping near $x=1/8$, these variational studies found period 4 states with half-filled stripes. The intertwined nature of the PDW states was also proposed on  phenomenological grounds in refs. \cite{Berg-2007,Berg-2009,Lee-2014}.

BCS mean-field methods have also been used to describe PDW phases. Loder and coworkers \cite{Loder-2010,Loder-2011} 
used BCS mean field theory for a 2D system with a $t-t'$ band structure widely used in the cuprates, 
and  an effective attractive interaction for electrons on nearest-neighboring sites of strength $V$ (assumed to originate from spin fluctuations).
 Within a BCS type mean field theory these authors find that  while in the weak coupling regime the ground state is an uniform $d$-wave superconductor, 
 for systems near $1/8$ doping,  for $V$ larger than a critical value $V_c \gtrsim t$, the preferred pairing state is locally $d$-wave but has a finite momentum, i.e. a pair-density-wave, 
 with or without an associated spin stripe state. 
A subsequent publication by W{\aa}rdth and Granath found that  
$V_c$ is significantly larger than this estimate \cite{Wardth-2017}. These authors 
 proposed a model with an interaction with local attraction and longer range phase slip pair hopping and found,
using a  BCS type theory, that the critical coupling for the PDW state is significantly reduced, 
and typically of the order of the kinetic energy bandwidth  \cite{Wardth-2017, Wardth-2018}.

Other microscopic mechanisms have also been proposed and studied within mean-field approaches. Lee proposed an Amperean pairing mechanism  \cite{Lee-2014}, 
due to local spin current fluctuations in a RVB type  state \cite{Lee-2007}, and  showed (in mean field theory) that it favors a PDW. 
Soto-Garrido and Fradkin \cite{Soto-Garrido-2014} studied the superconducting condensates arising in the vicinity of a (Pomeranchuk) 
instability in the quadrupolar spin triplet channel of a Fermi liquid \cite{Wu-2007}, and found that the PDW competes with a spin triplet $p$ and a spin singlet $d$ 
wave superconducting states. Soto-Garrido and coworkers \cite{Soto-Garrido-2015} used a quasi-1D approach based on a model of
stripe phases \cite{Granath-2001}  and found  PDW states.

At present, the only model that has been definitively shown to have a PDW state is in a model of strongly correlated systems in one dimension known as the Kondo-Heisenberg chain. 
This model consists   of a 1D system of mobile electrons, a Luttinger liquid,  coupled by a local Kondo exchange interaction $J_K$  
to a spin-1/2 quantum Heisenberg antiferromagnetic chain with exchange coupling $J_H$. 
Models of this type have been studied for a long time in the context of the physics of heavy-fermion superconductors. 
Bosonization and  density-matrix renormalization group (DMRG) studies by Sikkema, Affleck and White revealed that for $J_H \gtrsim J_K$ 
 this system has a spin gap  \cite{Sikkema-1997}, 
 which was subsequently interpreted as an exotic $\eta$-pairing superconducting state \cite{Zachar-2001,Zachar-2001b}. Specifically, the dominant long-range (power-law) correlations involve an oscillatory charge 2e order parameter, but it is a composite order in the sense that it cannot be simply expressed as a product of two electron creation operators.
 Berg, Fradkin and Kivelson \cite{Berg-2010} reexamined this system by DMRG and showed that it is 
 indeed a PDW state, albeit an exotic one. 
 Specifically, on open chains with different boundary conditions, they showed that in this state all the fermion bilinear operators decay 
 exponentially with distance. Only order parameters that are composite operators of the Luttinger liquid and the spin chain have (quasi) long range order, as suggested by the bosonization studies.
 For instance, the PDW order parameter is realized as the scalar product of the N\'eel order parameter of the spin chain with the spin triplet superconductor 
 of the Luttinger liquid, and has quasi-long-range order.  (Remarkably, later work showed that this incarnation of the PDW is actually a topological superconductor \cite{Cho-2014}.) Similar behavior was found for the expected charge 4e uniform superconducting order parameter. 
 Later on, similar results were found  for extended generalized Hubbard models on 2-leg ladders at special 
 filling fractions of the bonding band \cite{Jaefari-2012}.

A real challenge is establishing the existence of a PDW state in  two  dimensions using unbiased approaches, even numerically. Due to the notorious fermion sign problem, quantum Monte Carlo type methods are only useful at relatively high temperatures, and have not been able to reach low enough temperatures to see unambiguous signs of $d$-wave superconductivity in Hubbard models and its generalizations.  None-the-less, down to the lowest temperatures accessible, such calculations always find a superconducting susceptibility that is peaked at ${\bm P} = {\bm 0}$ \cite{Huang-2018}.  However, exact diagonalizations can only deal with systems which are too small to be useful to detect PDW states (even if they were ground states).  One option is DMRG simulations on relatively wide ladders, and  we will discuss  these in the next paragraphs. DMRG methods are known to be asymptotically exact matrix-product states (at least for gapped states), whose accuracy can be improved by increasing the bond dimension of the tensors. These approaches are known to generate large enough quantum entanglement to produce most states of interest  in one-dimensional systems, including quantum critical states. 

Other options  include  tensor network approaches which, conceptually, are extensions of DMRG to higher dimensions. However, while in 1D it is known that matrix product states are sufficient to describe most systems of interest, in higher dimensions this is an open question. The current most widely used tensor network approach is PEPS (projected entangled paired states) \cite{Verstraete-2008}. PEPS, and its relative iPEPS (infinite PEPS), consist of a variational ansatz in the form of a tensor network (a matrix product state) with many variational parameters that grow as a power of the ``bond dimension'' (the dimension of the tensor), of order $3D^4$ where $D$ is the bond dimension. Unlike conventional variational wavefunctions, which are essentially product states (and hence have only short-range entanglement), iPEPS can describe more complex states with large-scale entanglement. Corboz and coworkers \cite{Corboz-2011} initiated a study of of the $t-J$ model as a function of doping. In the most recent study of this type \cite{Corboz-2014} it was found that, for a wide range of parameters $J/t$ and doping $\delta$ the uniform $d$-wave superconductor, the striped superconductor, and the PDW (called the ``anti-phase'' stripe state by Corboz et al), are essentially degenerate. Although this result agrees with the simpler variational wave functions discussed above, in the iPEPS generated states the charge stripes are not half-filled and, in fact, their occupancy varies continuously with $J/t$. 
 (Were such a PDW state to be relevant in the cuprates, its period would not be particularly pinned to $8a$, even near 1/8 doping, in contrast with experiment in LBCO.)
The iPEPS results are highly encouraging  for the existence of PDW order. However, we should be careful to note that it is currently unclear  what  biases  are implied in this approach, and these results need to be ``benchmarked'' against other techniques. 

Quite recently, large-scale DMRG simulations of the $t-J$ model on 4-leg ladders in the doping range $5\% - 12.5 \%$ with $t/J=3$ (and other values as well) \cite{Dodaro-2016, Jiang-2018, Jiang-2018b} have been carried out.  These authors  find strong evidence for $d$-wave superconductivity (i.e. a sign change of the superconducting amplitude along two orthogonal directions) and 
charge-stripe phases with 
1/2 a hole per unit cell. These simulations kept a significantly larger number of states in the DMRG than earlier studies.  However, although the parameter range examined by these authors  is broad and overlaps with those used in the iPEPS simulations discussed above, no evidence for PDW states have (yet) been detected in the DMRG. This result is in apparent contradiction with the iPEPS results, and this discrepancy is quite puzzling. This is a  pressing problem a deeper understanding of the advantages and limitations of these methods is clearly required. A recent DMRG study of a doped  $t-J$ model  on a triangular lattice with ring exchange interactions found encouraging evidence 
of PDW-like superconducting correlations that change sign as a function of distance, but which fall sufficiently rapidly (at least like $r^{-4}$) that they do not give rise to a diverging susceptibility as $T\to 0$. \cite{Xu-2018}.

We close this section by noting that PDW states have been studied using methods of holography, the AdS-CFT correspondence. Although it is not clear what microscopic systems can be described with holography, these theories have the clear advantage of describing metallic states without well defined quasiparticles \cite{Hartnoll-2019}. Several holographic models have been published describing superconductors with striped phases \cite{Flauger-2011}, and systems with intertwined superconducting and PDW orders \cite{Cremonini-2017,Cremonini-2017b,Cai-2017}.


\subsection{Fulde-Ferrell-Larkin-Ovchinnikov state in degenerate atomic gases}
\label{sec:fflo}


One context in which a FFLO state has a well-understood microscopic
mechanism is a singlet superconductor with pairing frustrated by a
magnetic field \cite{Chandra,Clogston,Sarma,Fulde-1964,Larkin-1964}.
An ideal realization of such a system is a pseudo-spin imbalanced
Feshbach-resonant atomic Fermi
gas \cite{Zwierlein06Science,Partridge06Science,Shin2006prl,Navon2009prl},
where (in contrast to charged electronic superconductors in solid state), the Zeeman component of the magnetic field and the effective
magnetization can be tuned {\it independently} of the obscuring orbital
field effects, such as vortices.  This has rekindled extensive theoretical research
 \cite{Combescot01,Liu03,Bedaque03,Caldas04,CarlsonReddy05,
SRprl,SRaop,Pao06,Bulgac06pwavePRL,
Parish07nature,
ParishQuasi1dPRL2007}, reviewed in Refs.
\cite{RSreview,TormaReview2018}. 

\subsubsection{Model of a resonant Fermi gas}

A neutral fermionic atomic gas is well captured by a microscopic Hamiltonian
\begin{eqnarray}
  H = \sum_{\bk,\sigma}(\epsilon_k - \mu_\sigma) \ch_{\bk\sigma}^\dagger  
  \ch_{\bk\sigma}^\phdag 
  + g\sum_{\bk\bk'\bq} 
  \ch_{\bk\uparrow}^\dagger \ch_{-\bk+\bq\downarrow}^\dagger 
  \ch_{-\bk'+\bq\downarrow}^\phdag\ch_{\bk'\uparrow}^\phdag.
\label{eq:Hfermi}
\end{eqnarray}
with the single-particle energy $\epsilon_k =\hbar^2 k^2/2m$.  The
separately conserved number $N_\sigma = (N_\uparrow,N_\downarrow)$ of
atomic species (hyperfine states) $\sigma = (\uparrow,\downarrow$) is
imposed by two chemical potentials,
$\mu_\sigma=(\mu_\uparrow,\mu_\downarrow)$ or equivalently by the
average chemical potential $\mu=\oh(\mu_\uparrow+\mu_\downarrow)$ and
the Zeeman field $h=\oh(\mu_\uparrow-\mu_\downarrow)$, that
respectively tune the total atom number $N=N_\uparrow+N_\downarrow$ and
atom pseudo-spin imbalance $\Delta N = N_\uparrow-N_\downarrow$.

Key features  distinguishing this Fermi system from those familiar electronic ones in solid state contexts (discussed in other parts of this review) are ($a$) the fermions are neutral and thus do not couple to the electromagnetic vector potential, (i$b$) absence of a periodic ionic potential (though an optical lattice can be imposed by an off-resonant interfering laser fields) that explicitly breaks rotational and translational spatial symmetries, and ($c$) the {\it resonant} nature of the Feshbach interaction, parameterized by a short-range s-wave
pseudopotential, $g <0$. The resulting attractive interaction can be computed through an exact $T$-matrix scattering analysis \cite{GRaop}, with $g$ controlling the magnetic-field
tunable~\cite{RegalJinRF03} 3D scattering length
\begin{equation}
\label{eq:deltalength} 
\as(g)
=\frac{m}{4\pi}\frac{g}{1+g/g_c},
\end{equation}
that diverges above a critical attraction strength, $|g| = g_c\equiv 2\pi^2 d/m$,
corresponding to a threshold for a two-atom bound state of size $d$.

The Zeeman $h$ field-driven Fermi surface
mismatch (that, for these {\it neutral} fermions can be tuned independently of the orbital field, and can also be realized via atomic mass and other dispersion imbalance), energetically penalizes the
conventional BCS $-{\bf k}$ to ${\bf k}$ pairing at weak coupling.
However, as noted at the beginning of Sec. \ref{sec:evidence},
a finite momentum ${\bf P}$ pairing,
$\Delta_{\bf P} = \sum_{\bk} g\langle c_{-{\bf k}\downarrow}c_{{\bf k} + {\bf P}\uparrow}\rangle$, (set by the Fermi surface mismatch) allows for an
FFLO ground state, that compromises
between a fully paired superconductor and a magnetized Fermi liquid. \cite{Fulde-1964,Larkin-1964}

The simplest treatment is a mean-field analysis\cite{SRprl,SRaop} for
the PDW order parameter $\Delta(\rv)=\sum_{\bp}\Delta_{\bp} 
  e^{i{\bp}\cdot{\bf r}} = \fermiint\langle \ch_\downarrow(\rv)
\ch_\uparrow(\rv) \rangle$,
generalized to pair-condensation at a set of reciprocal lattice
vectors, $\bp$, with the amplitudes $\Delta_\bp$ and $\bp $
self-consistently determined by  minimizing the ground state
energy.
This gives a satisfactory qualitative description (quantitatively
valid deep in the weakly-coupled BCS regime, $k_F|\as|\ll 1$), as a
starting point of more sophisticated
large-$N_f$ ~\cite{Nikolic07largeN,Veillette07largeN} and
$\epsilon$-expansions~\cite{Nishida06eps} treatments.

\subsubsection{Ginzburg-Landau model and transitions to FFLO state}

Starting at a high Zeeman field, above the Chandrasekhar-Clogston-Pauli
limit \cite{Chandra,Clogston}, $h_c =\Delta_{BCS}/\sqrt{2}$ (a critical field for a direct first-order mean-field transition from a Fermi-liquid (FL) to a uniform BCS superconductor), inside the polarized Fermi liquid
and reducing $h$, one finds (in mean-field) a continuous transition at $h_{c2} \approx
\frac{3}{4}\Delta_{BCS} > h_c$ to a FFLO
superconductor\cite{Larkin-1964,Fulde-1964,SRprl,SRaop} (stable for $h_{c1} < h < h_{c2}$), most strongly paired at a
momentum with magnitude $p_0 \approx 1.2 h_{c2}/v_F \approx
1.81\Delta_{BCS}/v_F$. This is captured by a Ginzburg-Landau expansion
(derived by integrating out the fermions \cite{Rpra}) for the
ground-state energy density,
\begin{eqnarray}
\cH&\approx&\sum_\bp\eps_p|\Delta_\pv|^2 
+ \sum_{\{\bp_i\}}V_{\pv_1,\pv_2,\pv_3,\pv_4}
\Delta_{\pv_1}^*\Delta_{\pv_2}\Delta_{\pv_3}^*\Delta_{\pv_4},
\label{HDelta_qhc2}
\end{eqnarray}
which is valid for a weak finite-momentum pairing amplitude $\Delta_p$ near the
continuous FL to FFLO phase transition at $h_{c2}$. It is notable that this
expansion is analytic at small $\Delta_p$, in contrast to a vanishing
Zeeman field, that at zero temperature exhibits $|\Delta|^2\ln\Delta$
nonanalyticity. The finite momentum instability is captured by the
dispersion \cite{SRprl,SRaop,Rpra}
\begin{eqnarray}
\eps_p&\approx& \frac{3n}{4\epsilon_F}\left[-1
+ \frac{1}{2}\ln\frac{v_F^2p^2-4h^2}{\Delta_{BCS}^2}
+ \frac{h}{v_Fp}\ln\frac{v_F p+2h}{v_F p-2h}\right],\nonumber\\
&\approx& J(p^2-p_0^2)^2 + \eps_{p_0},
\label{epsSR}
\end{eqnarray}
whose minimum at a finite $p_0(h)\approx 1.2 h/v_F$ (near $h_{c2}$)
captures the polarized Fermi system's energetic tendency to pair at a
finite momentum, forming a FFLO state at a fundamental reciprocal lattice
vector with a magnitude $p_0$. The Zeeman energy $h_{c2}$ at which
$\eps_{p_0}(h)$ vanishes determines the corresponding mean-field
FL-FFLO phase transition point.

As in other problems of periodic ordering (e.g., crystallization), in
the absence of an underlying lattice (as in trapped atomic gases)
$\eps_{p}$ is rotationally invariant, and thus, the quadratic
$|\Delta_\pv|^2$ term only selects the fundamental {\it magnitude} of
the reciprocal lattice, $|\pv| = p_0$, with all degenerate orientations
becoming unstable simultaneously at $h_{c2}$. This contrasts qualitatively with PDW ordering in solid state, where the underlying ionic crystal explicitly breaks rotational and translational symmetries, selecting a discrete set of ${\bf p}$ momenta, as discussed in Sec. 2. In the rotationally and translationally invariant trapped atomic gases, it is the quartic and higher order nonlinearities in $\Delta_{\bf p}$ that select the structure of the FFLO state, characterized by reciprocal lattice of $\bf p$'s and the corresponding amplitudes,
$\Delta_{\bf p}$.  Near $h_{c2}$ it is the $(-{\bf p}, {\bf p})$ LO
state \cite{Larkin-1964} with $\Delta_{LO}({\bm r}) = \Delta_{L0}\cos{{\bm p}\cdot {\bm r}}$,
that is energetically preferred over the single plane-wave FF state,
$\Delta_{FF}({\bm r}) = \Delta_{FF}e^{i{\bm p}\cdot {\bm r}}$ \cite{Larkin-1964,Fulde-1964}. However, no study has conclusively determined the structure of the FFLO state
throughout the field-interaction ($h$ - $1/k_Fa$) phase diagram, despite
heroic efforts in the relativistic quantum chromodynamics (QCD) context \cite{Alford,Bowers}.

Near $h_{c2}$, the {\it unidirectional} pair-density wave (Cooper-pair
stripe) order, characterized by a {\it collinear} set of $\pv_n$'s is
well captured by focusing on long-wavelength fluctuations of these
most unstable modes. The state is well described by a Ginzburg-Landau
Hamiltonian density
\begin{equation}
\cH = J\left[|\nabla^2\Delta|^2 - 2p_0^2|\nabla\Delta|^2\right] +
r|\Delta|^2 
+ \frac{1}{2} \lambda_1|\Delta|^4 + \frac{1}{2}\lambda_2 \jv^2,
\label{H_GL} 
\end{equation}
where deep in the BCS limit, near $h_{c2}$ the model parameters are
given by 
\bse
\begin{eqnarray}
J&\approx&\frac{0.61n}{\epsilon_F p_0^4},\  
p_0\approx\frac{1.81\Delta_{BCS}}{\hbar v_F},\ 
r\approx\frac{3n}{4\epsilon_F}\ln\left[\frac{9h}{4h_{c2}}\right],\ \ \
\ \ \ \ 
\\
h_{c2}&\approx&\frac{3}{4}\Delta_{BCS},\ 
\lambda_1\approx\frac{3n}{4\epsilon_F\Delta_{BCS}^2},\  
\lambda_2\approx\frac{1.83n m^2}{\epsilon_F\Delta_{BCS}^2p_0^2},\ \ \
\ \ \ \ \
\end{eqnarray}
\label{Jmoduli}
\ese
and the inclusion of the current-current interaction, ${\bf
  j}=\frac{1}{m}\text{Re}
\left[-\Delta^*(\rv)i\nabla\Delta(\rv)\right]$ is necessary for a
complete description of the transverse superfluid stiffness.  

Well below $h_{c2}$, the PDW order parameter $\Delta_{\bf p}$ is no longer
small, invalidating the above Ginzburg-Landau expansion and requiring a
complementary weak $h$ Bogoliubov-deGennes (BdG) treatment, that is
fully nonlinear in $\Delta(\rv)$. However, it is challenging to handle
analytically for anything other than a single harmonic FF state,
$\Delta_{FF}(\rv) = \Delta_{FF} e^{i{\bf p}\cdot\rv}$, as it requires
a fully self-consistent BdG band-structure analysis, with energetics
strongly dependent on the details of the FFLO state. A single-harmonic BdG
calculation\cite{SRprl,SRaop} finds that a BCS singlet superconductor
is unstable to the FF state at $h_{c1}\approx 0.70\Delta_{BCS}$, thus
suggesting that PDW state is stable only over a very narrow range of $h$.

However, numerical BdG analyses
\cite{MachidaNakanishiLO1984,BurkhardtRainerLO,MatsuoLO} and a negative
domain-wall energy in an otherwise fully-paired singlet BCS superfluid
in a Zeeman field\cite{MatsuoLO,YoshidaYipLO} argue that a more generic pair-density wave state (that includes a larger set of collinear wavevectors) may be significantly more stable.
Well below $h_{c2}$ the FFLO state is thus more accurately described
as a periodic array of solitons,  well-paired $\pm\Delta_{BCS}$
stripes interrupted by ``normal'' gapless domain-walls that
accommodate the imposed fermion imbalance, driven by $h$.  This state
can be equivalently thought of as a periodically ordered 
  microphase separation between the normal and paired states, that
naturally replaces the macrophase
separation\cite{Combescot,Bedaque03} ubiquitously found in the BCS-BEC
detuning-imbalance phase
diagram \cite{SRprl,SRcomment,SRaop,Parish07nature} (see Fig.\ref{SRphasediagram}).  Upon increasing
$h$ above $h_{c1}$ the excess of the majority fermionic atoms
(polarization) in an imbalanced system can be {\it continuously}
accommodated by the sub-gap states localized on the self-consistently
induced domain-walls between $+\Delta$ and $-\Delta$. Thus the imbalance and density of domain-walls
continuously grows above $h_{c1}$ eventually overlapping at $h_{c2}$
and thereby interpolating between the two limiting forms of the LO
state. This picture
resembles the soliton mechanism for doping of
polyacetylene \cite{SuSchriefferHeeger}, and is explicitly realized in
one-dimension (1d) through exact BdG \cite{MachidaNakanishiLO1984} and
Bethe ansatz \cite{OrsoBA2007,HuBA2007} solutions and via
bosonization\cite{YangLL,ZhaoLiuLL}, that exhibits the commensurate-incommensurate (CI) Pokrovsky-Talapov (PT)
transition \cite{PokrovskyTalapov} from a fully paired s-wave
superfluid to a Larkin-Ovchinnikov
state. Such phenomenology also emerges from the numerical BdG studies in two
dimensions\cite{BurkhardtRainerLO,MatsuoLO}. This response to a Zeeman
field is quite analogous to the more familar phenomenology of type-II
superconductor in an orbital magnetic field, with fully-gapped BCS,
partially paired FFLO and fully depaired normal FL playing the role of
the Meissner, Abrikosov vortex lattice and normal states, respectively.

\subsubsection{Goldstone modes and topological excitations}

Trapped atomic gases (in the absence of an optical lattice) exhibit underlying translational and rotational symmetries. Thus, as we discuss below, in addition to the off-diagonal-long-range order, the FFLO states break {\it continuous} spatial symmetries, and hence exhibit unusual Goldstone modes and novel topological defects. This contrasts qualitatively with the putative solid state PDW realizations (formulated in Sec. 2 and discussed in the rest of this review), where spatial symmetries are broken {\it explicitly} by the underlying crystal, and, thus, orientational Goldstone modes are absent. 

Inspired by the one dimensional picture discussed above, a class of striped unidirectional FFLO, with Cooper pairs condensed at a co-linear set of wavevectors $\pv_n = n \pv_0$, has received considerable attention. The FF plane wave and LO standing wave states are qualitatively accurate representatives, that have been extensively explored \cite{RSreview,TormaReview2018}. 
In particular, beyond mean-field theory, the time-reversal breaking FF
state is characterized by an order parameter
$\Delta_{FF}(\rv)=\Delta_{p_0} e^{i\pv_0\cdot\rv +i\phi(\rv)}$, with a
single Goldstone mode $\phi(\rv)$, that in addition to superfluid phase
fluctuations, also describes local fluctuations in the orientation of
FF stripes. Because in a trapped atomic gas context, free of the
underlying lattice, FF state spontaneously breaks rotational (but not
translational) symmetry, the energetics of $\phi(\rv)$ is
qualitatively ``softer'', with the Hamiltonian (derivable from the microscopics, above GL theory, or deduced based on symmetry)
\begin{eqnarray}
\cH_{FF} &=& \oh\chi^{-1}n^2 + \oh K (\nabla^2\phi)^2 + 
\oh\rho_s^\parallel(\partial_\parallel\phi)^2,
\label{HGMff}
\end{eqnarray}
where $\partial_\parallel\equiv\hat\pv_0\cdot\grad$, $\rho^{||}_s =
8Jp^2|\Delta_{p_0}|^2$ is the superfluid stiffness along $p_0$, $K
= 2J|\Delta_{p_0}|$, and  $n$ the density operator (only well-defined on a lattice), canonically conjugate to the phase field $\phi$.  The spontaneous breaking of rotational symmetry requires a strict vanishing of FF's transverse superfluid stiffness,
$\rho^{\perp}_s=0$.\cite{Shimahara,RVprl,Rpra}

The time-reversal preserving LO state spontaneously breaks both rotational and
translational symmetries, with a ($-\pv_0,\pv_0$) order parameter 
\bse
\begin{eqnarray}
\Delta_{LO}(\rv)
&=&2\Delta_{p_0}e^{i\phi}
\cos\big[\pv_0\cdot\rv + \theta\big],
\end{eqnarray}
\label{DeltaLO}
\ese
that is a {\it product} of a superfluid and a unidirectional density
wave order parameters. These are respectively characterized by two
Goldstone modes $\phi(\rv)$ and $\theta(\rv)$, corresponding to the superfluid phase
and the smectic phonon $u(\rv) = -\theta(\rv)/p_0$ of the striped state.

Similarly to the FF state, the underlying rotational symmetry of the
LO state strongly restricts the form of the Goldstone-mode
Hamiltonian.  Namely, its $\theta(\rv)=-p_0 u(\rv)$ sector must be invariant
under a rotation of $\pv_0$, that defines the spontaneously-chosen
orientation of the pair-density wave, and therefore must be described
by a smectic form \cite{deGennesProst,ChaikinLubensky,GP}. On the other
hand because a rotation of the LO state leaves the superconducting
phase, $\phi(\rv)$ unchanged, the superfluid phase $\phi(\rv)$ sector of the
Hamiltonian is therefore expected to be of a conventional $xy$-model
type. Consistent with these symmetry-based expectations the LO
Goldstone-mode Hamiltonian was indeed found\cite{RVprl,Rpra,Samokhin-2010,Samokhin-2011} to be
given by
\begin{eqnarray}
\cH_{LO}&=&\oh\Pi^2 + \oh K(\nabla^2 u)^2 + 
\oh B(\partial_\parallel u)^2\nonumber\\
&&+\oh\chi^{-1} n^2 + \frac{1}{2}\rho_s^\parallel(\partial_\parallel\phi)^2 
+ \frac{1}{2}\rho_s^\perp(\nabla_\perp\phi)^2,
\label{HgmLO}
\end{eqnarray}
with $\Pi$ the momentum operator field, which is canonically conjugate to the phonon $u$. 
Thus, the LO state is a highly anisotropic superfluid (though less so
than the FF state, where $\rho_s^\perp=0$), with the ratio
\begin{equation}
\frac{\rho_s^\perp}{\rho_s^\parallel}=
\frac{3}{4}\left(\frac{\Delta_{p_0}}{\Delta_{BCS}}\right)^2
\approx\frac{1}{4}\ln\left(\frac{h_{c2}}{h}\right)\ll 1,
\label{derive:ratio}
\end{equation}
that vanishes for $h\rightarrow h_{c2}^-$.\cite{RVprl,Rpra}

We note that in the presence of underlying rotational invariance, at nonzero temperature, the collinear FFLO states exhibit a 3d quasi-long-range translational order. Thus, the translational symmetry is restored and somewhat oxymoronically, the FFLO order parameter $\Delta_{\bf P}$ vanishes inside a collinear FFLO phase. Consequently, the uniform $\Delta_{4e}$ is the fundamental nonzero order parameter at any nonzero temperature.

In addition to Goldstone modes, the low-energy phenomenology is also
controlled by topological defects, that in conventional superfluids
are limited to $2\pi$ vortices in the superfluid phase $\phi(\rv)$. In
stark contrast, the additional LO phonon Goldstone mode $\theta(\rv)$
also admits $2\pi$ vortices, corresponding to an integer $a$
dislocation in the striped LO order. Even more interestingly, as also discussed in Section 2.1, in
addition to these integer vortex $(\pm 2\pi,0)$ and dislocation $(0,\pm 2\pi)$ defects, the product nature of the LO order parameter,
\rf{DeltaLO} allows for half-integer vortex-dislocation composite
defects, $(\pm \pi,\pm \pi)$.\cite{Agterberg-2008,RVprl,Berg-2009b,Rpra,Lee-2014}  The sequential unbinding of this
larger class of defects leads to a rich variety of LO descendent phases. Many interesting consequences such as three-dimensional quasi-long-range order, importance of Goldstone-mode nonlinearities, charge 4e superconductivity, exotic topological phases and transitions of the enriched nature of the FF and LO states have been extensively explored in Refs.\cite{RVprl,Rpra,RSreview}.

Finally, we note that although the bosonic sector of the FFLO state, discussed above is well understood, the problem is seriously complicated by the gapless 
fermions confined to the $\pm\Delta$ domain-walls of the PDW. These will certainly lead to damping of the bosonic Goldstone modes. Coupling between the gapless fermions confined to strongly fluctuating FFLO phonons remain a challenging open problem, some aspects of which are discussed in Ref.\cite{Rpra}.


\subsection{Nonuniform pairing in non-centrosymmetric and Weyl superconductors} 
\label{sec:noncentro}


Nonuniform pairing states with  FFLO-like pairing mechanism have also been obtained in non-centrosymmetric systems~\cite{agterberg-2007,dimitrova-2007}. In general these systems involve pairing on Fermi surfaces (FS) whose centers do not sit at high symmetry points in the Brillouin zone, and thus the Cooper pairs forming from these FSs carry nonzero momentum.

\begin{figure}[hbt]
 \includegraphics[width=0.7\columnwidth]{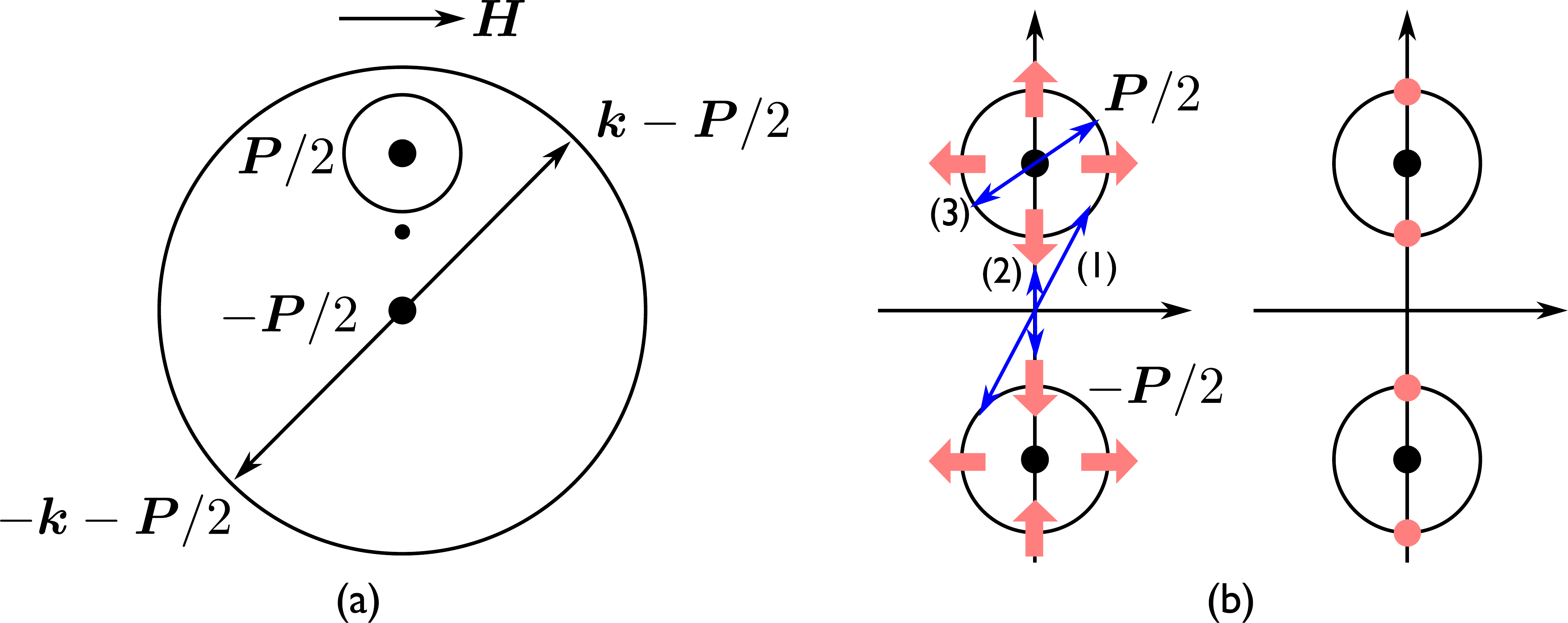}
 \caption{($a$):The split Fermi surfaces described by Eq.~\eqref{eq:rashbaH} Figure adapted from Ref.~\cite{agterberg-2007}. ($b$): Left: spin texture on the Weyl FS's, and various Cooper pairs. Arrows 1 and 2 show Cooper pairs in a uniform SC state, but for Arrow 2 the fermions have the same spin, and a singlet pairing gap vanishes here. Arrow 3 shows the Cooper pairs in a PDW state with intra-FS pairing. Right: The point nodes in the spin-singlet pairing state. A more generic reasoning for point nodes was given later; see main text. Panel \textit{a} adapted from Reference \cite{agterberg-2007}, and panel \textit{b}
adapted from Reference \cite{cho-2012}. }
 \label{fig:RashbaH}
 \end{figure}

One such scenario is for a metal with Rashba spin-orbit coupling in an in-plane Zeeman field $H$:
\begin{align}
H=\frac{\bm{k}^2}{2m}-\mu + \lambda({\bm k}\times{\bm\sigma})\cdot \hat{\bm z}-\mu_B H\sigma_x
\label{eq:rashbaH}
\end{align}
The superconducting instabilities of the concentric split FS's at $H=0$ have been analyzed in Ref.~\cite{gorkov-rashba-2001} (see also Ref.~\cite{frigeri-2004}). 
With $H\neq 0$, in the limit $\mu_B H \ll \lambda k_F$, the two split FS's approximately retain their shapes and get relatively shifted by $\pm \bm P/2$,  where 
$\bm P = \frac{2\mu_B H}{v_F} \hat{\bm{y}},$
 shown in Fig.~\ref{fig:RashbaH}. Naturally, the finite-momentum pairing order parameters that couples as $\Delta_{\pm {\bm P}}\,c^\dagger({\bm k}\pm{\bm P}/2)c^\dagger(-{\bm k}\pm{\bm P}/2)$, which separately gap out the two shifted split FS's.   Compared with the FFLO scenario in which only a small part of the FS is gapped, the present state has a larger condensation energy and can potentially be realized at relatively small coupling.
 
 Three pairing states have been theoretically investigated in Refs.~\cite{barzykin-2002,agterberg-2007,dimitrova-2007}: a uniform pairing order, an FF-like ``helical" pairing order with only one ordering momentum (say $\Delta_{+\bm{P}}$), and an LO-like stripe pairing order with both $\Delta_{\pm \bm P}$.  Note that unlike a canonical PDW state where $\Delta_P$ and $\Delta_{-P}$ are related by inversion and time-reversal symmetry, here both symmetries are explicitly broken already in the normal state. In this case, symmetry arguments on the Ginzburg-Landau free energy imply the superconducting ground state generally has finite-momentum pairing~\cite{Mineev-1994,agterberg-2002}.  In  microscopic studies, it was indeed obtained that the finite-momentum pairing orders including helical order and stripe order occupy sizable regions in the pairing phase diagram as a function of temperature and the in-plane Zeeman field~\cite{agterberg-2007,dimitrova-2007,Mineev-2008}.  Properties of these nonuniform pairing states in non-centrosymmetric materials have been reviewed in Ref.~\cite{smidman-2017}
 
Signatures of such a stripe PDW-like state were indeed observed in proximitized HgTe Quantum Wells~\cite{yacoby-2017}. When gated to electron-doped regime, HgTe quantum well exhibits split FS's with Rashba spin-orbit coupling. The quantum well is coupled to a conventional superconductor on one side and subject to an in-plane magnetic field $\bm B$. As $\bm B$ varies, the oscillation of the Josephson current across the quantum well has been observed, which is evidence for finite-momentum pairing order induced in HgTe.
The in-plane Zeeman field either can be an external field, or can be realized intrinsically. In Ref.~\cite{michaeli-2012} PDW pairing has been proposed for SrTiO$_3$-LaAlO$_3$ oxide interfaces, which exhibits coexistence between ferromagnetism and superconducting orders. It was shown that the effective interaction between the local moments drive the interface into a ferromagnetic phase. Together with the Rashba spin-orbit coupling, the system was proposed to realize a PDW-like state via the aforementioned mechanism.
  Another context of this mechanism for nonuniform pairing is the surface superconductivity on topological insulators (TI). Much of the attention towards TI-surface superconductivity has been focused on its realization of topological superconductivity, i.e., the well-known Fu-Kane superconductivity~\cite{Fu-2008}. However, with a nonzero chemical potential, the surfaces also hosts $\pm \bm{P}/2$-shifted FSs with a spin texture \cite{santos-2010}. However, unlike the FS in Fig.~\ref{fig:RashbaH}, the two FS's are located at opposite spatial surfaces. Moreover, inversion symmetry is not necessarily broken in the three-dimensional system, thus the two FCs can have the same size. Remarkably, there is recently experimental evidence observing finite-$\bm P$ pairing on TI surfaces with an in-plane Zeeman field.~\cite{mason-2018}

An interesting extension of this mechanism for the nonuniform pairing state is to 3d systems with FSs not centered around any high symmetry points $\pm\bm{P}/2$~\cite{cho-2012, burkov-2015,li-haldane-2018,wang-ye-2016}. The simplest way of obtaining these FS's is from doping a Weyl semimetal, which does not require any symmetry to stabilize.
A simple two-band lattice model given by Ref.~\cite{cho-2012} describing this situation is
\begin{align}
H_0 = t(\sigma^x\sin k_x + \sigma^y \sin k_y) + t_z[ \cos k_z -\cos (P/2)] \sigma^z + m(2-\cos k_x - \cos k_y)\sigma^z -\mu
\end{align}
For small chemical potential $\mu$, each spherical FS encloses a Weyl point at $\pm\bm{P}/2$ and is spin-textured, analogous to the 2d case with Rashba spin-orbit coupling (see Fig~\ref{fig:RashbaH}(a) for a $k_y=0$ slice).

Both intra-FS and inter-FS pairing can potentially occur, giving rise to a PDW-like state with ordering momenta $\pm\bm P$ and a uniform SC state respectively. The energetic interplay between PDW and a uniform superconductor depends on several factors. First, unlike the previous case where the FS's remain symmetric about their shifted centers for small $H$, here in general there is no symmetry relating $\epsilon(\bm{k+P}/2)$ and $\epsilon(-\bm{k+P}/2)$ unless $\mu$ is very small. Either way, the susceptibility towards a PDW state is reduced, while typically inversion symmetry (for the present case with two Weyl nodes) or time-reversal symmetry (e.g., for cases with four Weyl nodes) guarantees a weak-coupling instability towards a uniform superconductor. 
By contrast, there is a robust topological reason that the uniform SC order parameter has point nodes, which tends to suppress the uniform superconductor.  This was first observed in Ref.~\cite{cho-2012} and formulated in generic cases in Ref.~\cite{li-haldane-2018}. Weyl points are monopoles of the Berry curvature ${\bm {\mathcal{B}}} (\bm{k}) = i\langle\nabla_{\bm k} u({\bm k})|\times |\nabla_{\bm k} u(\bm k)\rangle$  in $\bm k$ space. Weyl points at $\pm \bm{P}/2$ carry monopole charges $\pm 1$, and this monopole charge is equal to the Berry flux through its enclosing FS. In the Nambu space of the pairing Hamiltonian, the Berry fluxes through the electron-like FS and hole-like shadow FS subject to pairing, and the monopole charges adds up. It was shown that~\cite{li-haldane-2018,wang-ye-2016} the total monopole charge inside the original FS is 2 for uniform (inter-FS) SC, and is 0 for (intra-FS) PDW. This means that the uniform SC state has to host {\it at least} two point nodes on each FS, independent of any microscopic details. The PDW state can be fully gapped, likely leading to a higher condensation energy.

The detailed interplay between these two opposite effects has been examined in Ref.~\cite{burkov-2015} with short-range attractive interactions in PDW and SC channels, and the authors found that PDW state is favored for a noncentrosymmetric Weyl metal. However, in more realistic systems a more careful examination on the band structure and the interaction is needed to pin down the superconducting ground state.


\section{BROADER RELEVANCE FOR THE CUPRATE SUPERCONDUCTORS}
\label{sec:broader}


In the preceding sections we presented evidence of the existence of pair-density-wave superconducting order in diverse systems 
ranging from cuprate high temperature superconductors to heavy fermionic materials to organics to topological materials, as well in cold atomic systems. Unlike its FFLO predecessors, the PDW discussed in the context of the Cuprate family does not require an external magnetic field for its existence. 

As reviewed in earlier sections, the PDW is a new state of matter with unique properties not encountered in other superconductors. 
It is a superconducting state with more than one complex order parameter. 
Its more complicated order parameter manifold allows this state to accommodate various charge orders, together with superconducting states, some even with an exotic flux quantization. This richness leads, in a natural way, to an explanation of several intriguing experimental effects, such as dynamical layer decoupling and a 
rich structure of superconducting vortices. 
These features also 
imply a complex phase diagram  
with a variety of broken symmetry phases.

\begin{figure}[hbt]
\begin{center}
\includegraphics[width=0.6\textwidth]{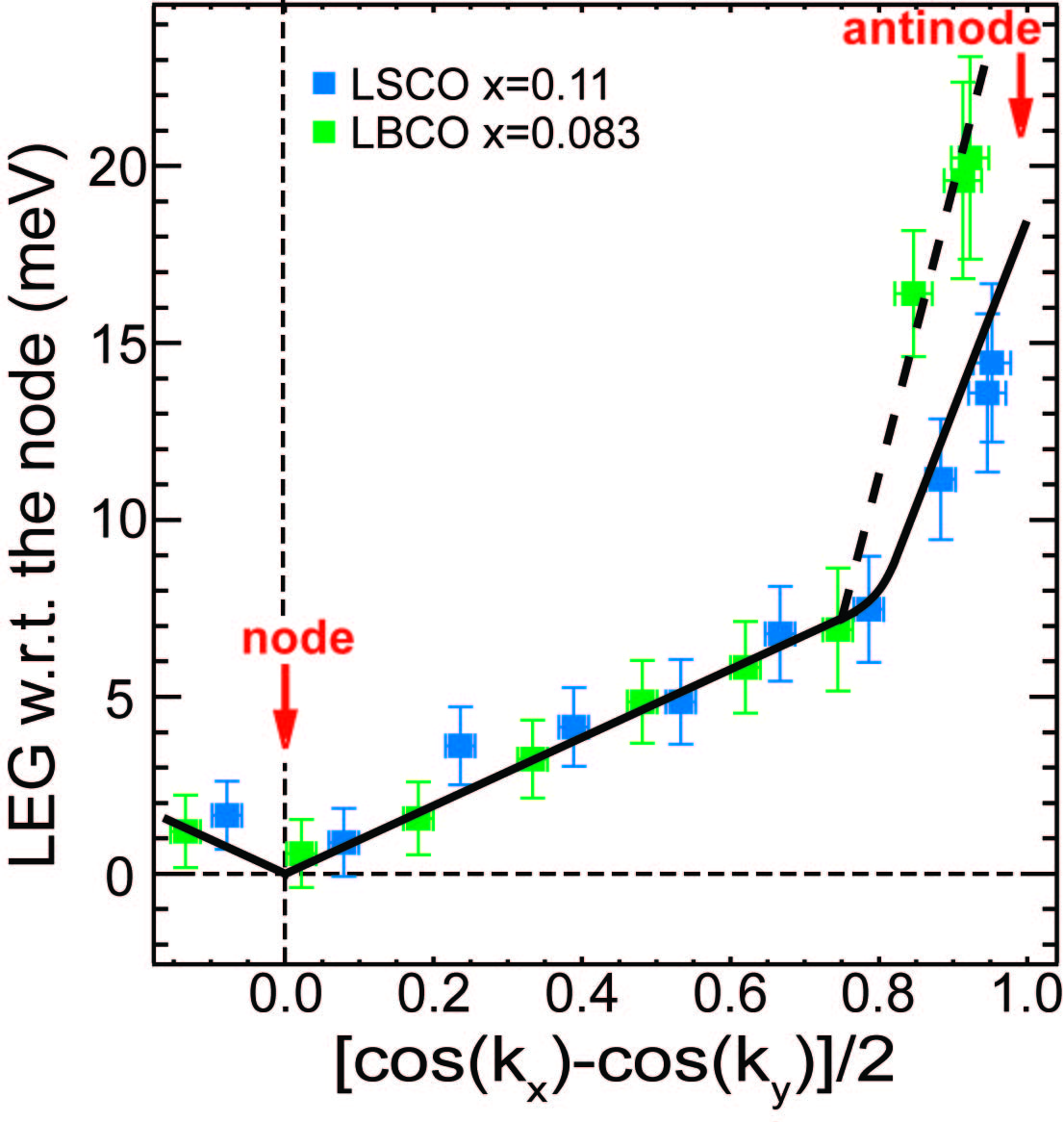}
\caption{ Momentum-dependent  gap of {\lbco} and {\lsco} from ARPES experiments by Vishik et al. \cite{Vishik-2010}. Abbreviations: ARPES, angle-resolved photoemission spectroscopy; LBCO, {\lbco} LEG, leading
edge gap; LEM, leading edge midpoint; LSCO,{\lsco}. Figure adapted from Reference \cite{Vishik-2010}.}
\label{Fig:lbco-arpes}
\end{center}
\end{figure}

Given the  evidence that PDW order appears  
in some places in the cuprate phase diagram, it remains to discuss the implication of this observation in the broader context of cuprate physics (and beyond).  Up to now the evidence 
comes primarily from two groups of experiments on two groups of materials. First, as reviewed in section 3.1, the layer decoupling observed in 1/8 doped LBCO which led to the PDW concept. 
This suggestion has been leant further credibility by the observation of a variety of other related phenomena, such as the detection of an anomalously large second harmonic in the Josephson relation in LBCO-Nb junctions and the appearance of the familiar signatures of layer decoupling in other 214 materials when  stripe order is enhanced, including LSCO in a magnetic field.  Second, a CDW with wave-vector equal to half of what is commonly observed was seen in the vicinity of the vortex core in underdoped Bi-2212 \cite{Edkins-2018}.
As such a CDW subharmonic  is expected as a  consequence of coexisting uniform and PDW order, its observation 
provides strong evidence of the existence of PDW in the vortex halo, as discussed in section 3.2.

It is worth mentioning that none of these pieces of evidence is entirely immune to the possibility of alternative explanations;  indeed, in all cases there are additional experimental observations that while not actually contradictory with the PDW interpretation, are also not entirely natural.  Most importantly, to date no diffraction experiments have detected the expected CDW subharmonic associated with the coexistence of PDW and uniform SC correlations, either in LBCO or in the magnetic field and temperature regimes that have been so far explored in YBCO and BSCCO. 
Conversely, in the range of $T$ in which layer decoupling gives strong evidence of dominantly PDW correlations in 1/8 doped LBCO,  ARPES data 
have been interpreted \cite{He-2008,Vishik-2010} (see fig. 14) as showing a nodal-d-wave-like  
one electron spectrum, rather than the nodal-arc spectrum expected for a pure PDW  (similar two-gap ARPES spectra has been seen in BSCCO \cite{Vishik-2010,leeWS-2007}; see fig. 15).
In all cases, there are multiple possible ways one can imagine reconciling these observations with the PDW interpretation. However, ultimately we will rely on further experiments (some of which are discussed below) to resolve these issues.

\begin{figure}[htb]
\begin{center}
\includegraphics[width=0.6\textwidth]{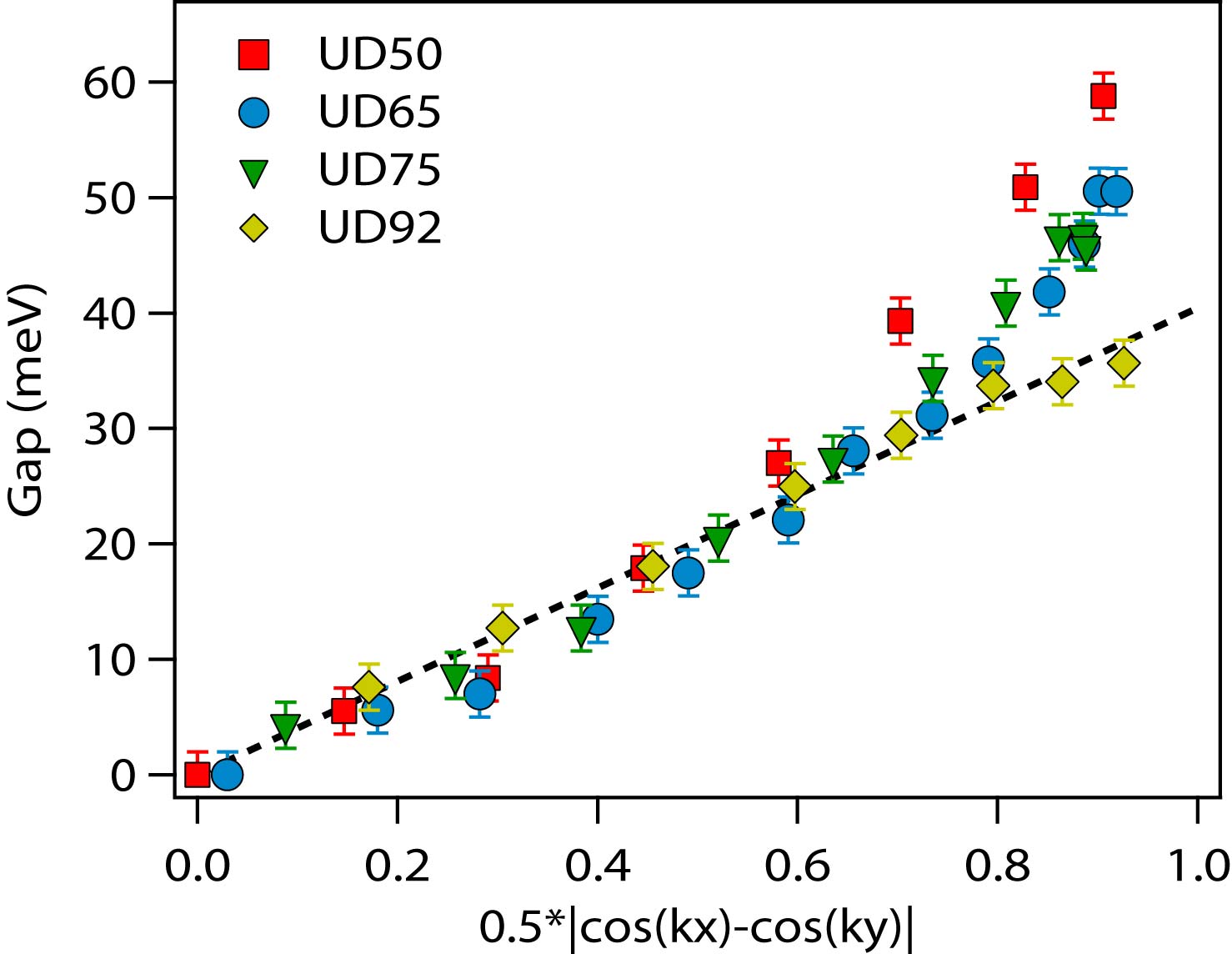}
\caption{ Momentum-dependent gap in {\bscco} from ARPES experiments by Vishik et al. \cite{Vishik-2010}. Abbreviation: ARPES, angle-resolved photoemission spectroscopy. Figure adapted from Reference \cite{Vishik-2010}.}
\label{Fig:bscco-arpes}
\end{center}
\end{figure}

\begin{figure}[b]
\begin{center}
\includegraphics[width=3in]{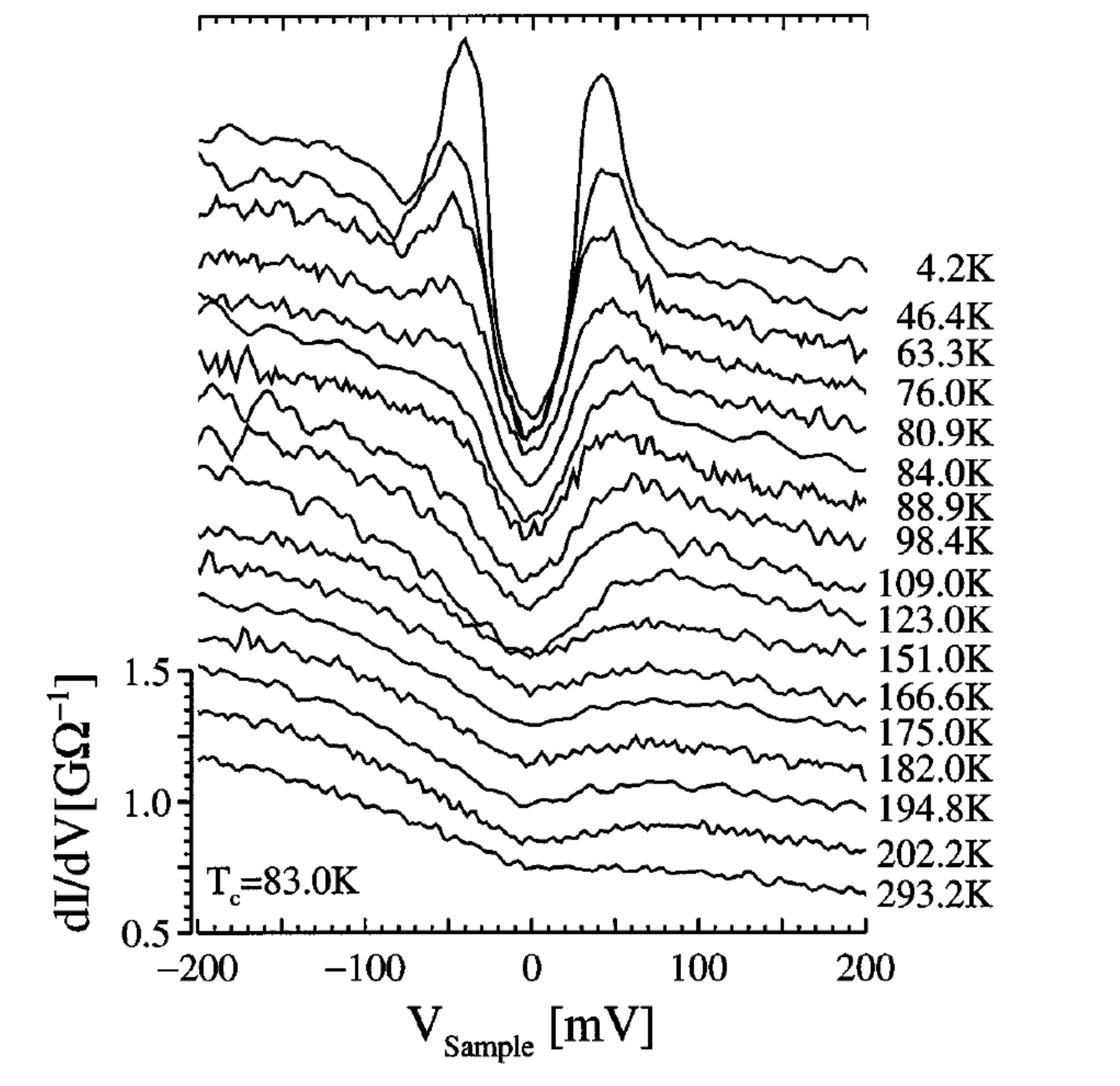}
\caption{Pseudogap in {\bscco} from STM experiments by Renner et al. Figure adapted from Reference \cite{renner-1998}.}
\label{Fig:renner}
\end{center}
\end{figure}

Putting these concerns aside, a key question 
is whether these PDW ``sightings'' 
are relevant only in the relatively narrow context of cuprates with certain structural peculiarities (such as LBCO) or in the vicinity of isolated vortex cores,
or whether they have broader implications to the physics of underdoped cuprates. 
In particular, there is the question of  whether PDW correlations are in any way responsible for the so-called ``pseudo-gap regime''
observed in underdoped cuprates. 
 For the purpose of this review, 
we will focus the discussion of this question on the range of 
hole doping from $p\sim$ 0.08 to 0.15 in which there is an identifiable temperature scale, $T^*$, below which various measurable properties show a depletion of the density of states at low energies. This 
thus pertains to  an intermediate doping range between
very low doping where physics is dominated by significant local antiferromagnetic order 
and the high doping region where a full Fermi surface enclosing  
$1+p$ holes forms. 
Note that 
$T^*$ 
 is at largest 300K-400K, which is still low compared with microscopic scales such as the
 exchange scale $J$. A phenomenon closely associated with this $T^*$ scale is the appearance
  of  a deep depression in the single-particle spectral weight ({\it i.e.} the eponymous pseudo-gap) in the anti-nodal portion of the  Fermi surface up to 
  $T^*$, which in turn can be much higher than $T_c$. This pseudo-gap has been seen in not only in ARPES data and but also by STM, as 
  shown in
  Fig. \ref{Fig:renner};  
  importantly, since STM  
  accesses both the unoccupied and occupied states, 
  it reveals an approximate particle-hole symmetry of this gap. This suggests that the gap is associated with some form of SC pairing. 
  On the other hand, certain features of the gap as inferred from ARPES evolve differently as a function of doping and temperature in the near nodal region where the gap is small and in the antinodal region where it is large -- the so-called nodal-antinodal dichotomy.  This can be seen to some extent in the data from Bi2212 shown in Fig. \ref{Fig:bscco-arpes}, and more dramatically
in the case of LBCO (Fig. \ref{Fig:lbco-arpes}) and Bi-2201 where detailed ARPES data are available \cite{He-2011}.
This apparent dichotomy has led many researchers to conclude that the pseudo-gap has a distinct (non-superconducting)  origin.

In addition to the energy gap, another striking feature of the underdoped cuprate phase diagram is the existence of ``intertwined'' order parameters corresponding to multiple distinct broken symmetry states that occur with similar energy and temperature scales, and which in some ways compete and in some ways cooperate strongly with each other.  In addition to the insulating N\'eel order and uniform d-wave superconducting order, this list includes a variety of other metallic or superconducting spin-density wave, CDW, and nematic orders.  To this list we now add PDW order.  It it is unreasonable that these materials should be accidentally fine-tuned close to an extraordinarily complex multicritical point at which all these orders are degenerate with each other, so it is reasonable to search for a description in which the observed orders derive from a smaller set of  ``primary'' order parameters. 

The underlying idea is that, under certain circumstances, some class of  ``soft'' fluctuations can partially melt a parent broken symmetry state in such a way that some but not all the underlying symmetries are restored, leaving behind a partially ordered state with some form of vestigial order \cite{Nie-2013}.  Formally, as discussed in Sec. 2, this corresponds to forming composite order parameters that are bilinear (or higher order) in the primary order parameter fields.  In the present context, starting from an assumed primary PDW order parameter, one can readily construct composite orders corresponding to both CDW order and nematic orders making it possible to view these as vestigial orders remaining when a fully ordered PDW is partially melted by strong phase fluctuations.  

There are several phenomenologically appealing aspects of this perspective.  With regards to nematic order, while there is suggestive evidence that it arises at temperature scales comparable to the pseudo-gap $T^*$, it is very difficult to see how a primary nematic order could be responsible for the pseudo-gap, but viewed as a composite order, it can be understood as an avatar of a more basic set of correlations.  Similarly, the observed CDW order is mostly weak in the sense that it causes unusually small magnitude lattice distortions ,which is why it was overlooked for so many years. (LBCO may be an exception.)  The CDW does not open gaps or fold the Fermi surface observed in ARPES in ways that are familiar from other classes of CDW materials. \cite{He-2011} Still, CDW order appears to be thermodynamically robust in both the sense that significant CDW correlations persist over a wide range of $T$ and $p$, and that it appears to compete on a more or less equal footing with the uniform d-wave SC state,  for instance in that there is a clear suppression of the superconducting $T_c$ in the range of dopings where the CDW correlations are strongest, while high $T_c$ can be restored with the application of modest pressure, presumably because this suppresses the CDW tendency.  Again, viewing the CDW as a composite order gives a rationale for viewing it as being simultaneously weak and strong in different aspects.

 At a more local level, there has long been strong intuitive appeal to viewing the pseudo-gap as a form of local pairing without any significant SC phase coherence.  The fact that the pseudo-gap shares so many obvious similarities with the d-wave SC gap is, of course, the strongest piece of evidence in favor of this interpretation.  For instance, at low temperatures, the gap magnitude along the FS seen in ARPES in Bi-2212 for a range of doping roughly in the range $0.12 < p < 0.19$, has the simplest $d$-wave form, $\Delta_0\{ [\cos(k_x) - \cos(k_y)]/2\} $. 
  This gap is  well understood as being due to the $d$-wave SC state.  However,  
  there is a distinct difference in the thermal evolution of the near nodal and antinodal gaps; this is an aspect of the previously mentioned  nodal anti-nodal dichotomy.  In particular,
  in the anti-nodal regime of the BZ, the gap magnitude is only very weakly $T$ dependent and it evolves smoothly 
  into the pseudo-gap that persists to well above $T_c$.  Moreover, as seen in Fig. \ref{Fig:bscco-arpes}  as a function of decreasing $p$, the energy gap below $T_c$ at the antinode starts to grow strongly but continuously, i.e. the gap structure increasingly looks like the sum of a near nodal superconducting gap (whose magnitude is in fact largely independent of $p$ down to $p \approx 0.07$) plus another gap which is large at the antinode and small or vanishing in an arc region near the nodal point.  In the context of the present work, it is 
  tempting to associate this evolution with an increasingly significant admixture of a PDW component in the gap structure. Also, at least in one material where detailed ARPES spectra  is available, Bi-2201, the top of the gap does not line up with the location of the Fermi momentum, as one would expect for a uniform superconductor, but is consistent with a momentum carrying condensate, such as the PDW \cite{He-2011,Lee-2014}. 
  While it may be possible to account for the differences in the thermal evolution of the gap in the nodal and antinodal regimes as a purely kinematic effect of phase fluctuations in a simple $d$-wave superconductor \cite{bergandaltman,dessau}, the nodal anti-nodal dichotomy has been taken as evidence against fluctuating d-wave as the origin of the pseudo-gap at the anti-node \cite{Vishik-2010,hash14b}. The PDW seems like a natural candidate that is next in line.
 
 To summarize, there remains the larger question of the broader significance of the PDW in the phase diagram of the cuprates and, particularly, of its role in the pseudogap regime. 
 We find it useful to focus on two extremal 
 perspectives which lie at the opposite ends of
 a spectrum of possibilities. 
 For ease of discussion, we will label the first view as ``competing order'' and the second view as ``mother state.''
 
   In the ``competing order" perspective, the PDW is a close competitor of uniform d-wave superconductivity and of the observed CDW orders, which 
   are to be regarded as independent (although strongly coupled) order parameters. Their interplay may change from one family of cuprates to another (or in a particular family as a function of doping) depending on other factors, e.g. the crystal structure, whether SDW order plays a role or not, etc.  This is especially clear in the case of LBCO, where the critical temperatures are comparable, which naturally leads to the notion that they may have a common microscopic origin (and, in this sense, are intertwined). Nevertheless, this does not necessarily imply that they should not be regarded as completely separate orders. 
This is because there are  terms in the Landau theory of the type presented in earlier sections, e.g.  trilinear couplings of the form $\rho_{2{\bm P}} \Delta_{{\bm P}}^* \Delta_{-{\bm P}}$. Physically, these terms can  interpreted as saying that there should be a non-vanishing value of the CDW order $\rho_{2{\bm P}}$ as soon as  the PDW order parameter is present, $|\Delta_{\pm {\bm P}}|\neq 0$, i.e. $\rho_{{\bm P}} \propto  \Delta_{-{\bm P}}^* \Delta_{{\bm P}}$, and the CDW order is a composite order. However, this coupling can also be interpreted as saying that as $\rho_{2{\bm P}}$ becomes sufficiently strong, the quadratic term of the PDW  order can aquire a negative coefficient which when large enough, can induce a PDW order parameter $\Delta_{\pm {\bm P}}$.
   
   A convenient mathematical language to describe competing order is the nonlinear $\sigma$ model. This has been employed to describe the appearance of static PDW in the vicinity of the vortex core, where d-wave order smoothly rotates to a PDW order \cite{Wang-2018}. Another situation  for its application may be the onset of layer decoupling driven by a magnetic field in LSCO \cite{Schafgans-2010}. In that case one may envision a rotation to a PDW from a d-wave under the influence of magnetic field driven spin order.
   
 A second perspective is to regard the PDW as the primary order (the ``mother state''), and  to regard other orders (e.g. CDW) as composite (or descendant) orders.  If the low temperature phase is indeed a PDW, then the other orders may appear through a partial melting cascade of phase transitions. 
 In these regimes, the CDW (and nematic) orders are vestigial orders of the PDW, and the PDW itself can be regarded as a ``fluctuating'' order. 
 
  As a fluctuating order, the PDW is locally defined on some  length scale  without ever being the ground state. The correlation length should be long enough to form the antinodal gap and to induce the composite orders, but short enough that the properties usually associated with superfluidity are not apparent. Taken to its logical conclusion, this point of view states that PDW fluctuations are pervasive over a large part of the doping, temperature and magnetic field range of the phase diagram and are the root cause of the pseudogap phenomenology. This view has been strongly advocated by one of us \cite{Lee-2014}, while a more guarded proposal  that fluctuating PDW may be responsible  for the pseudo-gap structure at the anti-node and the Fermi arc near the node have been made earlier \cite{Berg-2009}.
  The appeal of this picture is that  it reduces the explanation of the panoply of observed orders to the existence of local PDW orders. The rapid winding near the vortex  has been  suggested to pin the PDW near the vortex core, rendering it static and generating the short range  CDW  with wavevector ${\bm P}$ observed in STM experiments \cite{Dai-2017}. Just as in the competing order scenario, the PDW provides a mechanism for lowering  the vortex core energies and, hence, making $H_{c2}$ parametrically smaller \cite{Wang-2018}. What happens for fields greater than $H_{c2}$ is probably the most intriguing open question. One has to face the question of how to describe a pairing state that has been destroyed by quantum phase fluctuations, which remains an unsolved problem \cite{Kapitulnik-2019}. 
  The quantum disordering of a PDW which already breaks translational symmetry and is gapless  
  even in the absence of a magnetic field, is still more clearly {\it terra incognita}. Experimentally CDW  with longer range order appears with increasing magnetic field and a metallic state with small Fermi pocket emerges. Whether this state can be the result of a fluctuating PDW remains to be seen \cite{Zelli-2012,Lee-2014,Baruch-2008,Berg-2009,Vafek-2013,Norman-2018,Agterberg-2015,Kacmarcik-2018}. It is worth noting that the competing order scenario faces the same issue: in the non-linear sigma-model description \cite{Wang-2018}, an ordered PDW appears when the vortex halo's overlap and one needs to address the question of how this order is destroyed.
  
  Another important question for the ``mother state'' scenario is whether a coherence length large enough to create an energy gap will necessarily produce large observable consequences of superconducting fluctuations. Experimentally diamagnetic fluctuations have been observed up to two or three times $T_c$ and also above $H_{c2}$ at low temperatures \cite{Li-2010,Ong-2016}. However, there is little in the way of transport signatures. Can these observations be reconciled? The ultimate question is whether these ideas and the contrast between the two extreme perspectives can be either directly supported or falsified. In this sense the recent proposal of a method to directly measure PDW fluctuations in a tunnel junction using  Bi-2201 as one electrode and then taking advantage of the known momentum that is present in this material offers some hope for the future \cite{Lee-2018a}. 
  
 Finally, 
  it is important to determine the extent to which the 1${\bm P}$ CDW observed so far by STM in Bi-2212 can be 
  detected with other probes  and  in  other cuprate families.  We 
  note that several groups have searched for, but so far failed to find, 
  evidence of a subharmonic CDW ${\bm P}$ peak in  X-ray diffraction in the superconducting state of high $T_c$ 
  cuprates in which CDW order is known to be present.  
  So far unpublished X ray searches for
  a ${\bm P}$ peak  have been 
  carried out in $15.5 {\%} $ doped {\lbco} (P. Abbamonte, private communication) and in {\ybco} at relatively low temperatures in magnetic fields $\sim 6$T \cite{Blackburn-2019}. Data from other experiments in
   Bi$_2$(Sr, La)$_2$CuO$_{6+\delta}$ at zero magnetic field \cite{Peng-2016,Chaix-2017} also do not show evidence for this peak. It would be  interesting to  extend the X-ray searches  to  regimes 
in which the layer decoupling effect, an  indicator of PDW order, has already been seen, such as in underdoped {\lsco} \cite{Schafgans-2010} in a magnetic field and {\lbco}  \cite{Stengen-2013}, or in superconducting {\bscco} in the  regime where   STM  experiments see the 
peak at wavevector ${\bm P}$  peak in 
  vortex halos \cite{Edkins-2018}.

\section*{DISCLOSURE STATEMENT}
The authors are not aware of any affiliations, memberships, funding, or financial holdings that
might be perceived as affecting the objectivity of this review. 

\section*{ACKNOWLEDGMENTS}
J.C.S.D. acknowledges support from Science Foundation Ireland under Award SFI 17/RP/5445 and from the European Research Council (ERC) under Award DLV-788932. J.C.S.D and S.D.E acknowledge support, and the funding to carry out STM/SJTM studies of cuprate PDWs, from the Gordon and Betty Moore Foundation EPiQS Initiative through Grant GBMF4544. SDE acknowledges support from the Karel Urbanek Postdoctoral Fellowship at Stanford University. SAK was supported in part by NSF grant DMR-1608055. EF was supported in part by NSF grant DMR-1725401. This work was funded by the Office of
Basic Energy Sciences, Materials Sciences and Engineering Division, U.S. Department of Energy (DOE) under
contracts DE-SC0012368 (DVH,EF). DVH was supported in part by NSF grant DMR-1710437.
PAL acknowledges support by DOE grant DE-FG02-03ER46076. YW was supported by the Gordon and Betty Moore Foundation EPiQS Initiative through the grant GBMF 4305. LR was supported by the Simons Investigator Award from The James Simons Foundation. JMT was supported at Brookhaven by DOE Contract No.\ DE-SC0012704. This research was supported in part by the National Science Foundation under Grant No. NSF PHY-1748958 at the Kavli Institute for Theoretical Physics of the University of California Santa Barbara, and we  thank KITP for its hospitality.

\begin{thebibliography}{237}
\expandafter\ifx\csname natexlab\endcsname\relax\def\natexlab#1{#1}\fi

\bibitem{Larkin-1964}
Larkin AI, Ovchinnikov YN. 1965.
\textit{Sov. Phys. JETP} {\bf 20}, 762

\bibitem{Fulde-1964}
Fulde P, Ferrell RA. 1964.
\textit{Phys. Rev.} 135:A550

\bibitem{Himeda-2002}
Himeda A, Kato T, Ogata M. 2002.
\textit{Phys. Rev. Lett.} 88:117001

\bibitem{Berg-2007}
Berg E, Fradkin E, Kim EA, Kivelson SA, Oganesyan V, et~al. 2007.
\textit{Phys. Rev. Lett.} 99:127003

\bibitem{Agterberg-2008}
Agterberg DF, Tsunetsugu H. 2008.
\textit{Nature Phys.} 4:639

\bibitem{Berg-2009}
Berg E, Fradkin E, Kivelson SA, Tranquada JM. 2009.
\textit{New J. Phys.} 11:115004

\bibitem{Lee-2014}
Lee PA. 2014.
\textit{Phys. Rev. X} 4:031017

\bibitem{Casalbuoni-2004}
Casalbuoni R, Nardulli G. 2004.
\textit{Rev. Mod. Phys.} 76:263

\bibitem{TormaReview2018}
Kinnunen JJ, Baarsma JE, Martikainen JP, T{\"o}rma P. 2018.
\textit{Reports on Progress in Physics} 81:046401

\bibitem{Berg-2009b}
Berg E, Fradkin E, Kivelson SA. 2009.
\textit{Nat. Phys.} 5:830--33

\bibitem{Agterberg-2015}
Agterberg DF, Garaud J. 2015.
\textit{Phys. Rev. B} 91:104512

\bibitem{Wang-2018}
Wang Y, Edkins SD, Hamidian MH, Davis JCS, Fradkin E, Kivelson SA. 2018.
\textit{Phys. Rev. B} 97:174510

\bibitem{Lee-2018}
Dai Z, Zhang YH, Senthil T, Lee PA. 2018.
\textit{Phys. Rev. B} 97:174511

\bibitem{Edkins-2018}
Edkins SD, Kostin A, Fujita K, Mackenzie AP, Eisaki H, et~al. 2019.
\textit{Science} 364:976

\bibitem{Fradkin2015}
Fradkin E, Kivelson SA, Tranquada JM. 2015.
\textit{Rev. Mod. Phys.} 87:457--482

\bibitem{Agterberg-2009}
Agterberg DF, Sigrist M, Tsunetsugu H. 2009.
\textit{Phys. Rev. Lett.} 102:207004

\bibitem{Radzihovsky-2009}
Radzihovsky L, Vishwanath A. 2009{\natexlab{a}}.
\textit{Phys. Rev. Lett.} 103:010404

\bibitem{Radzihovsky_2011}
Radzihovsky L. 2011{\natexlab{a}}.
\textit{Physical Review A} 84:023677

\bibitem{Barci-2011}
Barci DG, Fradkin E. 2011.
\textit{Phys. Rev. B} 83:100509

\bibitem{Senthil-2015}
Mross DF, Senthil T. 2015.
\textit{Phys. Rev. X} 5:031008

\bibitem{BKT-1}
Berezinskii V. 1972.
\textit{Sov. Phys. JETP} 34:610-616

\bibitem{BKT-2}
Kosterlitz J, Thouless D. 1973.
\textit{J. Phys. C} 6:1181

\bibitem{Jose-1977}
Jose J, Kadanoff L, Kirkpatrick S, Nelson D. 1977.
\textit{Phys. Rev. B} 16:1217

\bibitem{Chen-2004-2}
Chen HD, Vafek O, Yazdani A, Zhang SC. 2004.
\textit{Phys. Rev. Lett.} 93:187002

\bibitem{Nie-2013}
Nie L, Tarjus G, Kivelson SA. 2014.
\textit{Proceedings of the National Academy of Sciences} 111:7980

\bibitem{Chan-2016}
Chan C. 2016.
\textit{Phys. Rev. B} 93:184514

\bibitem{He-2011}
He R, Hashimoto M, Karapetyan H, Koralek J, Hinton J, et~al. 2011.
\textit{Science} 331:1579

\bibitem{Baruch-2008}
Baruch S, Orgad D. 2008.
\textit{Phys. Rev. B.} 77:174502

\bibitem{Harrison-2011}
Harrison N, Sebastian S. 2011.
\textit{Phys. Rev. Lett.} 106:226402

\bibitem{Wang-2015}
Wang Y, Agterberg DF, Chubukov A. 2015.
\textit{Phys. Rev. Lett.} 114:197001

\bibitem{Tu-2019}
Tu WL, Lee TK. 2019.
\textit{Scientific Reports} 9:1719

\bibitem{mood88}
Moodenbaugh AR, Xu Y, Suenaga M, Folkerts TJ, Shelton RN. 1988.
\textit{Phys. Rev. B} 38:4596--4600

\bibitem{fuji04}
Fujita M, Goka H, Yamada K, Tranquada JM, Regnault LP. 2004.
\textit{Phys. Rev. B} 70:104517

\bibitem{huck11}
H\"ucker M, v.~Zimmermann M, Gu GD, Xu ZJ, Wen JS, et~al. 2011.
\textit{Phys. Rev. B} 83:104506

\bibitem{axe89}
Axe JD, Moudden AH, Hohlwein D, Cox DE, Mohanty KM, et~al. 1989.
\textit{Phys. Rev. Lett.} 62:2751

\bibitem{axe94}
Axe JD, Crawford MK. 1994.
\textit{J. Low Temp. Phys.} 95:271

\bibitem{Li-2007}
Li Q, H{\"u}cker M, Gu GD, Tsvelik AM, Tranquada JM. 2007{\natexlab{a}}.
\textit{Phys. Rev. Lett.} 99:067001

\bibitem{tran08}
Tranquada JM, Gu GD, H{\"u}cker M, Jie Q, Kang HJ, et~al. 2008.
\textit{Phys. Rev. B} 78:174529


\bibitem{tran95a}
Tranquada JM, Sternlieb BJ, Axe JD, Nakamura Y, Uchida S. 1995.
\textit{Nature} 375:561

\bibitem{taji01}
Tajima S, Noda T, Eisaki H, Uchida S. 2001.
\textit{Phys. Rev. Lett.} 86:500--503

\bibitem{baso05}
Basov DN, Timusk T. 2005.
\textit{Rev. Mod. Phys.} 77:721--779

\bibitem{crof14}
Croft TP, Lester C, Senn MS, Bombardi A, Hayden SM. 2014.
\textit{Phys. Rev. B} 89:224513

\bibitem{tham14}
Thampy V, Dean MPM, Christensen NB, Steinke L, Islam Z, et~al. 2014.
\textit{Phys. Rev. B} 90:100510

\bibitem{suzu98}
Suzuki T, Goto T, Chiba K, Shinoda T, Fukase T, et~al. 1998.
\textit{Phys. Rev. B} 57:R3229--R3232

\bibitem{kimu00}
Kimura H, Matsushita H, Hirota K, Endoh Y, Yamada K, et~al. 2000.
\textit{Phys. Rev. B} 61:14366--14369

\bibitem{lake02}
Lake B, {R\o nnow} HM, Christensen NB, Aeppli G, Lefmann K, et~al. 2002.
\textit{Nature} 415:299--301

\bibitem{Schafgans-2010}
Schafgans AA, LaForge AD, Dordevic SV, Qazilbash MM, Padilla WJ, et~al. 2010.
\textit{Phys. Rev. Lett.} 104:157002

\bibitem{wen12}
Wen J, Jie Q, Li Q, H\"ucker M, v.~Zimmermann M, et~al. 2012.
\textit{Phys. Rev. B} 85:134513

\bibitem{steg13}
Stegen Z, Han SJ, Wu J, Pramanik AK, H\"ucker M, et~al. 2013.
\textit{Phys. Rev. B} 87:064509

\bibitem{zhon18}
Zhong R, Schneeloch JA, Chi H, Li Q, Gu G, Tranquada JM. 2018.
\textit{Phys. Rev. B} 97:134520

\bibitem{he09}
He RH, Tanaka K, Mo SK, Sasagawa T, Fujita M, et~al. 2009.
\textit{Nat. Phys.} 5:119--123

\bibitem{vall06}
Valla T, Federov AV, Lee J, Davis JC, Gu GD. 2006.
\textit{Science} 314:1914

\bibitem{razz13}
Razzoli E, Drachuck G, Keren A, Radovic M, Plumb NC, et~al. 2013.
\textit{Phys. Rev. Lett.} 110:047004

\bibitem{Loder-2011}
Loder F, Graser S, Schmid M, Kampf AP, Kopp T. 2011.
\textit{Phys. Rev. Lett.} 107:187001

\bibitem{home12}
Homes CC, H\"ucker M, Li Q, Xu ZJ, Wen JS, et~al. 2012.
\textit{Phys. Rev. B} 85:134510

\bibitem{li19}
Li Y, Terzic J, Baity PG, Popovi\'c D, Gu GD, et~al. 2018.
\textit{Science Advances} 5:eaav7686

\bibitem{raja18}
Rajasekaran S, Okamoto J, Mathey L, Fechner M, Thampy V, et~al. 2018.
\textit{Science} 359:575--579

\bibitem{yuli_scanning_2007}
Yuli O, Asulin I, Millo O, Koren G. 2007.
\textit{Physical Review B} 75:184521

\bibitem{Yang-2009}
Yang KY, Chen WQ, Rice TM, Sigrist M, Zhang FC. 2009.
\textit{New Journal of Physics} 11:055053

\bibitem{yuli_spatial_2010}
Yuli O, Asulin I, Koren G, Millo O. 2010.
\textit{Physical Review B} 81:024516

\bibitem{yang13}
Yang K. 2013.
\textit{J. Supercond. Nov. Magn.} 26:2741--2742

\bibitem{shi19}
Shi Z, Baity PG, Terzic J, Sasagawa T, Popovi\'c D. 2019.
arXiv:1907.11708.

\bibitem{buzdin_periodic_2003}
Buzdin A, Koshelev AE. 2003.
\textit{Physical Review B} 67:220504

\bibitem{moshe_shapiro_2007}
Moshe M, Mints RG. 2007.
\textit{Physical Review B} 76:054518

\bibitem{stoutimore_second-harmonic_2018}
Stoutimore MJA, Rossolenko AN, Bolginov VV, Oboznov VA, Rusanov AY, et~al.
  2018.
\textit{Phys. Rev. Lett.} 121:177702

\bibitem{schneider_half-h/2e_2004}
Schneider CW, Hammerl G, Logvenov G, Kopp T, Kirtley JR, et~al. 2004.
\textit{EPL (Europhysics Letters)} 68:86

\bibitem{hamilton_signatures_2018}
Hamilton DR, Gu GD, Fradkin E, Van~Harlingen DJ. 2018.
arXiv:1811.02048

\bibitem{Fujita2015}
Fujita K, Hamidian M, Firmo I, Mukhopadhyay S, Kim CK, et~al. 2015.
\textit{Strongly Correlated Systems}, \textit{Springer Series in
  Solid-State Sciences}, vol. 180, ed. A Avella, F Mancini, pp. 73-109,
Berlin, Heidelberg: Springer.

\bibitem{Hoffman-2002}
Hoffman JE, Hudson EW, Lang KM, Madhavan V, Eisaki H, et~al. 2002.
\textit{Science} 295:466

\bibitem{Matsuba2007}
Matsuba K, Yoshizawa S, Mochizuki Y, Mochiku T, Hirata K, Nishida N. 2007.
\textit{Journal of the Physical Society of Japan} 76:063704

\bibitem{Yoshizawa2013}
Yoshizawa S, Koseki T, Matsuba K, Mochiku T, Hirata K, Nishida N. 2013.
\textit{Journal of the Physical Society of Japan} 82:083706

\bibitem{Machida2016}
Machida T, Kohsaka Y, Matsuoka K, Iwaya K, Hanaguri T, Tamegai T. 2016.
\textit{Nature Communications} 7:1--6

\bibitem{Pan1998}
Pan SH, Hudson EW, Davis JC. 1998.
\textit{Applied Physics Letters} 73:2992--2994

\bibitem{Naaman2001}
Naaman O, Teizer W, Dynes RC. 2001.
\textit{Phys. Rev. Lett.} 87:097004

\bibitem{Rodrigo2004}
{Rodrigo, J. G.}, {Suderow, H.}, {Vieira, S.} 2004.
\textit{Eur. Phys. J. B} 40:483--488

\bibitem{Proslier2006}
{Proslier, Th.}, {Kohen, A.}, {Noat, Y.}, {Cren, T.}, {Roditchev, D.}, {Sacks,
  W.} 2006.
\textit{Europhys. Lett.} 73:962--968

\bibitem{Randeria2016}
Randeria MT, Feldman BE, Drozdov IK, Yazdani A. 2016.
\textit{Phys. Rev. B} 93:161115

\bibitem{Hamidian-2016}
Hamidian M, Edkins S, Joo SH, Kostin A, Eisaki H, et~al. 2016.
\textit{Nature} 532:343--347

\bibitem{Hanaguri2004a}
Hanaguri T, Lupien C, Kohsaka Y, Lee DH, Azuma M, et~al. 2004.
\textit{Nature} 430:1001

\bibitem{McElroy2005}
McElroy K, Lee DH, Hoffman JE, Lang KM, Lee J, et~al. 2005.
\textit{Phys. Rev. Lett.} 94:197005

\bibitem{Kohsaka2007}
Kohsaka Y, Taylor C, Fujita K, Schmidt A, Lupien C, et~al. 2007.
\textit{Science} 315:1380--1385

\bibitem{Mesaros2016}
Mesaros A, Fujita K, Edkins SD, Hamidian MH, Eisaki H, et~al. 2016.
\textit{Proceedings of the National Academy of Sciences} 113:12661--12666

\bibitem{Zhang2018}
Zhang Y, Mesaros A, Fujita K, Edkins SD, Hamidian MH, et~al. 2019.
\textit{Nature} 570:484-490

\bibitem{Hamidian2015}
Hamidian MH, Edkins SD, Kim CK, Davis JC, Mackenzie AP, et~al. 2015.
\textit{Nature Physics} 12:150

\bibitem{Fujita2014}
Fujita K, Hamidian MH, Edkins SD, Kim CK, Kohsaka Y, et~al. 2014.
\textit{Proceedings of the National Academy of Sciences} 111:E3026--E3032

\bibitem{Agosta-2018}
Agosta C. 2018.
\textit{Crystals} 8:285

\bibitem{Matsuda-2007}
Matsuda Y, Shimahara H. 2007.
\textit{J. Phys. Soc. Jpn.} 76:051005

\bibitem{Kenzelmann-2017}
Kenzelmann M. 2017.
\textit{Rep. Prog. Phys.} 80:034501

\bibitem{Bergk-2011}
Bergk B, Demuer A, Sheikin I, Wang Y, Wosnitza J, et~al. 2011.
\textit{Phys. Rev. B} 83:064506

\bibitem{Tsuchiya-2015}
Tsuchiya S, Yamada JI, Sugii K, Graf D, Brooks J, et~al. 2015.
\textit{J. Phys. Soc. Jpn.} 84:034703

\bibitem{Lortz-2007}
Lortz R, Wang Y, Demuer A, Bottger PHM, Bergk B, et~al. 2007.
\textit{Phys. Rev. Lett.} 99:187002

\bibitem{Agosta-2017}
Agosta C, Fortune N, Hannahs S, Gu S, Liang L, et~al. 2017.
\textit{Phys. Rev. Lett.} 118:267001

\bibitem{Wright-2011}
Wright JA, Green E, Kuhns P, Reyes A, Brooks J, et~al 2011.
\textit{Phys. Rev. Lett.} 107:087002

\bibitem{Mayaffre-2014}
Mayaffre H, Kramer S, Horvatic M, Berthier C, Miyagawa K, et~al. 2014.
\textit{Nature Physics} 10:928

\bibitem{Agosta-2012}
Agosta C, Jin J, Coniglio WA, Smith BE, Cho K, et~al. 2012.
\textit{Phys. Rev. B} 85:214514

\bibitem{Tanatar-2002}
Tanatar M, Ishiguro T, Tanaka H, , Kobayashi H. 2002.
\textit{Phys. Rev. B} 66:134503

\bibitem{Coniglio-2011}
Coniglio W, Winter L, Cho K, Agosta C, Fravel B, Montgomery LK. 2011.
\textit{Phys. Rev. B} 83:224507

\bibitem{Koutroulaki-2016}
Koutroulakis G, Kuhne H, Schlueter J, Wosnitza J, Brown S. 2016.
\textit{Phys. Rev. Lett.} 116:067003

\bibitem{Cho-2009}
Cho K, Smith BE, Coniglio WA, Winter LE, Agosta CC, Schlueter JA. 2009.
\textit{Phys. Rev. B} 79:220507(R)

\bibitem{Yonezawa-2008}
Yonezawa S, Kusaba S, Maeno Y, Auban-Senzier P, Pasquier C, Jérome D. 2008.
\textit{Phys. Rev. Lett.} 100:117002

\bibitem{Gurevich-2010}
Gurevich A. 2010.
\textit{Phys. Rev. B} 82:184504

\bibitem{Cho-2017}
Cho CW, Yang J, Yuan N, Shen J, Wolf T, Lortz R. 2017.
\textit{Phys. Rev. Lett.} 119:217002

\bibitem{Gloos-1993}
Gloos K, Modler R, Schimanski H, Bredl CD, Geibel C, et~al. 1993.
\textit{Phys. Rev. Lett.} 70:501

\bibitem{Modler-1996}
Modler R, Gegenwart P, Lang M, Deppe M, Weiden M, et~al. 1996.
\textit{Phys. Rev. Lett.} 76:1292

\bibitem{Yamashita-1997}
Yamashita A, Ishii K, Yokoo T, Akimitsu J, Hedo M, et~al. 1997.
\textit{Phys. Rev. Lett.} 79:3771

\bibitem{Radovan-2003}
Radovan HA, Fortune NA, Murphy TP, Hannahs ST, Palm EC, et~al. 2003.
\textit{Nature} 425:51

\bibitem{Bianchi-2003}
Bianchi A, Movshovich R, Capan C, Pagliuso PG, Sarrao JL. 2003.
\textit{Phys. Rev. Lett.} 91:187004

\bibitem{Kenzelmann-2008}
Kenzelmann M, Strassle T, Niedermayer C, Sigrist M, Padmanabhan B, et~al. 2008.
\textit{Science} 321:1652

\bibitem{BlochReview}
Bloch I, Dalibard J, Zwerger W. 2008.
\textit{Rev. Mod. Phys.} 80:885--964

\bibitem{KetterleZwierleinReview}
Ketterle W, Zwierlein M. 2008.
\textit{Rivista del Nuovo Cimento} 164:247-422

\bibitem{GRaop}
Gurarie V, Radzihovsky L. 2007.
\textit{Annals of Physics} 322:2 -- 119.
January Special Issue 2007

\bibitem{GiorginiRMP}
Giorgini S, Pitaevskii LP, Stringari S. 2008.
\textit{Rev. Mod. Phys.} 80:1215--1274

\bibitem{RSreview}
Radzihovsky L, Sheehy D. 2010.
\textit{Reports on Progress in Physics} 73:076501

\bibitem{Regal2004prl}
Regal CA, Greiner M, Jin DS. 2004.
\textit{Phys. Rev. Lett.} 92:040403

\bibitem{Zwierlein2004prl}
Zwierlein MW, Stan CA, Schunck CH, Raupach SMF, Kerman AJ, Ketterle W. 2004.
\textit{Phys. Rev. Lett.} 92:120403

\bibitem{Kinast2004prl}
Kinast J, Hemmer SL, Gehm ME, Turlapov A, Thomas JE. 2004.
\textit{Phys. Rev. Lett.} 92:150402

\bibitem{Eagles}
Eagles DM. 1969.
\textit{Phys. Rev.} 186:456--463

\bibitem{Leggett}
Leggett A. 1980.
 In \textit{{Modern Trends in the Theory
  of Condensed Matter}}, eds. A~P\c{e}kalski, J~Przystawa, vol. 115 of
  \textit{Lecture Notes in Physics}. Proceedings of the XVI Karpacz Winter
  School of Theoretical Physics, February 19 -- March 3, 1979 Karpacz, Poland,
  Berlin, Heidelberg: Springer Berlin Heidelberg

\bibitem{NSR}
Nozieres P, Schmitt-Rink S. 1985.
\textit{Journal of Low Temperature Physics} 59:195--211

\bibitem{Zwierlein06Science}
Zwierlein MW, Schirotzek A, Schunck CH, Ketterle W. 2006.
\textit{Science} 311:492--496

\bibitem{Partridge06Science}
Partridge GB, Li W, Kamar RI, Liao Ya, Hulet RG. 2006.
\textit{Science} 314:54--54

\bibitem{Shin2006prl}
Shin Y, Zwierlein MW, Schunck CH, Schirotzek A, Ketterle W. 2006.
\textit{Phys. Rev. Lett.} 97:030401

\bibitem{Navon2009prl}
Nascimb{\`e}ne S, Navon N, Chevy F, Salomon C. 2010.
\textit{New Journal of Physics} 12:103026

\bibitem{Combescot01}
Combescot R. 2001.
\textit{Europhysics Letters (EPL)} 55:150--156

\bibitem{Liu03}
Liu WV, Wilczek F. 2003.
\textit{Phys. Rev. Lett.} 90:047002

\bibitem{Bedaque03}
Bedaque PF, Caldas H, Rupak G. 2003.
\textit{Phys. Rev. Lett.} 91:247002

\bibitem{Caldas04}
Caldas H. 2004.
\textit{Phys. Rev. A} 69:063602

\bibitem{Castorina05}
Castorina P, Grasso M, Oertel M, Urban M, Zappal\`a D. 2005.
\textit{Phys. Rev. A} 72:025601

\bibitem{Sedrakian05nematic}
Sedrakian A, Mur-Petit J, Polls A, M{\"u}ther H. 2005.
\textit{Physical Review A} 72:013613

\bibitem{SRprl}
Sheehy DE, Radzihovsky L. 2006.
\textit{Phys. Rev. Lett.} 96:060401

\bibitem{Pao06}
Pao CH, Wu ST, Yip SK. 2006.
\textit{Phys. Rev. B} 73:132506

\bibitem{Son06}
Son DT, Stephanov MA. 2006.
\textit{Phys. Rev. A} 74:013614

\bibitem{Bulgac06pwavePRL}
Bulgac A, Forbes MM, Schwenk A. 2006.
\textit{Phys. Rev. Lett.} 97:020402

\bibitem{SRaop}
Sheehy DE, Radzihovsky L. 2007{\natexlab{a}}.
\textit{Annals of Physics} 322:1790 -- 1924

\bibitem{Parish07nature}
Parish MM, Marchetti FM, Lamacraft A, Simons BD. 2007{\natexlab{a}}.
\textit{Nature Physics} 3:124 EP

\bibitem{RVprl}
Radzihovsky L, Vishwanath A. 2009{\natexlab{b}}.
\textit{Phys. Rev. Lett.} 103:010404

\bibitem{Rpra}
Radzihovsky L. 2011{\natexlab{b}}.
\textit{Phys. Rev. A} 84:023611

\bibitem{Hulet1dLO}
Liao Ya, Rittner ASC, Paprotta T, Li W, Partridge GB, et~al. 2010.
\textit{Nature} 467:567 EP --

\bibitem{1d3dCrossoverPRL2016}
Revelle MC, Fry JA, Olsen BA, Hulet RG. 2016.
\textit{Phys. Rev. Lett.} 117:235301

\bibitem{ParishQuasi1dPRL2007}
Parish MM, Baur SK, Mueller EJ, Huse DA. 2007{\natexlab{b}}.
\textit{Phys. Rev. Lett.} 99:250403

\bibitem{Zhang-1998}
Zhang SC. 1998.
\textit{Journal of Physics and Chemistry of Solids} 59:1774

\bibitem{Raczkowski-2007}
Raczkowski M, Capello M, Poilblanc D, Fr\'esard R, Ole\ifmmode~\acute{s}\else
  \'{s}\fi{} AM. 2007.
\textit{Phys. Rev. B} 76:140505

\bibitem{Capello-2008}
Capello M, Raczkowski M, Poilblanc D. 2008.
\textit{Phys. Rev. B} 77:224502

\bibitem{Loder-2010}
Loder F, Kampf AP, Kopp T. 2010.
\textit{Phys. Rev. B} 81:020511

\bibitem{Wardth-2017}
W\aa{}rdh J, Granath M. 2017.
\textit{Phys. Rev. B} 96:224503

\bibitem{Wardth-2018}
W\aa{}rdh J, Andersen BM, Granath M. 2018.
\textit{Phys. Rev. B} 98:224501

\bibitem{Lee-2007}
Lee SS, Lee PA, Senthil T. 2007.
\textit{Phys. Rev. Lett.} 98:067006

\bibitem{Soto-Garrido-2014}
Soto-Garrido R, Fradkin E. 2014.
\textit{Phys. Rev. B} 89:165126

\bibitem{Wu-2007}
Wu C, Sun K, Fradkin E, Zhang SC. 2007.
\textit{Phys. Rev. B} 75:115103

\bibitem{Soto-Garrido-2015}
Soto-Garrido R, Cho GY, Fradkin E. 2015.
\textit{Phys. Rev. B} 91:195102

\bibitem{Granath-2001}
Granath M, Oganesyan V, Kivelson SA, Fradkin E, Emery VJ. 2001.
\textit{Physical Review Letters} 87:167011

\bibitem{Sikkema-1997}
Sikkema AE, Affleck I, White SR. 1997.
\textit{Phys. Rev. Lett.} 79:929

\bibitem{Zachar-2001}
Zachar O, Tsvelik AM. 2001.
\textit{Phys. Rev. B} 64:033103

\bibitem{Zachar-2001b}
Zachar O. 2001.
\textit{Phys. Rev. B} 63:205104

\bibitem{Berg-2010}
Berg E, Fradkin E, Kivelson SA. 2010.
\textit{Phys. Rev. Lett.} 105:146403

\bibitem{Cho-2014}
Cho GY, Soto-Garrido R, Fradkin E. 2014.
\textit{Phys. Rev. Lett.} 113:256405

\bibitem{Jaefari-2012}
Jaefari A, Fradkin E. 2012.
\textit{Phys. Rev. B} 85:035104

\bibitem{Huang-2018}
Huang EW, Mendl CB, Jiang HC, Moritz B, Devereaux TP. 2018.
\textit{npj Quantum Materials} 3:22

\bibitem{Verstraete-2008}
Verstraete F, Murg V, Cirac J. 2008.
\textit{Advances in Physics} 57:143

\bibitem{Corboz-2011}
Corboz P, White SR, Vidal G, Troyer M. 2011.
\textit{Phys. Rev. B} 84:041108

\bibitem{Corboz-2014}
Corboz P, Rice TM, Troyer M. 2014.
\textit{Phys. Rev. Lett.} 113:046402

\bibitem{Dodaro-2016}
Dodaro JF, Jiang HC, Kivelson SA. 2017.
\textit{Phys. Rev. B.} 95:155116

\bibitem{Jiang-2018}
Jiang HC, Weng ZY, Kivelson SA. 2018.
\textit{Phys. Rev. B} 98:140505

\bibitem{Jiang-2018b}
Jiang HC, Devereaux T. 2018.
arXiv:1806.01465

\bibitem{Xu-2018}
Xu XY, Law KT, Lee PA. 2019.
\textit{Phys. Rev. Lett.} 122:167001

\bibitem{Hartnoll-2019}
Hartnoll SA, Lucas A, Sachdev S. 2019.
{Holographic Quantum Matter}.
Cambridge, MA: The MIT Press

\bibitem{Flauger-2011}
Flauger R, Pajer E, Papanikolaou S. 2011.
\textit{Phys. Rev. D} 83:064009

\bibitem{Cremonini-2017}
Cremonini S, Li L, Ren J. 2017{\natexlab{a}}.
\textit{Journal of High Energy Physics} 2017:81

\bibitem{Cremonini-2017b}
Cremonini S, Li L, Ren J. 2017{\natexlab{b}}.
\textit{Phys. Rev. D} 95:041901

\bibitem{Cai-2017}
Cai RG, Li L, Wang YQ, Zaanen J. 2017.
\textit{Phys. Rev. Lett.} 119:181601

\bibitem{Chandra}
Chandrasekhar BS. 1962.
\textit{Applied Physics Letters} 1:7

\bibitem{Clogston}
Clogston AM. 1962.
\textit{Phys. Rev. Lett.} 9:266--267

\bibitem{Sarma}
Sarma G. 2002.
\textit{Journal of Physics and Chemistry of Solids} 24:1029

\bibitem{CarlsonReddy05}
Carlson J, Reddy S. 2005.
\textit{Phys. Rev. Lett.} 95:060401

\bibitem{RegalJinRF03}
Regal CA, Jin DS. 2003.
\textit{Phys. Rev. Lett.} 90:230404

\bibitem{Nikolic07largeN}
Nikoli{\'c} P, Sachdev S. 2007.
\textit{Phys. Rev. A} 75:033608

\bibitem{Veillette07largeN}
Veillette MY, Sheehy DE, Radzihovsky L. 2007.
\textit{Phys. Rev. A} 75:043614

\bibitem{Nishida06eps}
Nishida Y, Son DT. 2006.
\textit{Phys. Rev. Lett.} 97:050403

\bibitem{Alford}
Alford M, Bowers JA, Rajagopal K. 2001.
\textit{Phys. Rev. D} 63:074016

\bibitem{Bowers}
Bowers JA, Rajagopal K. 2002.
\textit{Phys. Rev. D} 66:065002

\bibitem{MachidaNakanishiLO1984}
Machida K, Nakanishi H. 1984.
\textit{Phys. Rev. B} 30:122

\bibitem{BurkhardtRainerLO}
Burkhardt H, Rainer D. 1994.
\textit{Annalen der Physik} 506:181--194

\bibitem{MatsuoLO}
Matsuo S, Higashitani S, Nagato Y, Nagai K. 1998.
\textit{Journal of The Physical Society of Japan} 67:280--289

\bibitem{YoshidaYipLO}
Yoshida N, Yip SK. 2007.
\textit{Phys. Rev. A} 75:063601

\bibitem{Combescot}
Mora C, Combescot R. 2005.
\textit{Phys. Rev. B} 71:214504

\bibitem{SRcomment}
Sheehy DE, Radzihovsky L. 2007{\natexlab{b}}.
\textit{Phys. Rev. B} 75:136501

\bibitem{SuSchriefferHeeger}
Su WP, Schrieffer JR, Heeger AJ. 1979.
\textit{Phys. Rev. Lett.} 42:1698--1701

\bibitem{OrsoBA2007}
Orso G. 2007.
\textit{Phys. Rev. Lett.} 98:070402

\bibitem{HuBA2007}
Hu H, Liu XJ, Drummond PD. 2007.
\textit{Phys. Rev. Lett.} 98:070403

\bibitem{YangLL}
Yang K. 2001.
\textit{Phys. Rev. B} 63:140511

\bibitem{ZhaoLiuLL}
Zhao E, Liu WV. 2008.
\textit{Phys. Rev. A} 78:063605

\bibitem{PokrovskyTalapov}
Radzihovsky L. 2012.
\textit{Physica C: Superconductivity} 481:189--206

\bibitem{Shimahara}
Shimahara H. 1998.
\textit{Journal of the Physical Society of Japan} 67:1872--1875

\bibitem{deGennesProst}
deGennes PG, Prost J. 1993.
{The Physics of Liquid Crystals}.
Oxford, UK: Oxford University Press, 2nd ed.

\bibitem{ChaikinLubensky}
Chaikin PM, Lubensky TC. 1995.
{Principles of Condensed Matter Physics}.
Cambridge, UK: Cambridge University Press

\bibitem{GP}
Grinstein G, Pelcovits RA. 1981.
\textit{Phys. Rev. Lett.} 47:856--859

\bibitem{Samokhin-2010}
Samokhin K. 2010.
\textit{Phys. Rev. B} 81:224507

\bibitem{Samokhin-2011}
Samokhin K. 2011.
\textit{Phys. Rev. B} 83:094514

\bibitem{agterberg-2007}
Agterberg DF, Kaur RP. 2007.
\textit{Phys. Rev. B} 75:064511

\bibitem{dimitrova-2007}
Dimitrova O, Feigel'man MV. 2007.
\textit{Phys. Rev. B} 76:014522

\bibitem{gorkov-rashba-2001}
Gor'kov LP, Rashba EI. 2001.
\textit{Phys. Rev. Lett.} 87:037004

\bibitem{frigeri-2004}
Frigeri PA, Agterberg DF, Koga A, Sigrist M. 2004.
\textit{Phys. Rev. Lett.} 92:097001

\bibitem{cho-2012}
Cho GY, Bardarson JH, Lu YM, Moore JE. 2012.
\textit{Phys. Rev. B} 86:214514

\bibitem{barzykin-2002}
Barzykin V, Gor'kov LP. 2002.
\textit{Phys. Rev. Lett.} 89:227002

\bibitem{Mineev-1994}
Mineev V, Samokhin K. 1994.
\textit{Zh. Eksp. Teor. Fiz.} 105:747

\bibitem{agterberg-2002}
Agterberg D. 2002.
\textit{Physica C} 387:13

\bibitem{Mineev-2008}
Mineev V, Samokhin K. 2008.
\textit{Phys. Rev. B} 78:144503

\bibitem{smidman-2017}
Smidman M, Salamon M, Yuan H, Agterberg D. 2017.
\textit{Reports on Progress in Physics} 80:036501

\bibitem{yacoby-2017}
Hart S, Ren H, Kosowsky M, Ben-Shach G, Leubner P, et~al. 2016.
\textit{Nature Physics} 13:87 EP

\bibitem{michaeli-2012}
Michaeli K, Potter AC, Lee PA. 2012.
\textit{Phys. Rev. Lett.} 108:117003

\bibitem{Fu-2008}
Fu L, Kane CL. 2008.
\textit{Phys. Rev. Lett.} 100:096407

\bibitem{santos-2010}
Santos L, Neupert T, Chamon C, Mudry C. 2010.
\textit{Phys. Rev. B} 81:184502

\bibitem{mason-2018}
Chen AQ, Park MJ, Gill ST, Xiao Y, Reig-i Plessis D, et~al. 2018.
\textit{Nature Communications} 9:3478

\bibitem{burkov-2015}
Bednik G, Zyuzin AA, Burkov AA. 2015.
\textit{Phys. Rev. B} 92:035153

\bibitem{li-haldane-2018}
Li Y, Haldane FDM. 2018.
\textit{Phys. Rev. Lett.} 120:067003

\bibitem{wang-ye-2016}
Wang Y, Ye P. 2016.
\textit{Phys. Rev. B} 94:075115

\bibitem{He-2008}
He RH, Tanaka K, Mo SK, Sasagawa T, Fujita M, et~al. 2008.
\textit{Nature Physics} 5:119 EP 

\bibitem{Vishik-2010}
Vishik IM, Lee WS, He RH, Hashimoto M, Hussain Z, et~al. 2010.
\textit{New Journal of Physics} 12:105008

\bibitem{leeWS-2007}
Lee WS, Vishik IM, Tanaka K, Lu DH, Sasagawa T, et~al. 2007.
\textit{Nature} 450:81

\bibitem{renner-1998}
Renner C, Revaz B, Genoud JY, Kadowaki K, Fischer {\O}. 1998.
\textit{Physical Review Letters} 80:149

\bibitem{bergandaltman}
Berg E, Altman E. 2007.
\textit{Phys. Rev. Lett.} 99:247001

\bibitem{dessau}
Parham S, Li H, Nummy TJ, Waugh JA, Zhou XQ, et~al. 2017.
\textit{Phys. Rev. X} 7:041013



\bibitem{hash14b}
Hashimoto M, Nowadnick EA, He RH, Vishik IM, Moritz B, et~al. 2014.
\textit{Nature Materials} 14:37

\bibitem{Dai-2017}
Dai Z, Lee PA. 2017.
\textit{Phys. Rev. B} 95:014506

\bibitem{Kapitulnik-2019}
Kapitulnik A, Kivelson SA, Spivak B. 2019.
\textit{Rev. Mod. Phys.} 91:011002

\bibitem{Zelli-2012}
Zelli M, Kallin C, Berlinsky AJ. 2012.
\textit{Phys. Rev. B} 86:104507

\bibitem{Vafek-2013}
Wang L, Vafek O. 2013.
\textit{Phys. Rev. B} 88:024506

\bibitem{Norman-2018}
Norman MR, Davis JCS. 2018.
\textit{Proceedings of the National Academy of Sciences} 115:5389

\bibitem{Kacmarcik-2018}
Kacmarcik J, Vinograd I, Michon B, Rydh A, Demuer A, et~al. 2018.
\textit{Phys. Rev. Lett.} 121:167002

\bibitem{Li-2010}
Li L, Wang Y, Komiya S, Ono S, Ando Y, et~al. 2010.
\textit{Phys. Rev. B} 81:054510

\bibitem{Ong-2016}
Yu F, Hirschberger M, Loew G, Lawson B, Asaba T, et~al. 2016.
\textit{PNAS} 113:12667

\bibitem{Lee-2018a}
Lee PA. 2019.
\textit{Phys. Rev. B} 99:035132


\bibitem{Blackburn-2019}
Blackburn E. 2013.
{Talk at the 2019 Gordon Reserach Conference (Les Diablerets, Switzerland)}

\bibitem{Peng-2016}
Peng YY, Salluzzo M, Sun X, Ponti A, Betto D, et~al. 2016.
\textit{Phys. Rev. B} 94:184511

\bibitem{Chaix-2017}
Chaix L, Ghiringhelli G, Peng YY, Hashimoto M, Moritz B, et~al. 2017.
\textit{Nature Physics} 13:952--956

\bibitem{Stengen-2013}
Stengen Z, Han SJ, Wu J, Pramanik AK, {H\"ucker} M, et~al. 2013.
\textit{Phys. Rev. B} 87:064509

\end{thebibliography}

\end{document}